\documentclass[article,  
showkeys,
superscriptaddress,
 amsmath,amssymb,
 aps, physrev,
prx, twocolumn
]{revtex4-2}

\usepackage{graphicx,color}
\usepackage{dcolumn}
\usepackage{bm}
\usepackage{float}
\usepackage[dvipsnames]{xcolor}

\usepackage[draft=false, linktoc=none]{hyperref}
\hypersetup{
    colorlinks=true,    
    urlcolor=blue,      
    linkcolor=black,    
    citecolor=blue,    
    filecolor=black,    
    menucolor=black     
}

 \makeatletter

 \def\section{%
 \@startsection{subsection}{1}{\z@}%
 {1.5ex plus .5ex minus .2ex}%
 {0.8ex plus .2ex}%
 {\normalfont\bfseries\large\raggedright}}

 \def\subsection{%
 \@startsection{subsection}{2}{\z@}%
 {1.2ex plus .3ex minus .2ex}%
 {0.5ex plus .1ex}%
 {\normalfont\normalfont\bfseries\raggedright}}

 \def\subsubsection{%
 \@startsection{subsubsection}{3}{\z@}%
 {1.0ex plus .2ex minus .2ex}%
 {0.4ex plus .1ex}%
 {\normalfont\small\bfseries\raggedright}}

 \def\bibsection{%
  \@startsection{subsection}{1}{\z@}%
  {1.5ex plus .5ex minus .2ex}%
  {0.8ex plus .2ex}%
  {\normalfont\bfseries\large\raggedright}%
  *{References}}

 \makeatother
 
\usepackage{orcidlink}
\usepackage{soul} 

\begin{document}

\title{Tunable annealing of defects in self-aligning active crystals}

\author{Kanka Ghosh\,\orcidlink{0000-0003-1360-7699}}
\email{kanka.ghosh@ovgu.de}
 \affiliation{Otto-von-Guericke-Universit\"at Magdeburg, Institut f\"ur Physik, Universit\"atsplatz 2, 39106 Magdeburg, Germany}
\author{Andreas M. Menzel\,\orcidlink{0000-0003-0713-4979}}%
 \email{a.menzel@ovgu.de}
\affiliation{Otto-von-Guericke-Universit\"at Magdeburg, Institut f\"ur Physik, Universit\"atsplatz 2, 39106 Magdeburg, Germany}%

\begin{abstract} 

Obtaining defect-free single-crystals is of crucial importance in soft matter and materials science, spanning both fundamental interests and practical applications. Active crystals, encompassing diverse dense systems from living organisms to artificial metamaterials, can offer a promising platform for autonomous structural reorganization. We demonstrate that increasing spatiotemporal orientational fluctuations around the mean polarized state can enhance crystallographic order in active crystals. Poly-to-single-crystal transformations in defect-rich, self-aligned active Brownian crystals with coupled translational--rotational dynamics are investigated. We show that long-lived persistence with slow alignment controls the defect-annealing capability. By disentangling activity magnitude from associated timescales, we reveal activity-assisted annealing. Tuning the persistence length enables overcoming thermal energy barriers. Simultaneously adjusting persistence and alignment timescales can tune relaxation from `fast' to `slow', yielding defect-rich and defect-free crystals, respectively. Furthermore, active crystals of point particles are found to exhibit similar defect annealing when flocking is delayed by tuning the timescale of the velocity relaxation towards the polarized state. These findings suggest that the mechanism may extend to a broader class of active crystals. Our results intersect statistical physics of living soft matter with the mechanics of defects and microstructures, proposing a nonequilibrium route towards reconfigurable materials.
\end{abstract}

\maketitle

\section*{Introduction} \label{sec1}

\noindent From poly-crystalline classical solids to colloidal crystals, defects substantially affect material properties. Therefore, the ability to control and tune them is of fundamental and technological importance. Structural defects govern strength and ductility in materials \cite{hull2011introduction}, while their dynamics also influences pattern formation and self-organization \cite{jangizehi2020defects}. However, in many applications, the elimination of defects is highly desirable. For example, defects such as grain boundaries, often act as scattering centers for electrons and phonons, thereby impeding electrical \cite{yu2011control} and thermal \cite{klemens1955scattering} transport  in materials. In soft matter systems, defect-free colloidal crystals are particularly valuable as photonic crystals \cite{vlasov2001chip, vlasov2000manifestation, ye2001self}. Grain boundaries and structural imperfections disrupt long-range periodicity, increase optical scattering, and degrade performance of optoelectronic devices \cite{wong2003colloidal, rengarajan2005effect}.

To overcome thermal kinetic barriers, self-propulsion provides a local nonequilibrium drive. At high densities \cite{julian2012crystallization}, supported by hydrodynamic interactions \cite{singh2016universal, petroff2015fast}, or even at low densities \cite{jeong2025crystallization}, such active systems can combine crystalline order with dynamical softness. These active crystals encompass a diverse pool of systems across a range of scales, spanning biological living crystals \cite{petroff2015fast, petroff2018nucleation, tan2022odd} and synthetic soft materials \cite{palacci2013living, palacci2014light, mognetti2013living, singh2017non, ramananarivo2019activity, vutukuri2020light, tanuku2026active, mauleon2025slipping, massana2024multiple} to macroscopic self-propelled hard disks \cite{briand2016crystallization, briand2018spontaneously} and inanimate, human-made systems, such as artificial swarm robots \cite{wang2024robo, vansaders2021sculpting}. They offer a unique nonequilibrium platform to realize structural relaxation and defect reorganization. For example, living chiral crystals of starfish embryos, during their developmental phase, often modulate between higher and lower crystalline order, involving defect proliferation and annihilation \cite{tan2022odd}. Their autonomy of eliminating structural defects could make active crystals promising candidates for high-performance materials that can self-heal without conventional thermal annealing processing. The importance of understanding the underlying mechanisms and controlling activity-driven defect healing is therefore two-fold. Firstly, such mechanisms could provide a framework for tuning defects in view of performance. Secondly, beyond technological significance, new insight is generated into how dense living systems autonomously reorganize their structures far from equilibrium.

Active solids \cite{henkes2011active, menzel2013traveling, menzel2014active, giavazzi2018flocking, praetorius2018active, maitra2019oriented, scheibner2020odd, caprini2025odd, shi2023extreme, caprini2023entropons, huang2025anomalous, martinet2025emergent, shee2024emergent, keta2025long, binysh2026more}, apart from mean-field descriptions \cite{menzel2013traveling, menzel2014active, ophaus2021two, holl2025motility}, are frequently investigated using elastic-network models \cite{ferrante2013collective, woodhouse2018autonomous, kinoshita2025collective, baconnier2022selective, huang2021alignment, baconnier2023discontinuos, baconnier2024noise, veenstra2025adaptive, welker2025lattice, hernandez2024model, cocconi2025mechanical, baconnier2025reentrant, lie2025turbulent}. Here, elastic response is typically imposed by a prescribed connectivity of the particles, for instance, in terms of permanent elastic springs. Yet, such solids can also be realized as emergent active crystals, formed from unconnected particles that obey given mutual particle interactions \cite{paliwal2020role, caprini2020active, nath2026activated, das2024flocking, caprini2023inhomogeneous, musacchio2026flocking, briand2018spontaneously, briand2016crystallization}. Resulting lattices can accommodate both elastic deformation and plastic rearrangements through defect reorganization. Such structural defects, including dislocations and disclinations, had been analyzed, for instance, in crystals of self-propelled hard disks \cite{briand2018spontaneously}, active Brownian crystals \cite{paliwal2020role, nath2026activated}, and within the dense solid phases of active systems undergoing motility-induced phase separation \cite{redner2013structure, digregorio2022unified, caprini2020hidden}. At high densities, topological structural defects are found to crucially impact the melting behavior and elastic response in active Brownian particle crystals (ABP crystals) \cite{paliwal2020role}, indicating a melting scenario differing from the equilibrium BKTHNY theory \cite{berezinskii1971destruction, kosterlitz1973ordering, halperin1978theory, nelson1979dislocation, young1979melting}. When such motile defects are seen to move ballistically in a chiral host medium, they lead to dynamical poly-crystalline `whorl' states \cite{bililign2022motile}. Dense active systems have been shown to exhibit turbulent dynamics during the melting transition from the solid to the hexatic phase \cite{yang2026active}. Departure from BKTHNY melting is also discussed recently in active solids with odd elasticity and micropolar coupling \cite{davoyan2026topological}.

Activity-induced annealing has been discussed earlier, although in different contexts of active doping in an otherwise passive colloidal crystals \cite{van2016fabricating, van2016removing, bechinger2016active, ramananarivo2019activity}. Active dopants within a passive crystal were seen to initiate localized melting at grain boundaries, leading to a defect-free single-crystalline structure. A high-density defect-free hexatic phase was reported in ABP crystals \cite{paliwal2020role}, which was later argued to be a crystal phase  \cite{shi2023extreme}. Adopting a generalized Vicsek \cite{vicsek1995novel} model with mutual alignment and short-ranged repulsions, the poly-to-single-crystal transformation was studied in polar active crystals \cite{weber2014defect}. While strong alignment promotes a collectively migrating poly-crystalline state, strong repulsive interaction is found to favor a nonpolar, defect-free single-crystal, suggesting an antagonistic relation between flocking and defect annihilation. However, these conditions are not strictly necessary. A poly-to-single-crystal transformation has also been observed in flocking Wigner crystals with coupled rotational and translational dynamics of individual units, despite the presence of turning-away torques \cite{das2024flocking}. In that case, defect annealing was attributed to enhanced self-propulsion velocity. A recent study \cite{pingali2025orientational} showed that hydrodynamic interactions in self-driven spinner systems can spontaneously promote crystallization, while activity can progressively improve its structural order. 

Despite these efforts, the mechanisms underlying defect annihilation and poly-to-single-crystal ordering in active crystals, with coupled translational and rotational dynamics, remain elusive. This even concerns situations of particle orientation redirected by translational motion through so-called self-alignment \cite{fersula2024self, baconnier2025self, kammeraat2026active}. Elucidating defect annealing requires to disentangle the distinct roles of activity magnitude, persistence of active motion, and their interplay with the alignment dynamics, which is largely unexplored.

In this article, we explore the poly-to-single-crystalline transformation in defect-rich, self-aligned active crystals, modeled as active Brownian particles (ABP) with coupled translational–orientational dynamics. We observe collectively migrating crystals that remain dynamically fluctuating. Long-lived persistence combined with slow alignment appears to primarily control the defect-annealing capability of an active crystal. Quite remarkably, we observe that the residual spatiotemporal fluctuations, within a globally flocking crystal, facilitate structural reorganization and defect annihilation. Using protocols of varying the persistence length at fixed persistence time or vice versa, we disentangle the roles of activity magnitude and underlying timescales. Our two protocols generate notably distinct translational and orientational fluctuations both prior and after the onset of flocking. These fluctuations further manifest themselves at the scale of the grain-boundaries through predominantly activity-assisted migration. Occasionally, it is accompanied by grain-boundary sliding and coupling prior to defect annihilation. We interpret this behavior as activity-driven annealing, in which a threshold active force provides effective `heating' by driving particles out of metastable states. Simultaneously, persistence and alignment timescales provide tunable `cooling' that mimics either a `fast' or `slow' quench. It leads to defect-rich or defect-free final states, respectively.

Beyond clarifying collective phenomena in dense living crystals, activity-driven defect dynamics could provide new routes for controlled defect manipulation and healing, for instance, in colloidal crystals and swarm metamaterials \cite{vansaders2021sculpting}. This could further have potential implications for atomistic materials, where structural defects strongly influence functional properties. From a phenomenological point of view, our results could potentially address a broader class of flocking crystals, irrespective of alignment mediated by turn-towards or turn-away torques. As a counterexample, we present results of a collectively migrating active crystal composed of point particles, in the complete absence of rotational couplings. We show that defect annealing can still be tuned by varying the dissipative timescale of velocity relaxation towards the polarized state.

\section*{Results}\label{sec2}

\begin{figure*}[!htb]
\centering
\includegraphics[width=\linewidth]{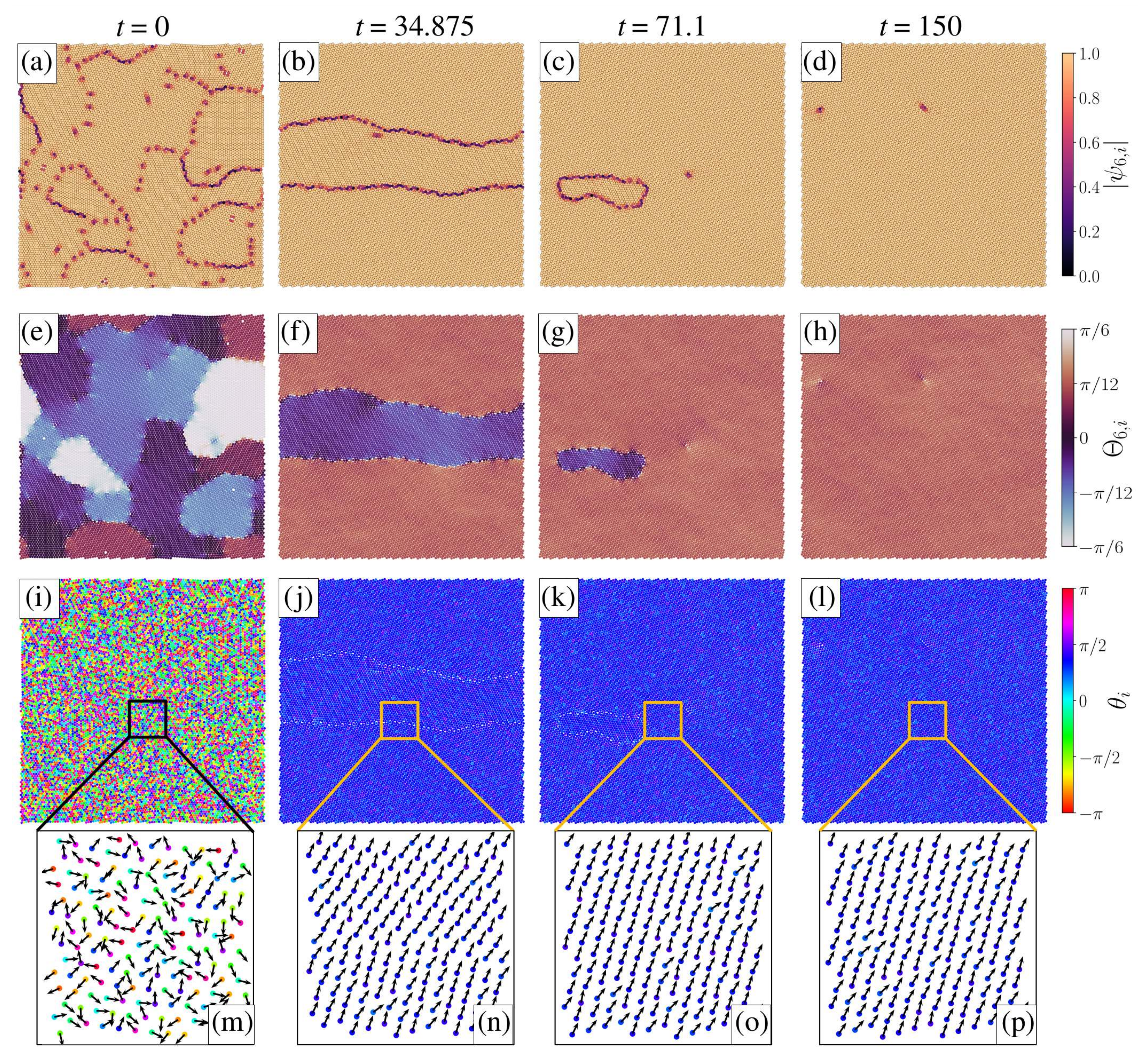}
\caption{\textbf{Progressive transformation from poly- to single-crystalline order in ABP crystals}. Snapshots at different time instances are shown from left to right in each row. Time is expressed in reduced Lennard-Jones units. The domains containing atoms are colored by \textbf{(a--d)} local hexatic order $|\psi_{6, i}|$, \textbf{(e--h)} local grain orientation $\Theta_{6,i}$, and \textbf{(i--l)} individual particle orientations $\theta_{i}$. \textbf{(m--p)} presents zoomed-in regions, respectively, from \textbf{(i--l)} with explicit arrows attached with active particles showing corresponding orientations of active forcing during these progressive snapshots. These simulations are performed on a domain of dimension $100\sigma \times 100\sigma$ at a fixed packing fraction $\phi$ = 0.87. Activity is set to Pe = 1000 (persistence length $l_{p}\approx 333$) with persistence time $\tau_{p} (= 1/D_{r}$)=3.33.}
\label{fig:poly_single}
\end{figure*}

\subsection*{Defect annihilation in self-aligned active poly-crystals} \label{subsec1}

\noindent We start from a two-dimensional system of passive Brownian particles of diameter $\sigma$. Initially, they are arranged in a hexagonal lattice, with a lattice constant of 1.02$\sigma$. The packing fraction is set to $\phi= 0.87$ throughout. Periodic boundary conditions are employed along both $x$- and $y$-directions. See Methods for further details, including the units of rescaling to dimensionless quantities for numerical analysis. 

Defect-rich Brownian passive crystals are numerically prepared by instantaneously quenching the system from higher ($T = 10$) to the desired lower ($T = 0.025$) temperature (in reduced units, see Methods) which is further equilibrated. These systems are activated by assigning a self-propulsion force to each particle, together with associated deterministic torques that affect the orientational state of each such active force. In this way, we initially obtain defect-rich self-aligned active Brownian particle crystals (ABP crystals).  

The overdamped Langevin equations of motion for the positions $\mathbf{r}_i$ and angles of orientation $\theta_{i}$ for each individual active Brownian particle (ABP) $i$ ($i=1,...,N$) are given by
\begin{eqnarray}\label{eq:trans}
    \gamma_{t} \dot{\mathbf{r}}_{i} &=&{} -\sum_{j\neq i}\bm{\nabla}_{i}U(r_{ij}) + f
    ^{a} \hat{\mathbf{e}}_{i} + \gamma_{t} \sqrt{2D_{t}} \bm{\xi}_{i}  \\
    \label{eq:rot}
    \gamma_{r}\dot{\theta}_{i} &=& \mathbf{T}_{i} \cdot \hat{\mathbf{e}}_{z} + \gamma_{r}\sqrt{2D_{r}} \xi_{\theta,i}
\end{eqnarray}
Here, as an interaction potential  $U(r_{ij})$ between the ABPs we set the repulsive Weeks-Chandler-Andersen (WCA) potential, see Methods, causing the conservative force $\mathbf{F}_{i}^{c} =-\sum_{j\neq i}\bm{\nabla}_{i}U(r_{ij})$. $\gamma_t$ and $\gamma_r$ are the translational and rotational friction coefficients, respectively. For each particle $i$, the active force is $\mathbf{F}_{i}^{a}$ = $f
^{a}\hat{\mathbf{e}}_{i}$ = $\gamma_{t} v_{0} \hat{\mathbf{e}}_{i}$, where $v_0$ is the magnitude of the active velocity in the absence of any other forcing. $\hat{\mathbf{e}}_{i}$ = (cos$\theta_i$, sin$\theta_i$) parametrizes the orientation of active forcing of particle $i$. Initially, each particle is assigned a random orientation $\theta_i$. Self-alignment is then introduced through a deterministic torque $\mathbf{T}_{i}$ on each individual particle, proportional to the velocity magnitude \cite{baconnier2025self, baconnier2022selective, baconnier2024noise, kinoshita2025collective, musacchio2026flocking}. It is represented as $\mathbf{T}_{i} = \zeta (\hat{\mathbf{e}}_{i} \times \mathbf{v}_{i})$, which, for positive values of $\zeta$ (set to 0.01 throughout our simulations), tends to align the orientation $\hat{\mathbf{e}}_{i}$ towards the direction of velocity $\mathbf{v}_i$. 
$\bm{\xi}_{i}$ and $\xi_{\theta,i}$ are Gaussian white noises with zero mean and equilibrium 
variance.
Thus, the translational and rotational diffusion coefficients satisfy the Stokes-Einstein relation $D_{t,r} = k_{B}T/\gamma_{t,r}$, involving Boltzmann's constant $k_B$. Their precise values depend on the nature of the Brownian particles. To reduce the parameter space, we assume their coupling \cite{brenner1967coupling, paliwal2020role, li2025interactions} $D_{r} = 3D_{t}/\sigma^2$. This is the relation that results for spheres in three-dimensional Stokes flow.  

Based on these quantities, the rotational persistence time is defined via $\tau_{p} = 1/D_r$. 
We quantify the degree of activity using the P\'eclet number Pe = $f^ {a}\sigma/k_{B} T $, and we modify Pe via controlling the active force magnitude $f^{a}$ within the simulations. The persistence length $l_p$ is related to persistence time $\tau_p$ via $l_{p} = v_{0} \tau_{p} = {\sigma}\,\mathrm{Pe}/{3}$, and thus is proportional to Pe. Moreover, the length scale $l_a= \gamma_{r}/\zeta$ of self-alignment via the torques $\mathbf{T}_i$ is chosen to be very small compared to the persistence length $l_p$ throughout our studies. See Methods for more details regarding the model and simulation protocols.

\begin{figure*}[!htb]
\centering
\includegraphics[width=0.95\linewidth]{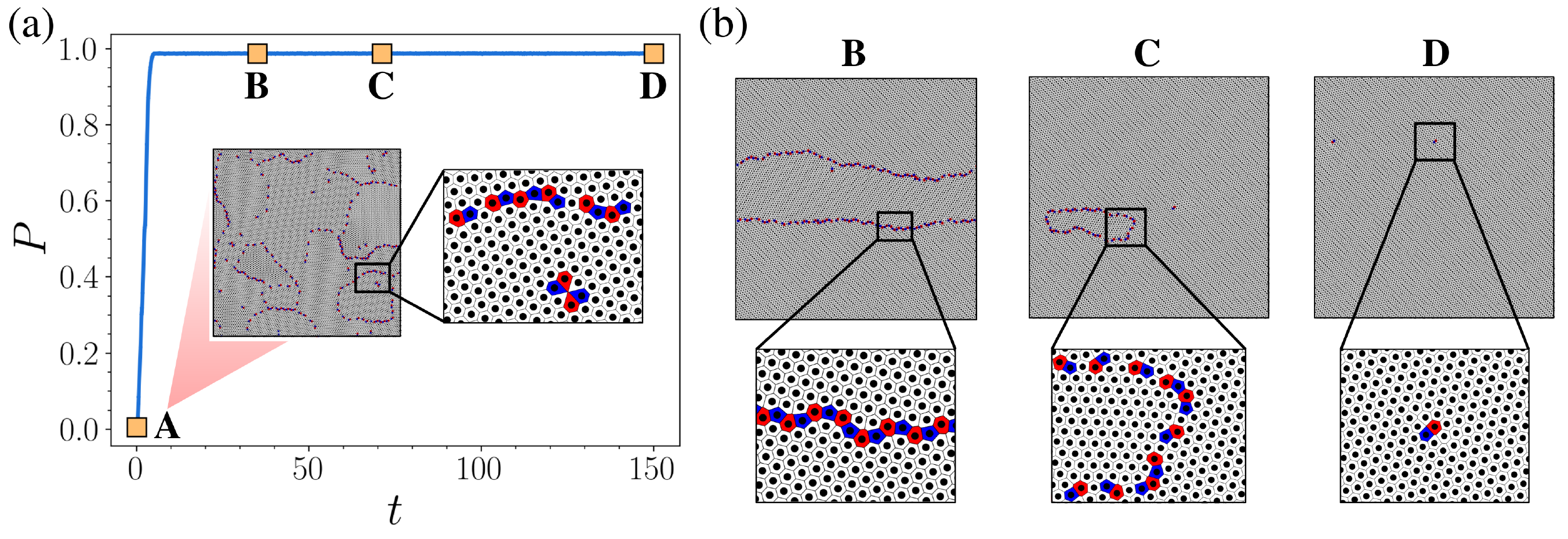}
\caption{\textbf{Structural positional reorganization towards a single-crystal continues beyond saturation of global polar orientational order}. \textbf{(a)} The global polar orientational order parameter $P$ is presented for the system in Fig.~\ref{fig:poly_single} as a function of time $t$. Four yellow squares $\textbf{A}$ ($t$=0), $\textbf{B}$ ($t$=34.875), $\textbf{C}$ ($t$=71.1), and $\textbf{D}$ ($t$=150) along the polarization curve relate to the four columns (times) in Fig.~\ref{fig:poly_single}. An inset shows the particle configuration at time $t = 0$,  with particles in a zoomed-in fragment colored according to their local coordination number inferred from Voronoi tessellation. \textbf{(b)} Voronoi-neighbor coordination maps for the three sequential snapshots labeled $\textbf{B}$, $\textbf{C}$, and $\textbf{D}$ in panel (a). The gradual annihilation of defects is evident from the evolution of the defect network within the configurations and their corresponding magnified views. Five-fold and seven-fold defect sites are shown in blue and red, respectively.}
\label{fig:polar_voronoi}
\end{figure*}

For $l_{a}\ll l_{p}$, the particles rapidly align towards their velocities at a distance much shorter than the persistence length of translational motion. This allows us to focus primarily on the role of persistence in structural reorganization. Also, the velocity dependence of the torques $\mathbf{T}_i$ indicates that the effective alignment is not controlled primarily by $\zeta$. Instead, it emerges from the particle dynamics themselves via velocity variations within the system. In fact, ABP crystals encounter motility-induced phase separation (MIPS) \cite{buttinoni2013dynamical, cates2015mips} at high packing fraction due to activity. MIPS is a manifestation of the simultaneous presence of crowded dense and dilute gas-like phases within the same system, therefore inducing large velocity variation. Thus, the  self-alignment rule enforces higher alignment torques on active particles with higher velocities. Interestingly, self-alignment has been found to suppress MIPS, both in the inertial and in the overdamped regime \cite{musacchio2025self}, thereby providing a realizable platform for studying activity-induced defect reorganization.

In Fig.~\ref{fig:poly_single}, we show a typical time evolution of a self-aligning ABP poly-crystal spontaneously transforming into an almost defect-free single-crystal due to activity. Corresponding activity is set to Pe = 1000 (persistence length $l_{p} \approx 333$) and persistence time to $\tau_p$ = 3.33. The system encounters MIPS upon introducing activity, which is suppressed via self-alignment. Three out of four columns of Fig.~\ref{fig:poly_single}, corresponding to $t =34.875$, $t = 71.1$, and $t = 150$, are time instances after self-alignment started to suppress MIPS. Consistently \cite{musacchio2026flocking}, as persistence length $l_p$ dominates over alignment length $l_a$, we eventually obtain a traveling crystal phase. In this typical state crystalline order is maintained, yet its constituent particles exhibit collective flocking \cite{menzel2013traveling, menzel2014active}.

This structural transformation is presented using the local hexatic order parameter defined as $\psi_{6, i}$ = $ \sum_{j=1}^{N_{i}} e^{i6\theta_{ij}}/{N_{i}}$, where $\theta_{ij}$ is the angle between the $x$-direction and the bond vector joining particle $i$ with its neighbor $j$ out of its $N_{i}$ nearest neighbors identified via Voronoi construction. The corresponding local grain orientation is obtained from the phase of the hexatic order parameter using $\Theta_{6,i}$ = $\arg(\psi_{6,i})/6$, which varies from $-\pi/6$ to $\pi/6$ due to hexagonal symmetry. See Methods for details regarding the Voronoi construction and the calculation of the hexatic order parameter.

$|\psi_{6, i}|$-maps show that defects and grain boundaries correspond to neighborhoods that substantially deviate from perfect hexagonal arrangements, implying $|\psi_{6, i}| < 1$, see Fig.~\ref{fig:poly_single}(a). Contrarily, inside grains nearly perfect hexagonal structural order is maintained, meaning $|\psi_{6, i}| \approx 1$. Different grain orientations of the initial configuration is highlighted in Fig.~\ref{fig:poly_single}(e). After MIPS suppression, formation of single-crystalline order is observed, with gradual annihilation of defects and collapse of isolated misoriented grains. The process is captured via $|\psi_{6, i}|$ in Fig.~\ref{fig:poly_single}(b--d) and $\Theta_{6,i}$ in Fig.~\ref{fig:poly_single}(f--h).

Tracking the same time evolution in individual orientation space is illustrated in Fig.~\ref{fig:poly_single}(i--l). The starting configuration of a defect-rich active crystal possesses randomly oriented constituents, see Fig.~\ref{fig:poly_single}(i) and its zoomed-in part in Fig.~\ref{fig:poly_single}(m). However, at $t = 34.875$, see Fig.~\ref{fig:poly_single}(j), and later, all individual particles within the crystal are globally aligned on average, see the maps of individual particle orientations $\theta_{i}$ in Fig.~\ref{fig:poly_single}(j--l). Explicit depictions of arrows pointing into the directions of individual active forcing further illustrate the scenario, see Fig.~\ref{fig:poly_single}(n--p). These figures indicate collective motion while retaining the underlying crystalline order.

Therefore, Fig.~\ref{fig:poly_single}(i--l) defies a simple, direct correlation between orientational alignment of individual particles and annihilation of positional defects. Only the latter leads to single-crystalline order in active crystals. Specifically, structural reorganization towards almost defect-free states continues even after the mean orientation reaches an approximate saturation, which implies a collectively migrating active crystal. 

The degree of collective orientational alignment is quantified through the global polar order parameter $P = |\sum_{i=1}^{N}\hat{\mathbf{e}}_{i}|/N $. Figure~\ref{fig:polar_voronoi}(a) locates the same four instances depicted in Fig.~\ref{fig:poly_single} along the curve of evolving polarization, see the yellow squares A, B, C, and D in Fig.~\ref{fig:polar_voronoi}(a). Alongside, these four configurations are now presented through their Voronoi-neighbor coordination maps. They show the defect networks in the initial state in Fig.~\ref{fig:polar_voronoi}(a) and their gradual annihilation in the three sequential snapshots in Fig.~\ref{fig:polar_voronoi}(b). Particularly, grain boundaries comprised of dislocations are seen to evolve as a loop-like structure. Eventually, it collapses with only very few isolated dislocations remaining. Supplementary Note~1 and Supplementary Figure~1 show an example corresponding to Pe = 640 ($l_{p} \approx 213$). There, a single-crystalline phase is recovered, devoid of any remaining defects. (For the transformed single-crystalline structure, stability is discussed in Supplementary Note~2 and Supplementary Figure~2, effects of larger system size in Supplementary Note~3 and Supplementary Figures~3 and 4, their crystalline character using corresponding orientational bond-order correlation functions in Supplementary Note~4 and Supplementary Figure~5, together with the additional Refs.~\onlinecite{mermin1966absence,hohenberg1967existence,mermin1968crystalline,klamser2018thermodynamic}.)

At this point, it is tempting to presume an annealing-like mechanism behind such defect annihilation. Although the bath temperature $T$ remains fixed, increasing activity $f^a$ may imply such a scenario. Particles can escape local kinetic barriers that would remain inaccessible in the passive crystal. This mimics the `heating' counterpart of annealing. However, the crucial part would be to identify the associated `cooling' mechanism that leads to reorganization and relaxation of the crystal to a defect-free structural state. This cooling would operate even after global orientational order saturates, as is evidenced in Fig.~\ref{fig:polar_voronoi}.

\begin{figure*}[!htb]
\centering
\includegraphics[width=\linewidth]{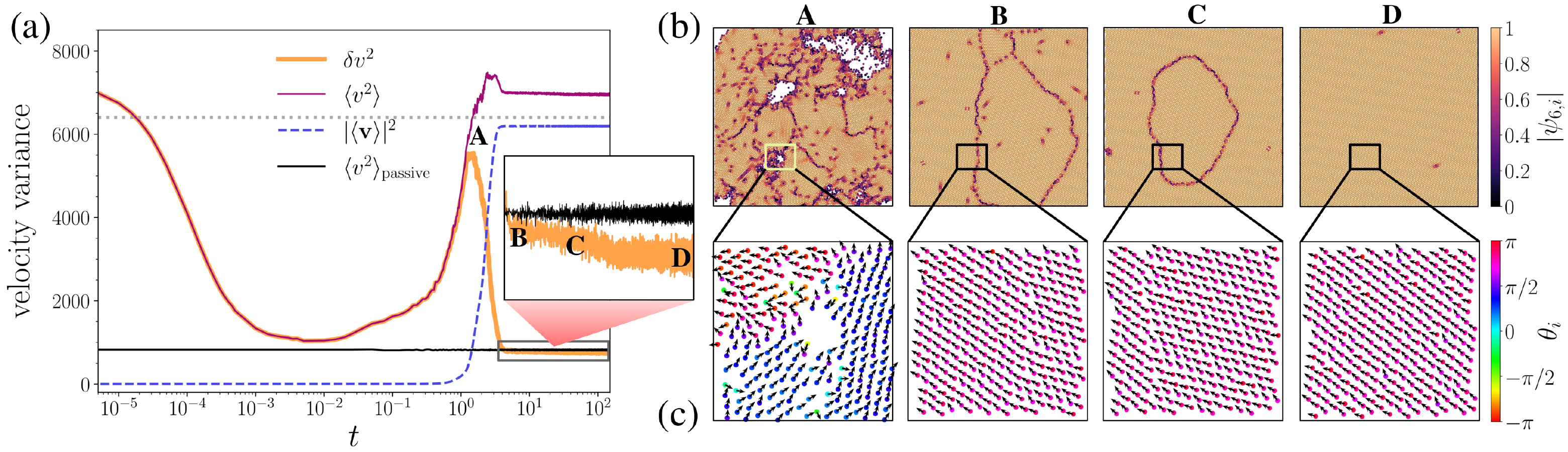}
\caption{\textbf{A typical demonstration of the time evolution of the velocity variance during poly-to-single-crystal transformation.} \textbf{(a)} Time evolution of the velocity variance $\delta v^{2}$(=$\sum_{i=1}^N (\mathbf{v}_{i}-\langle \mathbf{v} \rangle)^{2}/N $), presented together with the contributions from the mean squared velocity $\langle v^2 \rangle$($=\sum_{i=1}^N |\mathbf{v}_i|^2/N$), the squared mean velocity $ |\langle \mathbf{v} \rangle|^{2}$($ = | \sum_{i} \mathbf{v}_{i}/N |^{2}$ ), and the average squared velocity $\langle v^2 \rangle_\mathrm{passive}$ of the particles corresponding to the passive crystal at the same value of $D_t$. We indicate the magnitude of $v_{0}^2$ by the gray dotted line. This example situation corresponds to Pe = 800 ($l_{p} \approx 266.67$) and $\tau_{p} =3.33$. The inset shows a magnified view of the evolving velocity variance beyond the major relaxation process (orange) together with its passive counterpart (black). \textbf{(b)} Particle configurations corresponding to different time instances $t = 1.4$ (\textbf{A}), $t =5$ (\textbf{B}), $t =16.15$ (\textbf{C}), and $t = 150$ (\textbf{D}), marked along the velocity variance curve in \textbf{(a)}. The panels display the local hexatic order $|\psi_{6, i}|$. \textbf{(c)} Particle orientations $\theta_{i}$ of zoomed-in parts of the configurations shown in \textbf{(b)} are presented using colormaps and additional arrows.}
\label{fig:vel_fluc_example}
\end{figure*}

\subsection*{Tuned annealing: controlling persistence length versus persistence time} \label{subsec2}

\noindent To elucidate the microscopic mechanism underlying defect reorganization and annihilation during the proposed active annealing process, here we systematically investigate the role of persistence. To this end, we consider two complementary protocols that independently tune the temporal and spatial characteristics of activity: (i) a persistence length-controlled protocol, in which $\tau_{p}$ (= 1/$D_r$) is kept fixed while $l_p$ (= $\sigma$Pe/3) is varied; and (ii) a persistence time-controlled protocol, in which the persistence length $l_p$ is held constant while the persistence time $\tau_{p}$ is varied by changing the friction coefficients $\gamma_{t,r}$. See also Methods. This separation enables us to disentangle the distinct roles of the spatial extent of persistent active motion and temporal memory in driving defect annealing.

First, we focus on the translational dynamics of the active crystal. We reveal the consequences of the two distinct protocols. For this purpose, we estimate the velocity fluctuations by calculating the time evolution of the velocity variance during poly-to-single-crystal transformation. Frequently, the translational noise is neglected ($D_{t} = 0$) in studies related to dense overdamped interacting active systems \cite{caprini2020hidden, caprini2020active}, but here we explicitly keep it. We calculate the velocity variance $\delta v^{2} = \sum_{i=1}^N (\mathbf{v}_{i}-\langle \mathbf{v} \rangle)^{2}/N $, where $\langle\mathbf{v}\rangle=\sum_{i=1}^N \mathbf{v}_{i}/N$. Each particle velocity is obtained from the conservative interaction between the particles and the active forces via $\mathbf{v}_{i} = (\mathbf{F}_{i}^{\mathrm{c}}+\mathbf{F}_{i}^{a})/\gamma_{t}$. This approach avoids suppression of the features of velocity fluctuations in overdamped systems in the presence of translational noise. Still, it accounts for their effects on the dynamics.

Figure~\ref{fig:vel_fluc_example} demonstrates an example situation with Pe = 800 ($l_{p} \approx 266.67$) and $\tau_{p}$ = 3.33 of such velocity fluctuations during the poly-to-single-crystal transformation of  self-aligning active crystals. At short times, active forcing is expected to dominate. Initial velocity fluctuations show slightly elevated values $\gtrsim$ $v_{0}^2$. The excess contribution can be attributed to interaction forces. After a transient period, frequent collisions due to repulsive interactions at high densities constrain particle motion. Thereby, they suppress the overall velocity fluctuations within the crystalline structure. Meanwhile, at high Pe (hence high $l_p$), active persistence produces phase separated dense and dilute phases (MIPS), corresponding to low- and high-velocity regions, respectively. 
Eventually, this leads to a peak in the velocity variance averaged across the system, see point \textbf{A} in Fig.~\ref{fig:vel_fluc_example}(a). 

Separating the contributions to $\delta v^2$ reveals that, up to the peak \textbf{A}, the velocity fluctuations are dominated by the mean squared velocity $\langle v^2\rangle$, see the solid magenta curve in Fig.~\ref{fig:vel_fluc_example}(a). Shortly after the peak, self-alignment-assisted collective motion begins to develop. It leads to a gradual increase in the collective contribution $|\langle\mathbf{v}\rangle|^2$, see the blue dashed curve in Fig.~\ref{fig:vel_fluc_example}(a). Consistently \cite{musacchio2025self}, it saturates towards $v_{0}^2$, where the small deficit from $v_{0}^2$ accounts for an incomplete perfect polarization due to ongoing collisions between the particles. This onset of collective motion marks the start of post-alignment relaxation dynamics of the migrating active crystals. Simultaneously, the velocity variance starts to decrease towards its passive variance baseline, see the black curve in Fig.~\ref{fig:vel_fluc_example}(a). Eventually, $|\langle \mathbf{v} \rangle|^2$ appears to saturate, indicating a steady-state polarization. Interestingly, beyond the period of major alignment, the velocity variance drops below that of the corresponding passive crystal with identical $D_t$. The inset of Fig.~\ref{fig:vel_fluc_example}(a) shows a magnified part of this post-alignment relaxation dynamics. In this significantly aligned state, $\delta v ^2$ and  $\langle v^2\rangle$ are still found to decrease, as marked by \textbf{C} in the inset of Fig.~\ref{fig:vel_fluc_example}(a). They reach a steady state only after single-crystalline order is achieved, as marked by \textbf{D} in the inset of Fig.~\ref{fig:vel_fluc_example}(a).

Figure~\ref{fig:vel_fluc_example}(b) elucidates the relation between velocity fluctuations and corresponding progressive changes in the crystal structure. We mark four instances in Fig.~\ref{fig:vel_fluc_example}(a) at times $t = 1.4$, $5$, $16.15$, and $150$ via $\textbf{A}$, $\textbf{B}$, $\textbf{C}$, and $\textbf{D}$, respectively. The peak of the velocity variance curve and thus the peak in spatial velocity fluctuations corresponds to a typical MIPS state, shown in panel $\textbf{A}$ of Fig.~\ref{fig:vel_fluc_example}(b). The corresponding zoomed-in orientation map for the individual particle orientations $\theta_i$ in Fig.~\ref{fig:vel_fluc_example}(c), panel $\textbf{A}$, confirms associated large variations in directions of active particle forcing.

The structural order of the crystal and the particle orientations corresponding to $\textbf{B}$, $\textbf{C}$, and $\textbf{D}$ are presented in Fig.~\ref{fig:vel_fluc_example}(b) and Fig.~\ref{fig:vel_fluc_example}(c), respectively. Gradually from $\textbf{B}$ to $\textbf{D}$, defect networks are seen to reorganize. Transformation into an almost defect-free single-crystal occurs, while particles overall being aligned along their global collective migration direction, see Fig.~\ref{fig:vel_fluc_example}(c). Specifically, we note here the transformation from state \textbf{C} ($t = 16.15$) to state \textbf{D} ($t = 150$), by observing velocity variance and the corresponding structural transformation. The annihilation of defects, shown via the $|\psi_{6, i}|$-maps in Fig.~\ref{fig:vel_fluc_example}(b), directly correlates to a slight further drop in velocity variance, see the inset of Fig.~\ref{fig:vel_fluc_example}(a). Thus, velocity fluctuations, reflecting the action of active and conservative interaction forces, capture the signature of the dynamic process of defect annealing and single-crystal formation. 

We remark that focusing on the velocity variance resembles closely the concept of kinetic temperature \cite{hecht2024define}, although the latter is defined for inertial active systems. Moreover, in dense ABP systems, the steady-state average $\langle v^2\rangle$, excluding translational noise, had been used as a measure of kinetic temperature \cite{caprini2020active}. Using the curve of temporal evolution of the velocity variance, we can provide connections to the different stages of typical annealing-like behavior. After the initial drop in Fig.~\ref{fig:vel_fluc_example}(a), the increase up to the variance peak indicates `\emph{heating}'. Subsequently, the relaxation and decay below the corresponding passive-crystal fluctuations indicate `\emph{cooling}'. Thus, the process of obtaining a defect-free single-crystal from a defect-rich poly-crystal can be described within the framework of active annealing.  

\begin{figure*}[!htb]
\centering
\includegraphics[width=\linewidth]{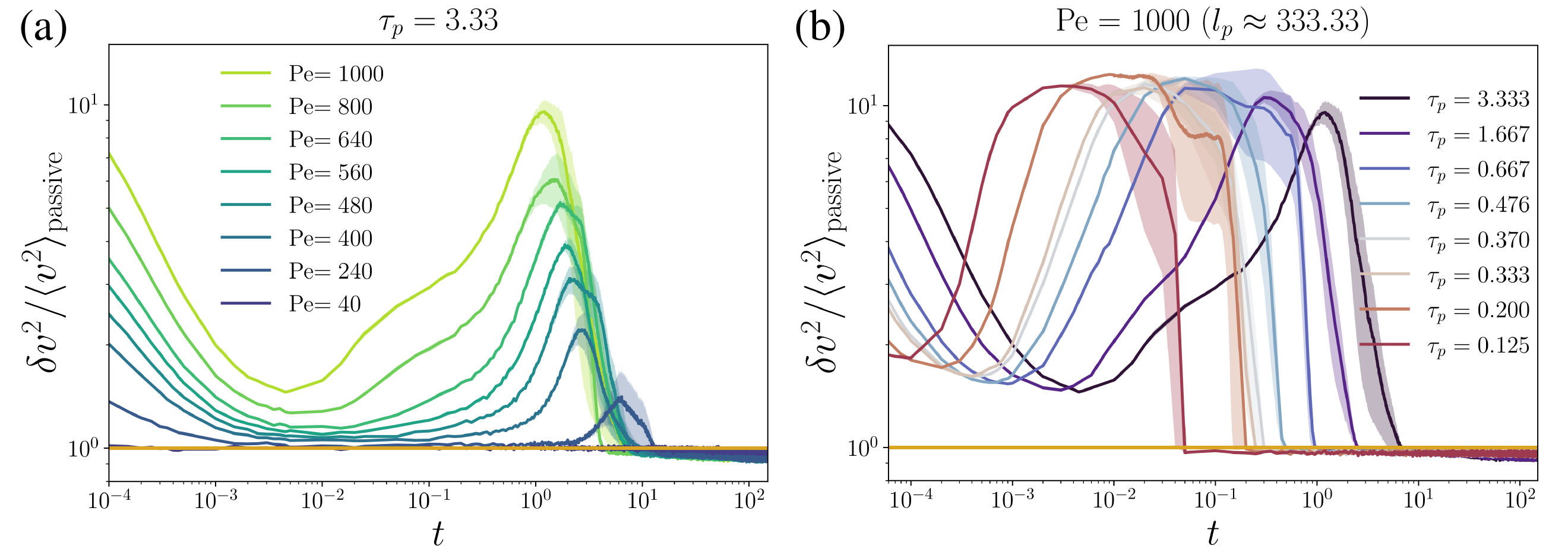}
\caption{\textbf{Velocity variances in persistence length- and persistence time-controlled protocols}. \textbf{(a)} {Persistence length $l_p$-controlled protocol.} Velocity variances $\delta v^2$ are shown at fixed persistence time ($\tau_{p} = 3.33$) but for varying Pe (changing $l_p$ from about 13.33 to 333.33), normalized by the mean squared velocity of the passive systems $\langle v^{2} \rangle_{\mathrm{passive}}$ at identical $D_t$. \textbf{(b)} {Persistence time $\tau_p$-controlled protocol.} Velocity variances $\delta v^2$ are shown at fixed Pe = 1000 ($l_{p} \approx 333.33$) but now with varying $\tau_p$, from 0.125 ($D_{r} = 8$) to about 3.33 ($D_{r} = 0.3$), again normalized by $\langle v^{2} \rangle_{\mathrm{passive}}$ at identical $D_t$. In both panels (a) and (b), the passive baseline corresponding to $\langle v^{2} \rangle_{\mathrm{passive}}$ is indicated by the horizontal orange line. All curves are averaged over 3$-$4 independent realizations, with sample-to-sample variations indicated by the shaded error bands. They represent the standard error of the mean.}
\label{fig:vel_fluc_lp_tau}
\end{figure*}

To delve deeper into the underlying physical process behind such poly-to-single-crystal formation, as shown in Fig.~\ref{fig:vel_fluc_example}, we distinguish between two annealing protocols. Figure~\ref{fig:vel_fluc_lp_tau} presents the time evolution of the velocity variances, normalized by their corresponding passive values. We use persistence length $l_{p}$- and persistence time $\tau_{p}$-controlled protocols, respectively. In the $l_p$-controlled protocol in Fig.~\ref{fig:vel_fluc_lp_tau}(a), $\tau_{p}$ is held fixed. $l_{p}$ is varied through Pe (or, more specifically, through the active force $f^{a}$ in our simulations). See also Methods for further details regarding the explored parameter space corresponding to these two protocols.

Firstly, increasing $l_{p}$ at fixed $\tau_p$ increases active forcing. Therefore, it injects more energy into the active crystal. This is evident from the higher transient velocity fluctuations, marked by increasingly larger peaks as  Pe increases from 40 to 1000 (and thus $l_p$ increases from around 13.33 to 333.33), see Fig.~\ref{fig:vel_fluc_lp_tau}(a). The passive velocity-variance baseline is marked by an orange horizontal line. Except for the case of Pe = 40 ($l_{p} \approx 13.33$), which almost lies along the corresponding passive baseline fluctuations in a broad regime, all other curves feature a gradually increasing trend with increasing Pe. This includes rising peak fluctuations with subsequent relaxations.

Secondly, persistence time of the constituents controls the memory of the particle orientations. Therefore, it influences the relaxation time of the velocity variances. Figure~\ref{fig:vel_fluc_lp_tau}(b) demonstrates this effect in the $\tau_p$-controlled protocol. Pe is kept constant at 1000 ($l_{p}\approx333.33$). $\tau_{p}$ is varied from 0.125 ($D_{r} = 8$) to 3.33 ($D_{r} = 0.3$) by modifying both friction coefficients $\gamma_{t,r}$. Here, the peaks of the normalized velocity variances are bound by the fixed Pe (hence fixed $l_p$). However, even with a relatively modest reduction of $\tau_p$ by a factor of around 30, we observe a significantly faster relaxation of velocity variances. We demonstrate in Fig.~\ref{fig:vel_fluc_lp_tau}(b) that increasing $\tau_{p}$ from $0.125$ to $3.33$ shifts the drop in velocity fluctuations by approximately two decades in timescales.

Hence, compared to the corresponding passive counterparts, $l_p$-controlled protocols govern the magnitude of the velocity variance. They allow for higher transient kinetic fluctuations. Conversely, $\tau_p$-controlled protocols primarily allow to modify the relaxation timescale of ordering. The peak magnitudes of the fluctuations remain less affected, and so are the interpretations of `\emph{heating}' left to the peak and `\emph{cooling}' right to the peak in terms of the proposed active annealing process.  

We note that the velocity variances in Fig.~\ref{fig:vel_fluc_lp_tau} are calculated for independent ensembles. Sample-to-sample variations are captured by shaded error bands. As we can see, these variations start developing mostly when the velocity variances reach their peaks. This strong sample-to-sample variability is related to the sequence of events, for instance, MIPS, partial melting, self-alignment-driven suppression of MIPS, and eventual emergence of a transient poly-crystal with variable defect networks. These systems reach either poly- or single-crystalline states. See Supplementary Note~5 and Supplementary Figure~6 for illustrations of explicit variations between different realizations after MIPS suppression. Thus, different realizations undergo different transient defect dynamics to reach a steady state with or without remaining defects.

\begin{figure*}[!htb]
\centering
\includegraphics[width=0.9\linewidth]{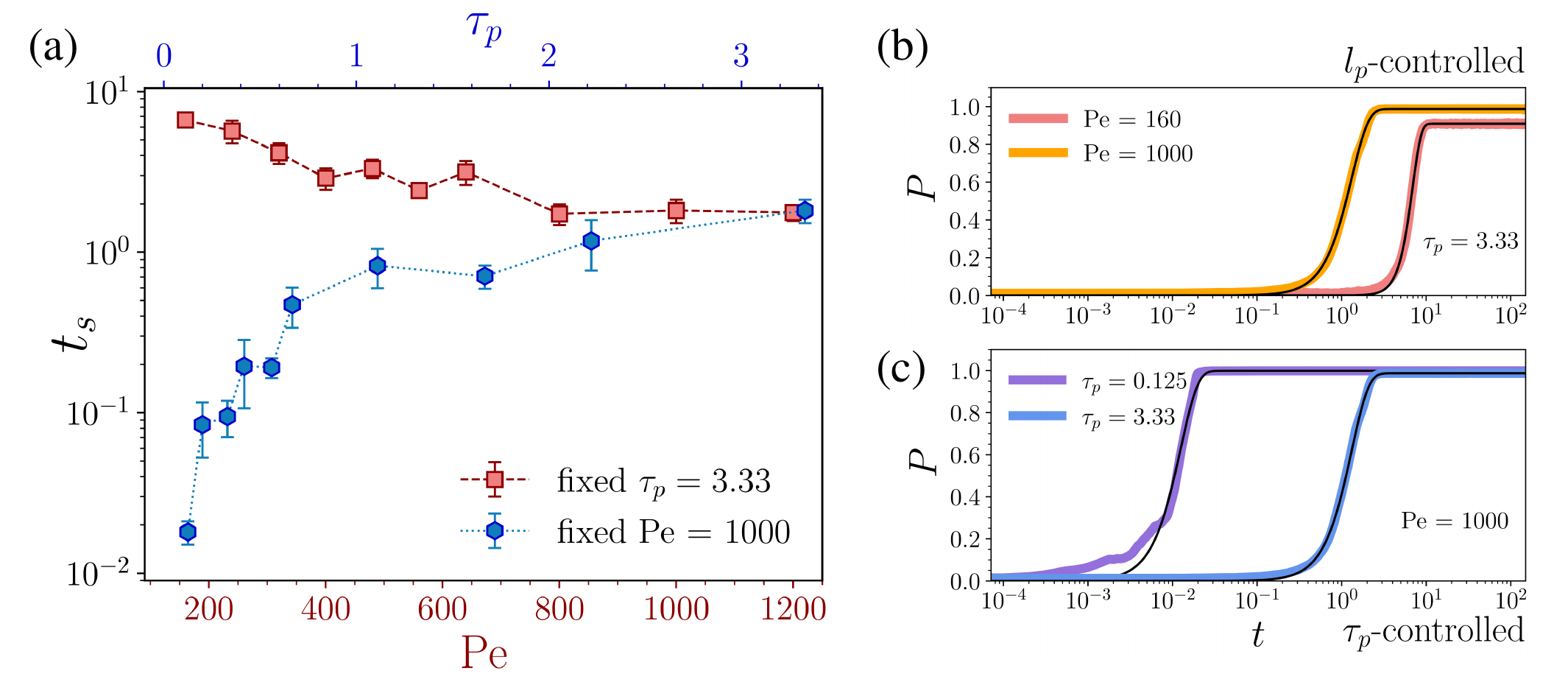}
\caption{\textbf{Polarization relaxation in persistence length- and persistence time-controlled protocols}. \textbf{(a)} Polarization relaxation time $t_s$, computed as the timescale for the global polar orientational order parameter $P(t)$ to reach its steady state. The bottom axis shows the dependence of $t_s$ on Pe (hence $l_p$), for fixed persistence time $\tau_{p}$ = 3.33, demonstrating the $l_p$-controlled protocol (red square symbols). Conversely, the top axis indicates $t_s$ as a function of persistence time, at a fixed persistence length $l_{p} \approx 333.33$ (Pe = 1000), representing the $\tau_p$-controlled protocol (blue hexagonal symbols). Error bars represent the standard error of the mean across 4$-$5 independent realizations of the ensemble. For each of these, every value of $t_s$ was extracted by fitting the global polarization curve $P(t)$ with a compressed exponential function. \textbf{(b)} Time evolution of the global polar order parameter $P(t)$ in $l_p$-controlled protocols at fixed $\tau_{p} = 3.33$ for two different example values of Pe (hence $l_p$). \textbf{(c)} Time evolution of $P(t)$ in $\tau_p$-controlled protocols at fixed Pe=1000 ($l_{p} \approx 333.33$) for two different example values of $\tau_p$. Both panels \textbf{(b)} and \textbf{(c)} include data from individual ensembles, with the resulting fitted curves indicated in black.}
\label{fig:orient_sig}
\end{figure*}

Next, we turn to the related orientational signatures of these two annealing protocols. As was shown earlier in Fig.~\ref{fig:vel_fluc_example}, the relaxation of velocity variances correlates to the corresponding onset of collective motion driven by the self-alignment of the particles. A particle experiences a torque when its direction of active driving (``its orientation'') is misaligned from the direction of its actual velocity. Such misalignment is caused by mutual collisions and stochastic noise. Here, we focus on the timescale at which the particles collectively reach a globally orientationally ordered migrating state. It leads to saturation of the polar order parameter $P (t)$. We determine the polarization relaxation time $t_s$ as the time for $P(t)$ to reach its steady state by fitting the time evolution curves of $P(t)$ with a compressed exponential function. 

Figure~\ref{fig:orient_sig}(a) illustrates the variation of the obtained polarization relaxation time $t_s$ as a function of Pe (hence $l_p$), for fixed $\tau_{p} = 3.33$. This $l_p$-controlled protocol is represented by the bottom axis as abscissa and red square symbols in Fig.~\ref{fig:orient_sig}(a). Moreover, we include the variation of $t_s$ as a function of $\tau_{p}$ at fixed Pe = 1000 ($l_{p} \approx 333.33$). This $\tau_p$-controlled protocol is connected to the top axis as abscissa and blue hexagonal symbols in Fig.~\ref{fig:orient_sig}(a). The $l_p$-controlled protocol allows for a modest decrease (from $ t_{s} \approx 6.65$ to $1.76$) with increasing Pe (hence increasing $l_p$). Conversely, for a fixed Pe, the $\tau_p$-controlled protocol enables a strong reduction, spanning almost two decades of timescale (from $ t_{s} = 1.82$ to $0.018$) to reach global polarization. These differences are explicitly visualized in Fig.~\ref{fig:orient_sig}(b) and Fig.~\ref{fig:orient_sig}(c). They show, respectively, the time evolution of $P(t)$ for $l_p$- and $\tau_p$-controlled protocols, together with their fitting curves to extract $t_s$. At fixed $\tau_p$, the potential to tune the polarization timescale $t_s$ is seen to be limited, even when Pe varies from 160 to 1000 ($l_p$ from around $53.33$ to $333.33$), see Fig.~\ref{fig:orient_sig}(b). Conversely, at fixed Pe in Fig.~\ref{fig:orient_sig}(c), $\tau_p$ varies from 3.33 ($D_{r} = 0.3$) to 0.125 ($D_{r} = 8$). Thus, there, significant potential emerges for tuning $t_s$. We remark that lowering Pe also reduces the degree of collective alignment. Therefore, $\langle P_{s}\rangle$, the average saturation value of $P$, decreases. Supplementary Note~6 and Supplementary Figure~7 show variations of $\langle P_{s}\rangle$ with Pe and $\tau_p$ in $l_p$- and $\tau_p$-controlled protocols, respectively.

We note that the sole purpose of fitting the polarization curves by compressed exponential functions of the form $P(t) = P_{s} \mathrm{exp}[-(t/t_{s})^{\eta}]$ is to extract a reliable timescale $t_s$ for global polarization relaxation. Indeed, for smaller $\tau_{p}$, the short-time variation of $P(t)$ is often not entirely captured by the fit, see, for example, the initial time variation of $P(t)$ in Fig.~\ref{fig:orient_sig}(c) for $\tau_p=0.125$. Nevertheless, for all cases studied in this article, we obtain from the fits $\eta > 1$, representing a compressed exponential.   

\begin{figure*}[!htb]
\centering
\includegraphics[width=\linewidth]{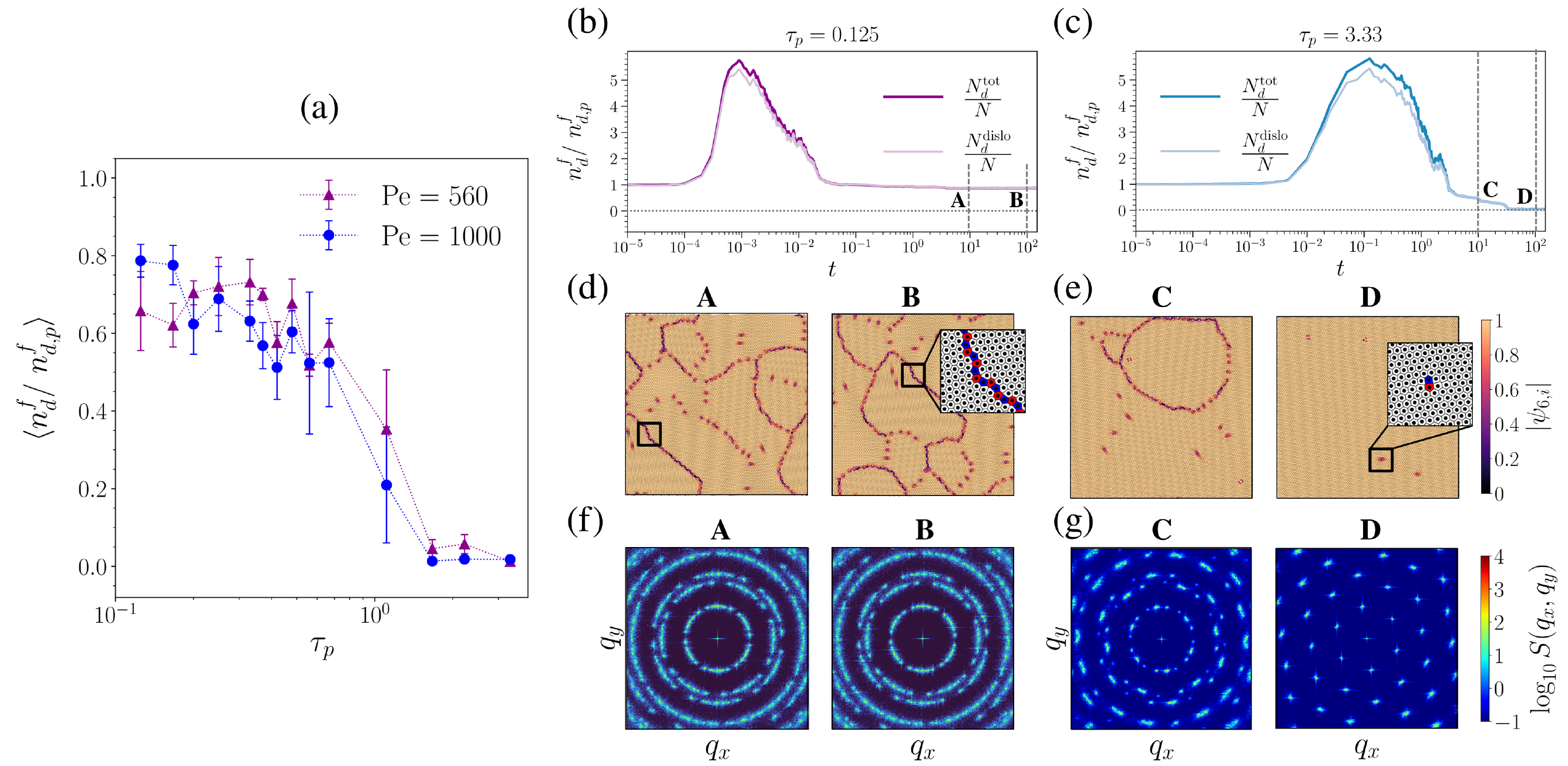}
\caption{\textbf{Tuned defect annihilation in persistence time $\tau_p$-controlled protocols}. \textbf{(a)} Steady-state defect fraction $n_{d}^{f}$($ = N_{d}^{\mathrm{tot}}/N$) in the active crystal relative to that of the passive poly-crystal $n_{d,p}^{f}$ as a function of persistence time $\tau_p$ for two different Pe of 560 ($l_{p} \approx 186.67$) and 1000 ($l_{p} \approx 333.33$). $N_{d}$ and $N$ denote total number of defects and total number of particles, respectively, in the system. Error bars correspond to the standard error of the mean across four independent realizations of each ensemble. \textbf{(b)} and \textbf{(c)} present the time evolution of the ratio of defect fractions $n_{d}^{f}/n_{d,p}^{f}$ (darker curves) for a particular realization for $\tau_{p} = 0.125$ ($D_{r} = 8$) and $\tau_{p} = 3.33$ ($D_{r} = 0.3$), respectively, at Pe = 1000 ($l_{p} \approx 333.33$). In both panels, lighter curves provide the corresponding values solely due to pairs of defects (dislocations). Zero defect fractions are indicated by horizontal dotted lines. \textbf{(d)} shows in two panels maps of the hexatic order parameter $|\psi_{6,i}|$ of the system in \textbf{(b)} at the two times $t = 10$ and $t = 100$, marked by \textbf{A} and \textbf{B}, respectively. These two panels represent a collectively traveling active crystal. The black squares indicate identical parts located at different positions due to collective migration. The inset of panel \textbf{B} shows a magnified view using Voronoi tessellation, revealing dislocation-rich grain boundaries. Similarly, \textbf{(e)} includes $|\psi_{6,i}|$-maps, now for the realization in \textbf{(c)} and marked by \textbf{C} and \textbf{D}, but again at times $t = 10$ and $t = 100$, respectively. Panel \textbf{D} identifies an almost defect-free active single-crystal. The magnification in the inset focuses on an isolated dislocation. \textbf{(f)} and \textbf{(g)} show maps of instantaneous static structure factors $S(\mathbf{q})$ corresponding to the configurations in \textbf{(d)} and \textbf{(e)}, respectively. Sharp Bragg peaks are shown to emerge in panel \textbf{D}. They confirm the formation of a collectively migrating active single-crystal.}
\label{fig:fast_slow_quench}
\end{figure*}

In conclusion, the collective polarization timescale of the system is found to be primarily controlled by the persistence time $\tau_p$ of the individual constituents. Further data are provided in Supplementary Note~7 and in Supplementary Figure~8, where the variation of $t_s$ with $\tau_p$ is presented for various Pe (from 160 to 1000). There, plotting $t_s(\tau_p)$, all of the curves follow the same trend. They cover a similar range of two decades of variation of $t_s$. At smaller $\tau_p$, almost a collapse occurs for $t_s$ obtained for different Pe. This indicates that the $\tau_p$-controlled protocol can significantly tune fast to slow relaxation towards global orientational polarization, see also Fig.~\ref{fig:vel_fluc_lp_tau}(b). This process goes hand in hand with a tuned fast to slow translational relaxation.

\subsection*{Tuning fast to slow `quenches' via persistence time-controlled protocols} \label{subsec3}

\noindent As noted before, the $\tau_p$-controlled protocol closely resembles an annealing process. Varying the persistence time of the particles effectively tunes the orientational and translational relaxation timescales. Thereby, it produces the dynamics analogous to either a slow or a fast `quench'. In this protocol, particles are allowed to explore the same spatial extent on the given timescales, owing to a fixed $l_p$. Yet, their orientational memory varies from short to long-lived. At constant bath temperature $T$ and active forcing $f^a$, this amounts to varying the friction coefficients through the Stokes-Einstein relations, thus varying the active velocity $v_0$ and $\tau_p$ simultaneously.

To further elucidate how these fast- and slow-quench analogs relate to defect annihilation and transformation from poly- to single-crystalline order in active crystals, we examine the defect dynamics under the $\tau_p$-controlled protocol. We denote by $N^\mathrm{tot}_d$ the total number of defects in the system. Figure~\ref{fig:fast_slow_quench}(a) presents the variation of the steady-state total defect fractions $n_{d}^{f} = N_{d}^{\mathrm{tot}}/N$ in the active crystal as a function of persistence time $\tau_p$. These fractions are shown relative to those of the passive poly-crystal $n_{d,p}^{f}$.  We focus on two different Pe (hence two different $l_p$). Each data point is obtained from averaging over several realizations. The average defect fractions decay towards zero with larger persistence time, thus relating fast- to slow-quench scenarios to decreasing resulting presence of defects. In-between the depicted high and low values of $\tau_p$, the systems adopt different relaxation channels 
for different realizations, therefore showing larger sample-to-sample variations. For example, we note the large error bars at $\tau_{p} \approx 1.11$ in Fig.~\ref{fig:fast_slow_quench}(a). Supplementary Note~8 and Supplementary Figure~9 correspond to such intermediate values of $\tau_p$.
Depending on the individual relaxation scenarios, the different samples can either produce complete defect-free or partially defect-free states within the simulation time. 

To inspect the dynamic behavior of such defects, we show the time evolution of the ratio of defect fractions $n_{d}^{f}/n_{d,p}^{f}$ for the lowest and the highest $\tau_p$ values at constant Pe = 1000 ($l_{p} \approx 333.33$) in Fig.~\ref{fig:fast_slow_quench}(b) for small $\tau_{p} = 0.125$ ($D_{r}$ = 8) and in Fig.~\ref{fig:fast_slow_quench}(c) for large $\tau_{p} = 3.33$ ($D_{r} = 0.3$). Indeed, within the $\tau_p$-controlled protocol, small $\tau_p$ accelerates relaxation. It  results in defect-jammed, collectively migrating steady states, see the two time instances marked as \textbf{A} and \textbf{B} at times $t = 10$ and $t = 100$, respectively. Their corresponding $|\psi_{6,i}|$-mapped configurations are depicted in Fig.~\ref{fig:fast_slow_quench}(d).  

The peaks at both small and large $\tau_p$ in Fig.~\ref{fig:fast_slow_quench}(b,c), respectively, coincide with defect proliferation around the transient MIPS states. Specifically around the peaks, MIPS enhances splitting of pairs of 5- and 7-defects (dislocations), so that they do not account for the total numbers of defects. Nevertheless, still, for both small and large $\tau_p$, dislocations dominate the defect fractions, shown via light magenta and light blue curves in Figs.~\ref{fig:fast_slow_quench}(b,c). A magnified fragment of the configuration in the inset of panel \textbf{B} within Fig.~\ref{fig:fast_slow_quench}(d) shows the bound dislocations inside a typical grain boundary. We stress that such configurations at steady state are parts of collectively traveling crystals within the periodic simulation domain. Therefore, repeating patterns appear at different spatial coordinates in panels \textbf{A} and \textbf{B} of Fig.~\ref{fig:fast_slow_quench}(d). For example, the square box marks the same part of the grain boundary in both panels, but located at different positions with proceeding time. Corresponding structure factors $S(\mathbf{q})$ for these poly-crystalline configurations, see Fig.~\ref{fig:fast_slow_quench}(f), show ringlike patterns. They reflect the superposition of scattering from grains of different orientations \cite{weber2014defect}. See Methods for details regarding the calculation of the static structure factor. 

At large $\tau_p = 3.33$, defect dynamics is markedly slower. However, the traveling states show misoriented grains that are gradually shrinking. The curve in Fig.~\ref{fig:fast_slow_quench}(c) drops substantially lower than in Fig.~\ref{fig:fast_slow_quench}(b), leaving a basically single-crystalline order. The corresponding transformation from $t = 10$ to $t = 100$ is marked by \textbf{C} and \textbf{D} in Fig.~\ref{fig:fast_slow_quench}(c). Corresponding configurations are illustrated by maps of local hexatic order $|\psi_{6,i}|$ in Fig.~\ref{fig:fast_slow_quench}(e). 
In other words, larger $\tau_p$ correspond to protocols of slow `{cooling}'. Effectively, they cause misoriented grains, see panel \textbf{C} in Fig.~\ref{fig:fast_slow_quench}(e), to be annealed. An almost defect-free active single-crystal results. Only few dislocations remain, see the inset of panel \textbf{D} in Fig.~\ref{fig:fast_slow_quench}(e). Consequently, in contrast to Fig.~\ref{fig:fast_slow_quench}(f), the $S(\mathbf{q})$-maps of the configurations in Fig.~\ref{fig:fast_slow_quench}(g) exhibit more spot-like patterns. They reflect the presence of comparatively fewer grain orientations in the system. At $t = 100$, the $S(\mathbf{q})$-map in panel \textbf{D} features sharp peaks of hexagonal symmetry in the reciprocal space. They confirm the presence of a well-ordered single-crystalline structure. Nevertheless, consistent with earlier observations of active crystals \cite{shi2023extreme, weber2014defect}, these Bragg peaks embody a noticeable broadening. We refer to Supplementary Figure~5 for illustration of such extended quasi-Bragg peaks in $S(\mathbf{q})$. Persistence time therefore controls the effective annealing of the active crystal. Larger $\tau_p$ leads to more efficient annealing of defects.

\subsection*{Mechanism of defect annealing in collectively migrating crystals}\label{subsec4}

\noindent Consequently, efficient defect annealing implies efficient structural reorganization in defect-rich active crystals.  Rather than the amplitude of fluctuations alone, what matters most is the coupling of active fluctuations to the structural degrees of freedom of the crystal. We recall from our observations, see Figs.~\ref{fig:poly_single}--\ref{fig:vel_fluc_example}, that significant structural reorganization occurs even after the crystal has reached a basically aligned, collectively migrating state of established global polar orientational order $P \sim 1$. 

For the moment, we consider an active poly-crystal of purely repulsive internal interactions and high active forcing in the absence of any orientational diffusion. In that case, perfect flocking would strictly imply the saturation of $P$ around $1$ with all particles featuring basically identical propulsion directions. In this scenario, active particles would lack any mechanism to affect an existing underlying lattice structure. ABP crystals of finite orientational diffusion, however, maintain residual fluctuations around the globally polarized state. Interparticle collisions due to fluctuations in self-propulsion directions enable a scattering pathway to reorganize the positional arrangements within the crystal corresponding to imperfect coordinations. Thus, to investigate the defect annealing mechanism, we address the following question: How do these excess orientational fluctuations facilitate defect annealing within the collectively migrating state of the active crystal?

\begin{figure*}[p]
\centering
\includegraphics[width=13cm]{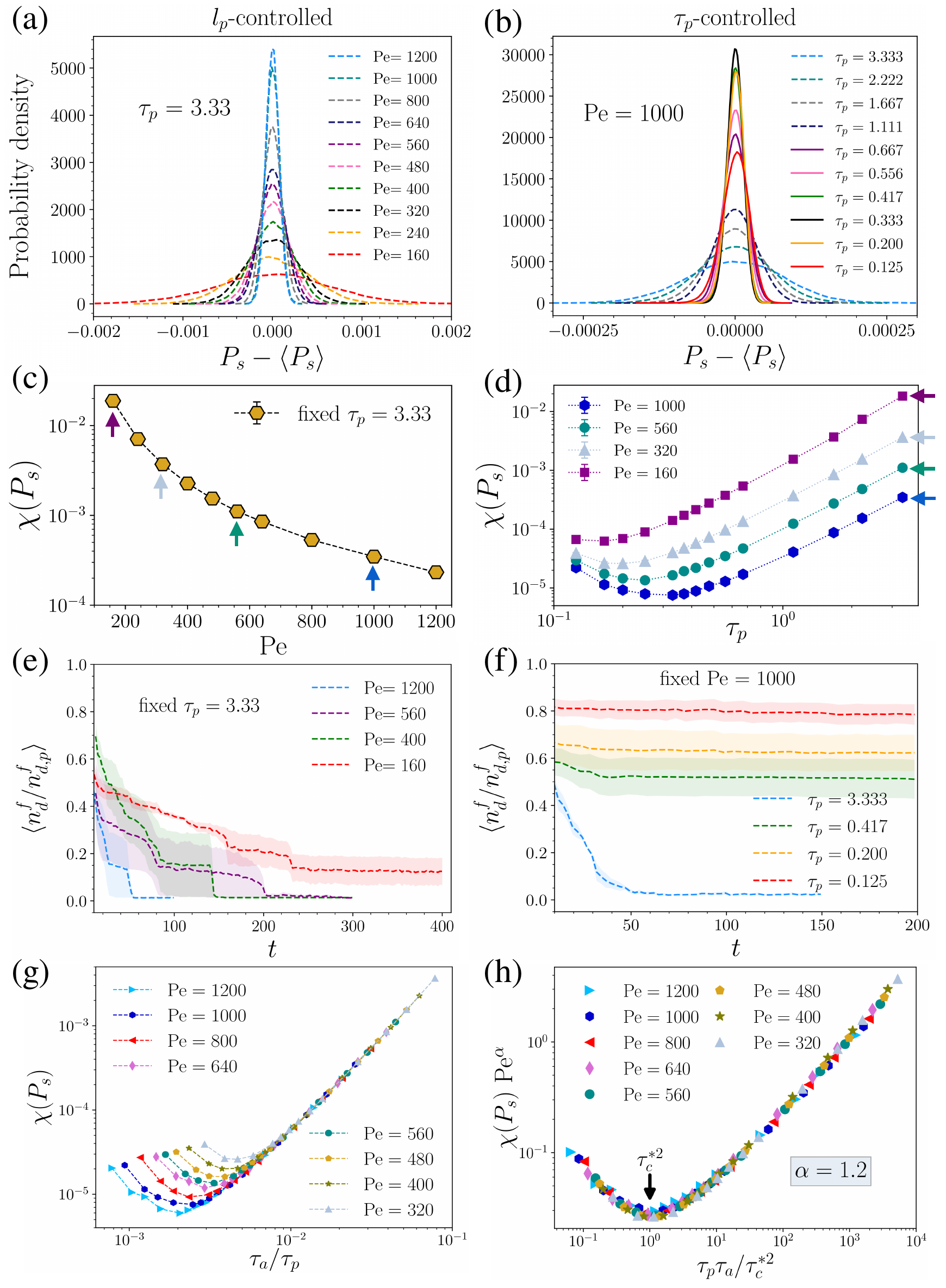}
\caption{\textbf{Relation between orientational fluctuations and structural reorganization in collectively migrating states for $l_p$- and $\tau_p$-controlled protocols}. \textbf{(a)} Probability density of the fluctuations of the steady-state polar orientational order parameter $P_s$, centered around its mean $\langle P_{s}\rangle$, for various values of Pe (hence $l_p$) at fixed $\tau_{p}=3.33$ in $l_p$-controlled protocols. \textbf{(b)} The same steady-state probability density, but now for different $\tau_p$ at constant Pe = 1000 ($l_{p} \approx 333.33$) in $\tau_p$-controlled protocols. Dashed lines in both (a) and (b) indicate defect-annealed states, while solid lines in (b) correspond to defect-rich crystals in seemingly steady states. \textbf{(c)} and \textbf{(d)} present polarization susceptibilities $\chi(P_s)$ (see text) corresponding to (a) $l_p$-controlled and (b) $\tau_p$-controlled protocols, respectively. (d) presents the $\tau_p$-variation of $\chi(P_s)$ for various constant Pe ($l_p$) values marked by arrows in (c). These results are averaged over four independent realizations and error bars are smaller than the marker size. \textbf{(e)} and \textbf{(f)} show the time evolution of the fractions of defects in the active crystal relative to that of its passive counterpart, $n_{d}^{f}/n_{d,p}^{f}$, in the collectively migrating states for $l_p$- and $\tau_p$-controlled protocols, respectively. All curves in (e) and (f) are averaged over four independent realizations in each case. Shaded error bands represent standard errors of the mean. Within the $\tau_p$-controlled protocols, \textbf{(g)} presents $\chi(P_s)$ as a function of the ratio of alignment time $\tau_a$ relative to persistence time $\tau_p$, leading to data collapse at elevated $\tau_a/\tau_p$. \textbf{(h)} illustrates a complete collapse of the data associated with the different values of Pe in (e) upon rescaling. Both quantities $\chi(P_s)$Pe$^\alpha$ and $\tau_{p}\tau_{a}/\tau_{c}^{*2}$ are dimensionless, where we found $\alpha = 1.2$. The minimum is located at $\tau_{c}^{*2} \approx \tau_{p}^{*}\tau_{a}^{*}$ (see text for details). }
\label{fig:post_rel}
\end{figure*}

\subsubsection*{Orientational fluctuations and structural reorganization}\label{subsubsec1}

\noindent Figure~\ref{fig:post_rel}(a) and (b) illustrate the probability density of the polarization fluctuations centered around its mean for $l_{p}$- and $\tau_{p}$-controlled protocols, respectively. These distributions correspond to temporal fluctuations, drawn from the polarized orientational dynamics of the collectively migrating crystals. See Methods for more details. For clarity, here the saturated polar order parameter is denoted as $P_s$, and the fluctuations as $P_{s} - \langle P_s \rangle$. 

At fixed $\tau_p$ (= 3.33), increasing Pe (and hence $l_p$) sharpens the fluctuations, leading to progressively narrower widths of the peaks, see Fig.~\ref{fig:post_rel}(a), from red to blue curves. Conversely, under the $\tau_{p}$-controlled protocol, increasing $\tau_p$ at fixed $l_p$ progressively broadens these fluctuations, see Fig.~\ref{fig:post_rel}(b), from red to blue curves, for Pe = 1000 ($l_{p} \approx 333.33$).  

We recall the translational signature of the active crystals, shown via velocity variances in Fig.~\ref{fig:vel_fluc_lp_tau}(a, b) for the $l_p$- and $\tau_{p}$-controlled protocols. In the first protocol, the magnitude is controlled, while in the latter the relaxation timescale can be tuned. Therefore, gradually increasing the magnitude of velocity fluctuations by increasing Pe (hence $l_p$) at fixed $\tau_p$ leads to progressively sharper temporal distributions of orientations.  This yields more and more stable collectively migrating crystals. Conversely, changing $\tau_{p}$ affects the relaxation timescale of the translational dynamics. Gradually elevating $\tau_p$ produces more and more temporally fluctuating particle orientations.

To further quantify these temporal fluctuations, we present the corresponding susceptibilities via $\chi(P_s) = N \mathrm{var}(P_s) $ for $l_p$- and $\tau_p$-controlled protocols in Fig.~\ref{fig:post_rel}(c) and (d), respectively. At fixed $\tau_{p} = 3.33$, $\chi(P_{s})$ gradually decreases with increasing Pe (hence $l_p$), see Fig.~\ref{fig:post_rel}(c). However, for each of these values of Pe and $l_p$, $\chi(P_{s})$ can be tuned along a wide range, implying a wide range of fluctuations. Examples are marked by arrows in Fig.~\ref{fig:post_rel}(c) and illustrated in Fig.~\ref{fig:post_rel}(d). Along each curve, $\chi(P_{s})$, quantifying underlying fluctuations, varies more than one (high $l_p$, Pe = 1000) up to close to three (low $l_p$, Pe = 160) orders of magnitude.

As we can infer from Fig.~\ref{fig:post_rel}(e) and (f), the tunability from lower to higher orientation fluctuations, estimated using $\chi(P_{s})$, has direct consequences on defect annealing. Figure~\ref{fig:post_rel}(e) depicts the time evolution of the steady-state defect fraction in active crystals scaled by its corresponding passive counterpart, $\langle n_{d}^{f}/n_{d,p}^{f}\rangle$. Each value is averaged over different realizations in the $l_p$-controlled protocol. We stress that the depicted time evolution takes place within the regime of collective migration with basically saturated $P_s$. At fixed $\tau_{p} = 3.33$, we observe that $\langle n_{d}^{f}/n_{d,p}^{f}\rangle$ drops to zero significantly quicker when Pe is high, corresponding to larger $l_p$ and stronger active forcing. For the lowest value of $l_p$, here corresponding to Pe = 160, some realizations do not attain a complete defect-free state at all within the simulation time, see the red curve in Fig.~\ref{fig:post_rel}(e). Thus, even at a fixed, considerably large persistence time $\tau_p$, larger spatial exploration, enabled by higher $l_p$, may contribute to faster, efficient defect annealing. Conversely, depending on the post-MIPS relaxation dynamics, lower $l_p$-values limit the spatial exploration, potentially leading to less efficient annealing.

From the temporal perspective, for an elevated persistence length, varying $\tau_{p}$ is observed to control the defect annealing process more efficiently, with substantially larger tunability. We consider an active crystal that reached a collectively migrating state of high global orientational order. In this case, it was possible to tune the excess orientation fluctuations such that the defects were either jammed at lower $\tau_p$, see the red, orange, and green curves in Fig.~\ref{fig:post_rel}(f), or annealed completely to reach a defect-free single-crystalline phase at higher $\tau_{p}$, see the blue curve in Fig.~\ref{fig:post_rel}(f). For a fixed, large $l_{p}$, small $\tau_{p}$ gives rise to narrower fluctuation distributions, see the solid curves in Fig.~\ref{fig:post_rel}(b). They imply temporally stable polar order. Conversely, keeping $l_{p}$ (and thus Pe) fixed, larger $\tau_{p}$ of the individual particles induces stronger temporal fluctuations. The polarization distribution broadens, see the dashed curves in Fig.~\ref{fig:post_rel}(b). Consequently, $\chi(P_{s})$ increases, see Fig.~\ref{fig:post_rel}(d). This behavior can be attributed to active particles exploring a fixed persistence length $l_{p} (= v_{0}\tau_{p}$) with either lower $v_{0}$ (higher $\tau_p$) or higher $v_{0}$ (lower $\tau_p$). 

Therefore, the possibility of realizing larger or smaller temporal fluctuations of orientations arises, respectively, specifically for a velocity-dependent alignment mechanism. This feature links to `slow' or `fast' quench scenarios, respectively, within the `cooling' counterpart of activity-induced annealing. Explicit illustrations of the $\tau_{p}$-dependent variations of instantaneous and temporal angular variances in $\tau_{p}$-controlled protocols are included in Supplementary Note~9 and Supplementary Figure~10. Moreover, we recall that system-size analysis shows the robustness of efficient defect annealing at larger $\tau_p$ in the $\tau_p$-controlled protocol and its underlying mechanism with increasing system size, see Supplementary Note~3, including Supplementary Figures~3 and 4.  Taken together, although $\tau_{p}$-controlled protocols provide greater tunability, efficient annealing of defects in active crystals still requires a sufficiently large $l_{p}$.

The subtle, yet crucial role of both $l_p$ and $\tau_{p}$, and their interplay with self-alignment in the governing orientation fluctuations during the relaxation dynamics after initial global alignment, are elaborated on in Figs.~\ref{fig:post_rel}(g) and (h). We first focus on Fig.~\ref{fig:post_rel}(d), where $\chi(P_s)$ exhibits a minimum at $\tau_p$ = $\tau_p^*$ in the small temporal fluctuation regime (higher $v_{0}$, lower $\tau_p$). This minimum systematically shifts towards lower $\tau_p$ in Fig.~\ref{fig:post_rel}(d) with decreasing Pe (hence decreasing $l_p$). Eventually, it disappears, as seen, for example, for Pe = 160 ($l_{p} \approx 53.33$) in Fig.~\ref{fig:post_rel}(d), where it is hardly visible any more. Physically, these minima indicate states of the active crystal with minimum temporal fluctuations in global polar orientational order. 

We recall that the velocity-dependent torque $\mathbf{T}_{i} = \zeta (\hat{\mathbf{e}}_{i} \times \mathbf{v}_{i})$ of self-alignment acting on each particle introduces a lengthscale $l_{a} = \gamma_{r}/\zeta$ and a timescale $\tau_{a} = \gamma_{r}/\zeta v_{0}$. These alignment lengths of trajectories and alignment times are typically needed for a particle to align its self-propulsion direction with its velocity. As mentioned earlier, ordered collective motion warrants $l_{a} \ll l_{p}$, and therefore throughout our study $\tau_{a} \ll \tau_{p}$. Thus, persistent motion lasting $\tau_{p} (= 1/D_{r})$ could introduce misalignment between the self-propulsion direction of a particle and its velocity direction. 
Yet, as $\tau_{a} \ll \tau_{p}$,  $\mathbf{T}_{i}$ aligns the self-propulsion direction quickly compared to its persistence time.

Figure~\ref{fig:post_rel}(g) presents $\chi(P_s)$ as a function of $\tau_p$, rescaled by the alignment timescale $\tau_a$ within $\tau_p$-controlled protocols, for various fixed Pe (hence $l_p$). The susceptibility in the collectively migrating state exhibits a nonmonotonic dependence on the ratio $\tau_{a}/\tau_{p}$ between alignment- and persistence-time. A minimum in $\chi(P_s)$ is observed for each considered curve. 
We remark that $\tau_{a} = \gamma_{r}/\zeta v_{0} = {3} (\frac{k_{B}T}{\sigma})^{2} \tau_{p}^{2}/{\zeta f^{a}}$. Thus, in view of the dimensionless parameters $k_{B}$, $\sigma$, fixed $T = 0.025$, and fixed $\zeta = 0.01$ in this study, $\tau_{a}/\tau_{p} \sim \tau_{p}/f^{a}$. Thus, $\tau_{a}/\tau_{p}$ remains proportional to $\tau_{p}$ as for the abscissa in Fig.~\ref{fig:post_rel}(d). 

Although $\tau_{a}/\tau_{p} \ll 1$ for the whole range of $\tau_p$, 
we recover two distinct regimes below and above $\tau_{a}/\tau_{p} \approx 10^{-2}$. At $\tau_{a}/\tau_{p} \gtrsim 10^{-2}$, self-alignment becomes slow enough so that the particles retain their orientational memory for a substantially longer time before they reorient back towards their velocity. The collapse of all data corresponding to different Pe (and $l_{p}$) in this regime, see the range $\tau_{a}/\tau_{p} \gtrsim 10^{-2}$ in Fig.~\ref{fig:post_rel}(g), indicates that polarization fluctuations here are primarily controlled by $\tau_{a}/\tau_{p}$, not by Pe (or $l_p$). As $\tau_{a}/\tau_{p}$ grows, self-alignment progressively slows down. This process seems beneficial for defect annealing, a process that requires a certain time by itself, under more pronounced fluctuations in overall alignment. 

When self-alignment becomes faster, see the data for $\tau_{a}/\tau_{p} < 10^{-2}$ in Fig.~\ref{fig:post_rel}(g), we observe two distinct features. First, minima emerge. They indicate an optimal separation between these two timescales $\tau_a$ and $\tau_p$ to reduce overall orientational fluctuations. As a result, a collectively polarized state is produced that is most stable concerning orientational ordering. This regime is also found to be sensitive to persistence length. Hence, Pe becomes relevant, with a systematic shift of  the minima towards lower $\tau_{a}/\tau_{p}$ for higher Pe (hence higher $l_p$). We recall that $\tau_{a}/\tau_{p} \sim 1/f^a\sim1/\mathrm{Pe}$. Second, at smaller $\tau_{a}/\tau_{p}$ below the minima, alignment becomes increasingly rapid. This implies that small random or deterministic local disturbances can more effectively result in direct local orientational fluctuations. In other words, decreasing $\tau_p$ (increasing $D_r$) enhances local misalignment through stronger orientational noise, resulting in an increase in polarization fluctuations.

We found that the residual dependence on Pe is eliminated by rescaling the susceptibility by $\mathrm{Pe}^{\alpha}$, where we obtained $\alpha \approx 1.2$ from fitting. Moreover, we replace the persistence time by the combined timescale $\tau_p\tau_a$. In this way, we yield a complete collapse of the data onto a common master curve across $\tau_p$-controlled protocols with different persistence lengths (hence different Pe), see Fig.~\ref{fig:post_rel}(h). The reduced form of this scaling collapse is expressed as 
\begin{equation}
   \chi(P_s) = \mathrm{Pe}^{-1.2} \mathcal{F}\left(\frac{\tau_{p}\tau_{a}}{\tau_{c}^{*2}}\right),
\end{equation}
where $\tau_{c}^{*2} = \tau_{p}^{*}\tau_{a}^{*}$ denotes the location of the minima of $\chi = \chi (\tau_p \tau_a)$. Technical details of extracting these minima are provided in Supplementary Note~10 and Supplementary Figure~11. The scaling collapse under the $\tau_p$-controlled protocol indicates that the polarization susceptibility has an activity-dependent amplitude scaling as $\mathrm{Pe}^{-1.2}$. Its remaining functional form is governed by the combined timescale $\tau_p\tau_a$.

Large fluctuations, owing to large $D_r$ and implying large local misalignment, with very fast self-alignment resembles a `fast'-quench scenario that is unable to anneal defects. Conversely, very small orientation fluctuations with strong self-alignment can develop orientationally rigid alignment without much scope for structural rearrangement. Therefore, effective annealing of defects is expected for long-lived, strong intermediate fluctuations, which are favored by $\tau_{p}\tau_{a}/\tau_{c}^{*2} \gg 1$ in Fig.~\ref{fig:post_rel}(h). In this regime, comparatively slow self-alignment is insufficient to strongly suppress persistence-supported orientational fluctuations. It allows the elevated persistence time to preserve orientational memory and sustain intermediate fluctuations in particle orientations. Nevertheless, at the same time, this combination maintains the global background polarization. Thus, this scenario reveals the characteristics of a `slow'-quench annealing process facilitating structural rearrangement. Effective annealing of defects (especially the `cooling' counterpart of annealing) in polar active crystals apparently warrants an optimal balance between orientation fluctuations and overall alignment. 

We remark that the nominal alignment timescale $\tau_a$ is a global concept. Still, the velocity dependence of the torque introduces microscopic heterogeneity in the alignment dynamics. Generally, the local velocity varies from particle to particle. This heterogeneity, in turn, influences the effective self-alignment of individual particles. It emerges intrinsically from the dynamics and cannot be independently controlled within the present description. An alternative process of self-alignment with velocity-independent coupling would therefore provide a useful framework for systematically disentangling these effects. It constitutes an interesting direction for future work.

\begin{figure*}[p]
\centering
\includegraphics[width=\linewidth]{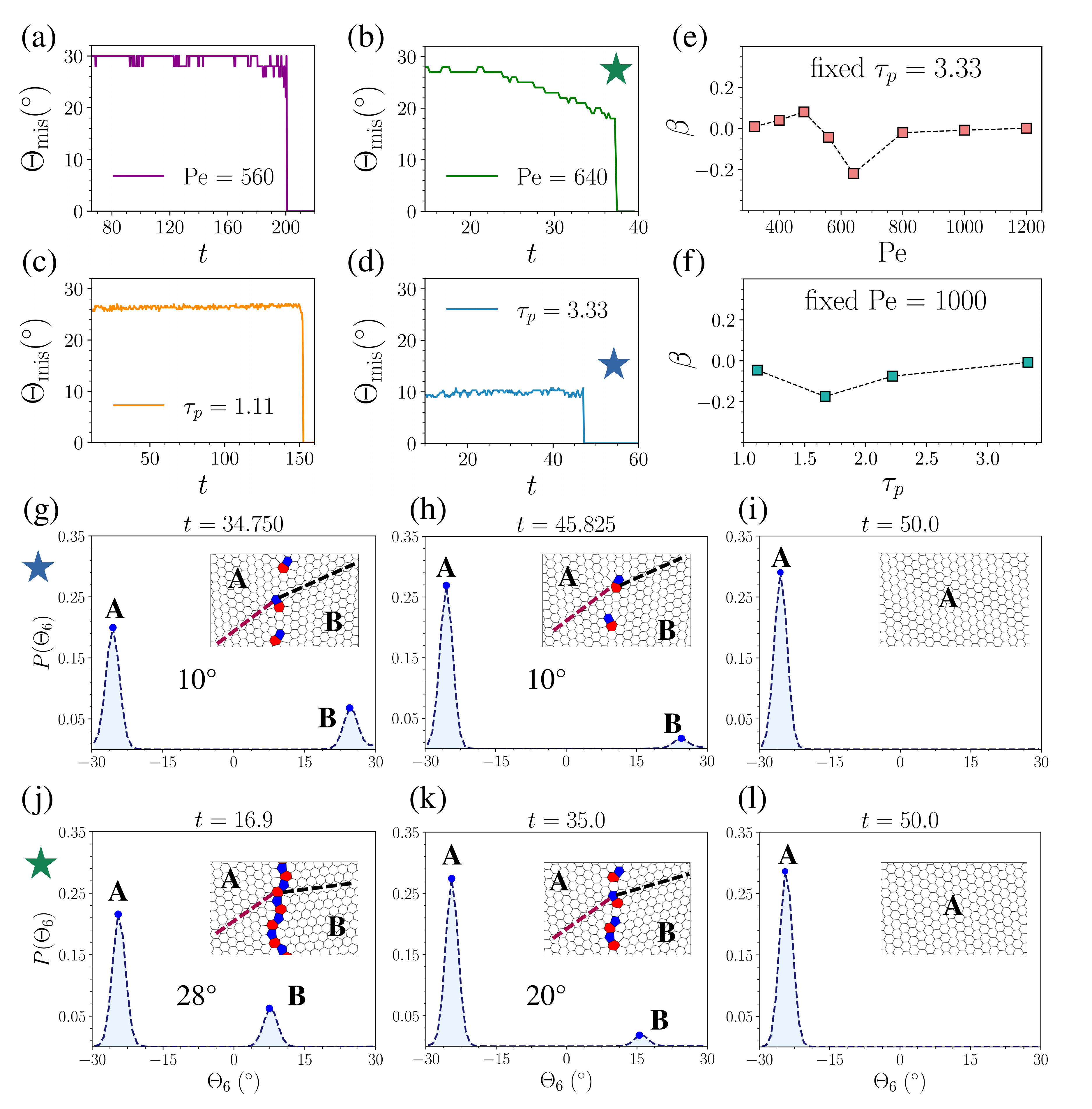}
\caption{\textbf{Grain-boundary (GB) shrinkage during defect annealing in globally polarized states.} \textbf{(a,b)} Time evolution of the misorientation angle $\Theta_{\mathrm{mis}}$ between the grain enclosed by the GB loop and the surrounding crystal within $l_p$-controlled protocols at fixed $\tau_{p} = 3.33$ for (a) Pe = 560 ($l_{p} \approx 186.67$) and (b) Pe = 640 ($l_{p} \approx 213.33$). \textbf{(c,d)} Same time evolution of $\Theta_{\mathrm{mis}}$ but for $\tau_{p}$-controlled protocols at fixed Pe = 1000 for (c) $\tau_{p} = 1.11$ and (d) $\tau_{p} =3.33$. \textbf{(e,f)} Time evolution of the shear-coupling factor $\beta$ for (e) $l_p$-controlled protocols at fixed $\tau_{p} = 3.33$ and (f) $\tau_p$-controlled protocols at fixed Pe = 1000 ($l_{p} \approx 333.33$). The selected state points exhibit a single GB loop during the final stage of defect annihilation. Thus $\beta$ is obtained from the associated shrinkage dynamics, prior to the formation of the final single-crystal (see text). \textbf{(g,h,i)} Illustration of the time evolution of the probability distribution of local grain orientation $P(\Theta_6)$ for three time instances for the scenario depicted in (d), namely at (g) $t = 34.75$, (h) $t=45.825$, and (i) $t = 50$. \textbf{(j,k,l)} Similarly, time evolution of $P(\Theta_6)$ for the scenario depicted in (b) at three instances in time, namely at (j) $t = 16.9$, (k) $t=35$, and (l) $t = 50$. The insets in (g--l) show magnified parts of the corresponding configurations. They highlight the misorientations between the grain, marked by $\textbf{B}$, and the surrounding crystal, marked by $\textbf{A}$. Defects are identified via Voronoi coordination maps. The corresponding misorientation angles are included in each panel in degrees, indicating pure shrinkage in (g--i) and grain rotation in (j--l).}
\label{fig:gb}
\end{figure*}

\subsubsection*{Mechanism of grain-boundary shrinkage}\label{subsubsec2}

\noindent Next, we elucidate how particle-level annealing assisted by orientational fluctuations manifests itself at the coarser scale of grain boundaries. We examine the  mechanisms of grain-boundary shrinkage underlying the poly-to-single-crystalline transformation. Interestingly, in both $l_p$- and $\tau_p$-controlled protocols, active crystals mostly form grain boundary (GB) loops before adopting the single-crystalline order. These GB loops correspond to various angles of misorientation $\Theta_{\mathrm{mis}}$ between the grain and the surrounding crystal matrix. To study the shrinkage mechanisms of GB loops, first we isolate the cases of obtaining the single-crystal after complete defect annihilation. We include both $l_p$- and $\tau_p$-controlled protocols. Second, we focus only on late-stage configurations, after global polarization has basically saturated, in which the microstructure has evolved into two dominant competing grains, with the smaller grain bounded by a GB loop. Still, we mention that defect annealing can also proceed through complex multi-grain interactions, even at the last few stages before single-crystal formation. See Supplementary Note~11 and Supplementary Figure~12 for one such instance.

Misorientation angles $\Theta_{\mathrm{mis}}$ between the grain and the surrounding crystal matrix are calculated as $\Theta_{\mathrm{mis}} = \mathrm{min} (|\Theta_{6}^{\mathrm{G}}-\Theta_{6}^{\mathrm{C}}|, 60^{\circ} - |\Theta_{6}^{\mathrm{G}}-\Theta_{6}^{\mathrm{C}}|)$ \cite{lavergne2018dislocation, lavergne2019shrinkage}. Here, $\Theta_{6}^{\mathrm{G}}$ and $\Theta_{6}^{\mathrm{C}}$ are the average orientations of the particles inside the grain and in the outer crystal matrix, respectively. For each particle, they are calculated from the hexagonal bond-orientational order parameter, see Methods. Due to the hexagonal symmetry of the crystal, misorientation angles are within the range $0^{\circ} < \Theta_{\mathrm{mis}} < 30^{\circ}$. 

Figures~\ref{fig:gb}(a) and (b) present the time evolution of $\Theta_{\mathrm{mis}}$ at the last stage of defect-annealing for shrinkage of a GB loop for two different $l_p$ (at fixed $\tau_{p} = 3.33$), corresponding to Pe = 560 ($l_{p} \approx 186.67$) and Pe = 640 ($l_{p} \approx 213.33$), respectively. Similarly, Figs.~\ref{fig:gb}(c) and (d) show the time evolution of $\Theta_{\mathrm{mis}}$ for two different $\tau_{p} = 1.11$ and $\tau_{p} = 3.33$, respectively, at fixed Pe = 1000 ($l_{p} \approx 333.33$). The nearly constant crystallographic misorientation in Figs.~\ref{fig:gb}(a,c,d) indicates that the final transformation to a single-crystal occurs predominantly through activity-assisted GB migration leading to grain shrinkage. Conversely, in Fig.~\ref{fig:gb}(b) misorientation is seen to decrease with time, indicating grain rotation. 

To further distinguish whether the GB shrinkage involves grain rotation or not, we calculate the shear coupling factor \cite{cahn2004unified, cahn2006coupling, trautt2012grain, camerin2025disentangling} $\beta = - d\Theta_{\mathrm{mis}}/d \mathrm{ln}(R_{g}/\sigma)$, where $\Theta_{\mathrm{mis}}$  is expressed in radians $R_{g}$ denotes the GB loop radius, see below. Each value of $\beta$ is extracted from a linear fit between $\Theta_{\mathrm{mis}}$ and the logarithm of $R_g$. Once we have calculated the local orientations of the particles $\Theta_{6,i}$, probability distributions of these orientations $P(\Theta_{6})$ are obtained as histograms. They further help to identify whether a given particle resides inside or outside the grain enclosed by the GB loop. The number of particles inside the grain $N_g$ can then be used to estimate the approximate GB loop radius \cite{lavergne2019shrinkage, camerin2025disentangling} $R_{g} = a\sqrt{N_{g}/\pi}$, where $a$ (= 1.02$\sigma$) is the lattice constant of the crystal. See Methods for further details on grain identification and calculations of misorientations.

Figures~\ref{fig:gb}(e) and (f) present the variation of $\beta$ with Pe (hence $l_p$)  within $l_p$-controlled protocols at fixed $\tau_{p} = 3.33$ and with $\tau_p$ within $\tau_p$-controlled protocols at fixed Pe = 1000 ($l_p\approx333.33$), respectively. Once the competing grains are established, subsequent GB shrinkage is mostly seen to follow activity-assisted GB migration. Misorientations are approximately conserved in most cases, yielding $\beta \approx 0$, see the corresponding data points in Figs.~\ref{fig:gb}(e) and (f), respectively. However, occasionally, grain rotations occur. They are identified from nonvanishing shear-coupling factors ($\beta \neq 0$). Figure~\ref{fig:gb}(b) provides a corresponding example. A gradual decrease in misorientation implies grain rotation, reminiscing GB sliding with $\beta < 0$. Similarly, for Pe = 480 ($l_{p} = 160$), see Fig.~\ref{fig:gb}(e), the grain rotates towards a marginally higher misorientation, $\beta \gtrapprox 0$, signaling a more involved GB migration mechanism. We point to Supplementary Note~11 and Supplementary Figure~13 for the time evolution of the misorientation for all displayed instances in $l_p$- or $\tau_p$-controlled protocols. 

Activity-induced GB migration with almost constant misorientation, see Fig.~\ref{fig:gb}(d), throughout GB loop shrinkage is illustrated in Figs.~\ref{fig:gb}(g,h,i). We display the evolution of the probability distribution of the orientations $P(\Theta_6)$ at three progressive instances in time. The locations of the maxima of $P(\Theta_6)$ remain approximately constant. Insets show zoomed-in views of the configuration space with Voronoi-neighbor-maps, indicating a constant $\Theta_{\mathrm{mis}} = 10^\circ$ between grains $\textbf{A}$ and $\textbf{B}$. Similarly, Figs.~\ref{fig:gb}(j,k,l) elaborate on the gradually decreasing $\Theta_{\mathrm{mis}}$ shown in Fig.~\ref{fig:gb}(b). Here, the time evolution of $P(\Theta_6)$ at three progressive time instances indicates their shift in position. From the magnified views of the Voronoi coordination-maps shown in the insets, grain $\textbf{B}$ is seen to rotate from approximately $7.8^\circ$, see Fig.~\ref{fig:gb}(j), to about $ 16^\circ$, see Fig.~\ref{fig:gb}(k). $\Theta_{\mathrm{mis}}$ changes from $28^\circ$ to $20^\circ$, while the dominant orientation of the outer crystal remains almost constant at about $-24^\circ$. 
GB shrinkage and collapse of the smaller grain can be inferred from the gradual disappearance of peak $\textbf{B}$ at the expense of the growing peak $\textbf{A}$ in Figs.~\ref{fig:gb}(g--i) and (j--l). 

Thus, activity seems to unlock a rich GB dynamics in the defect-rich ABP crystal. Different existing GB shrinkage mechanisms are at work. Yet, unlike passive systems, these mechanisms seem to be rather insensitive to the initial misorientation of the grain, as can be inferred from the above. Still, we note the limitation of our analysis. The emergence of a GB loop and a misoriented grain is rather uncontrolled with respect to our preparation of the system and can only be accessed at the last stage of defect annealing. However, this strategy follows their natural appearance and does not necessitate artificial grain preparation. 

Finally, defect annihilation is completed via complex dislocation interactions. It  involves the recombination of oppositely signed dislocations and local neighbor rearrangements that progressively restore sixfold coordination and single-crystalline order. An example is illustrated in Supplementary Note~11 and Supplementary Figure~14.

\section*{Discussion}\label{sec3}

\noindent To summarize, we have demonstrated that defect-rich, self-aligning active crystals undergo a poly-to-single-crystalline transformation. It takes place within a coupled translational–orientational dynamics setting. Ultimately, it gives rise to collectively traveling crystal-like states. However, structural relaxation persists well beyond completing the major part of conventional orientational ordering. The collectively migrating crystal remains dynamically fluctuating. Spatiotemporal fluctuations of polarization drive structural reorganization and defect annihilation. Persistence time thus emerges as a key control parameter for tuning the defect-annihilation capability of an active crystal. It contributes to produce single-crystalline order, despite an orientationally fluctuating finite active forcing of each particle.

\begin{figure}
\centering
\includegraphics[width=8cm]{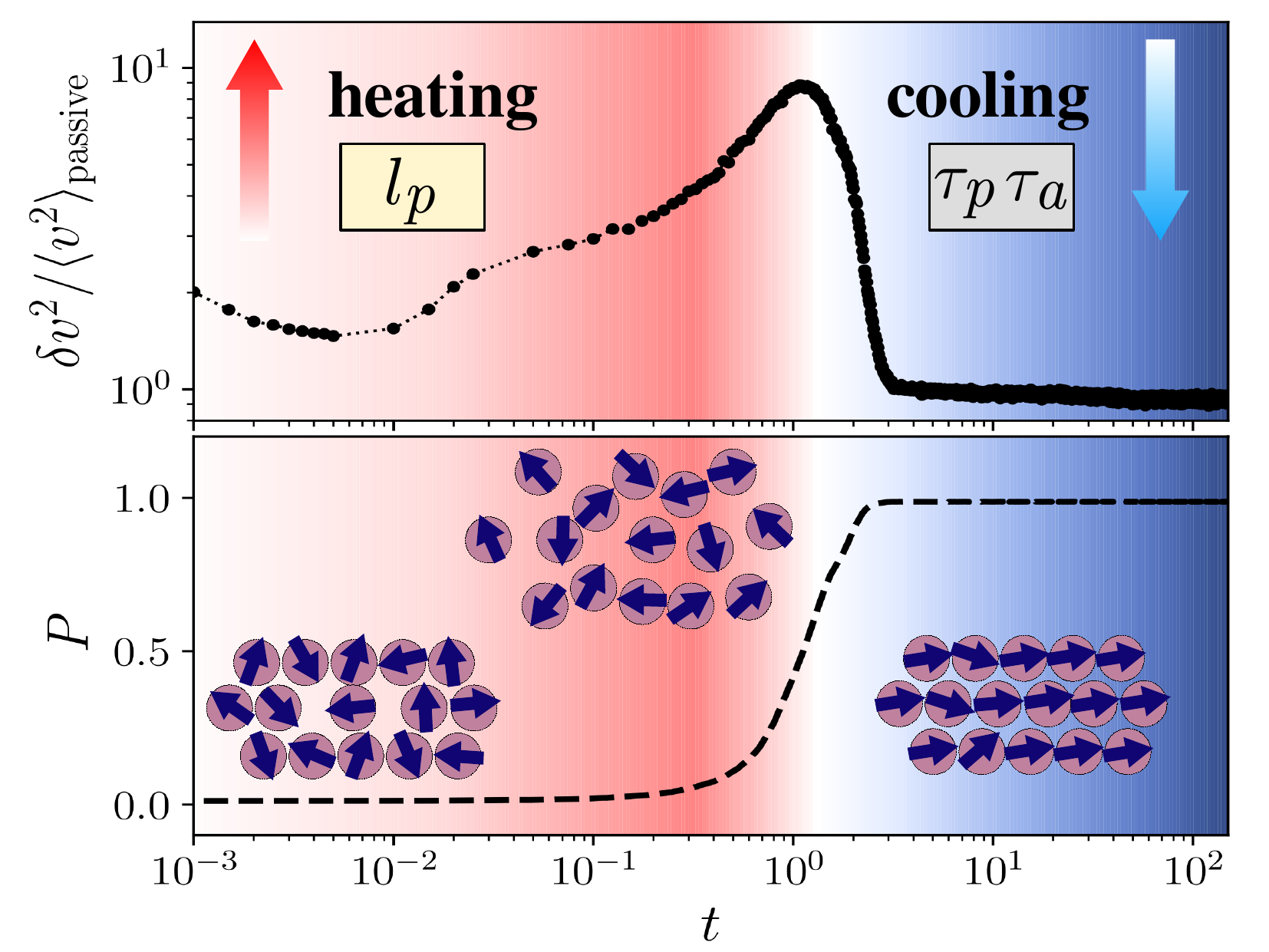}
\caption{\textbf{Schematic representation of activity-driven defect annealing in ABP crystals}. Top panel: Time evolution of the velocity variance of the active crystal $\delta v^{2} = \sum_{i=1}^N (\mathbf{v}_{i}-\langle \mathbf{v} \rangle)^2/N$ relative to its passive counterpart $\langle v^{2}\rangle_{\mathrm{passive}}$. Bottom panel: Corresponding evolution of the global polar orientational order parameter $P$. Increasing Pe (hence persistence length $l_p$) enhances `heating'. Conversely, choosing an optimized value of combined persistence and alignment timescale $\tau_p\tau_a$, see Fig.~\ref{fig:post_rel}(h), serves to control `cooling' in a desired way. Sketches to illustrate corresponding states of the particles in the ABP crystals are illustrated by the insets in the bottom panel. Therein, orientations of individual active driving forces are indicated by arrows within the particles. The background color visually represents progressive heating (white to red) and subsequent cooling (red to white to blue).}
\label{fig:scheme_annealing}
\end{figure}

This process of defect annihilation can be further elucidated using the framework of activity-driven annealing, governed by two distinct protocols. One is varying persistence length $l_p$ while fixing persistence time $\tau_p$, the other is varying $\tau_p$ while keeping $l_p$ constant. These protocols instigate distinct translational and orientational fluctuations, both before and after the onset of collective motion. The $l_p$-controlled protocols primarily enhance velocity fluctuations, with limited tunability of the translational and orientational relaxation timescales. Conversely, the $\tau_p$-controlled protocols provide substantially greater control over these timescales. Once collective motion is established, the residual orientational fluctuations, encoded in the temporal variance of the polarization, drive particle collisions and structural reorganization. Thus, active forcing, correlated with persistence length $l_p$, provides an effective `heating' mechanism. It drives particles out of metastable, defect-rich states. Conversely, for fixed $l_p$, varying the combined persistence and alignment timescales $\tau_p\tau_a/\tau_c^{*2}$ enables tunable `cooling'. It mimics and allows for `fast' or `slow' quenching. A schematic illustration of the activity-driven annealing process is presented in Fig.~\ref{fig:scheme_annealing}. 

Across different Pe (hence $l_p$), the polarization fluctuations, encoded by the susceptibility, collapse onto a single master curve, see Fig.~\ref{fig:post_rel}(h). This collapse reveals an activity-dependent amplitude scaling as approximately Pe$^{-1.2}$. Its remaining functional dependence is governed by the combination $\tau_p\tau_a/\tau_c^{*2}$. In particular, defect annihilation leads to poly-to-single-crystalline transformation. It occurs for $\tau_a\tau_p/\tau_c^{*2} \gg 1$ and corresponds to long-lived persistence combined with slow residual alignment. 

These residual changes in global polarization manifest themselves at the scale of the grain boundaries as rich shrinkage dynamics. Mainly, we find relatively pure shrinkage of misaligned grains in the collectively traveling state, but occasionally GB sliding and grain rotation are observed. 

We remark that annealing as a method of heat treatment in its original form involves several cycles of rising temperature and slow cooling. A recent study, although focused on amorphous glasses \cite{sharma2025activity}, addressed the correspondence. It draws parallels between activity-driven annealing and yielding of glasses under oscillatory shear. Reconsidering these statements in view of our results, they suggest that activity-driven annealing can mimic the effectiveness of cyclic transformations.

Previous activity-induced annealing discussed in the literature was mostly concerned with doping few active particles inside an otherwise passive system. This strategy revealed profound impact on colloidal suspensions, from promoting crystallization of hard-sphere glasses \cite{ni2014crystallizing} to the formation of metastable clusters, including surface melting between crystalline domains \cite{kummel2015formation}. Surface melting occurs when doped active particles accumulate at passive grain boundaries, enhancing their mobility and eliminating them in both two- and three-dimensional colloidal crystals \cite{van2016fabricating, van2016removing}. Such localized melting may be understood as a manifestation of the attraction between `hot' particles in a `cold' bath. The highly diffusive `hot' particles generate a circular depletion region that mediates an effectively attractive interaction \cite{tanaka2017hot}. Experiments demonstrated accelerated annealing in passive colloidal monolayers using the same strategy of introducing active dopants \cite{ramananarivo2019activity}. They showed that short-$l_p$ dopants accumulate at grain boundaries, whereas long-$l_p$ dopants disperse uniformly throughout the system.

Instead, our work considers a fully active crystal and distinguishes between $l_p$- and $\tau_p$-controlled annealing protocols. Thereby, spatially and temporally controlled annealing are separated from each other. Defect annihilation in a fully active crystal, had been addressed by a generalized Vicsek model containing mutual polar alignment and short-ranged repulsion \cite{weber2014defect}. Defect-free states were obtained when repulsion dominated over polar order. Conversely, alignment-dominated regimes lead to collectively migrating poly-crystalline, defect-rich crystals. Distinctive from imposed mutual Vicsek-type alignment \cite{vicsek1995novel}, we use the intrinsic mechanism of self-alignment \cite{baconnier2025self}. It originates at the individual particle level and can lead to collectively migrating \textit{single}-crystalline, \textit{defect-free} states. Thus, seemingly unlike previously studied mutual alignment, 
self-alignment is not antagonistic to defect annihilation. Rather, the residual orientational fluctuations borne out of misalignment facilitate the emergence of single-crystalline order in active crystals.

\begin{figure*}[!htb]
\centering
\includegraphics[width=17.5cm]{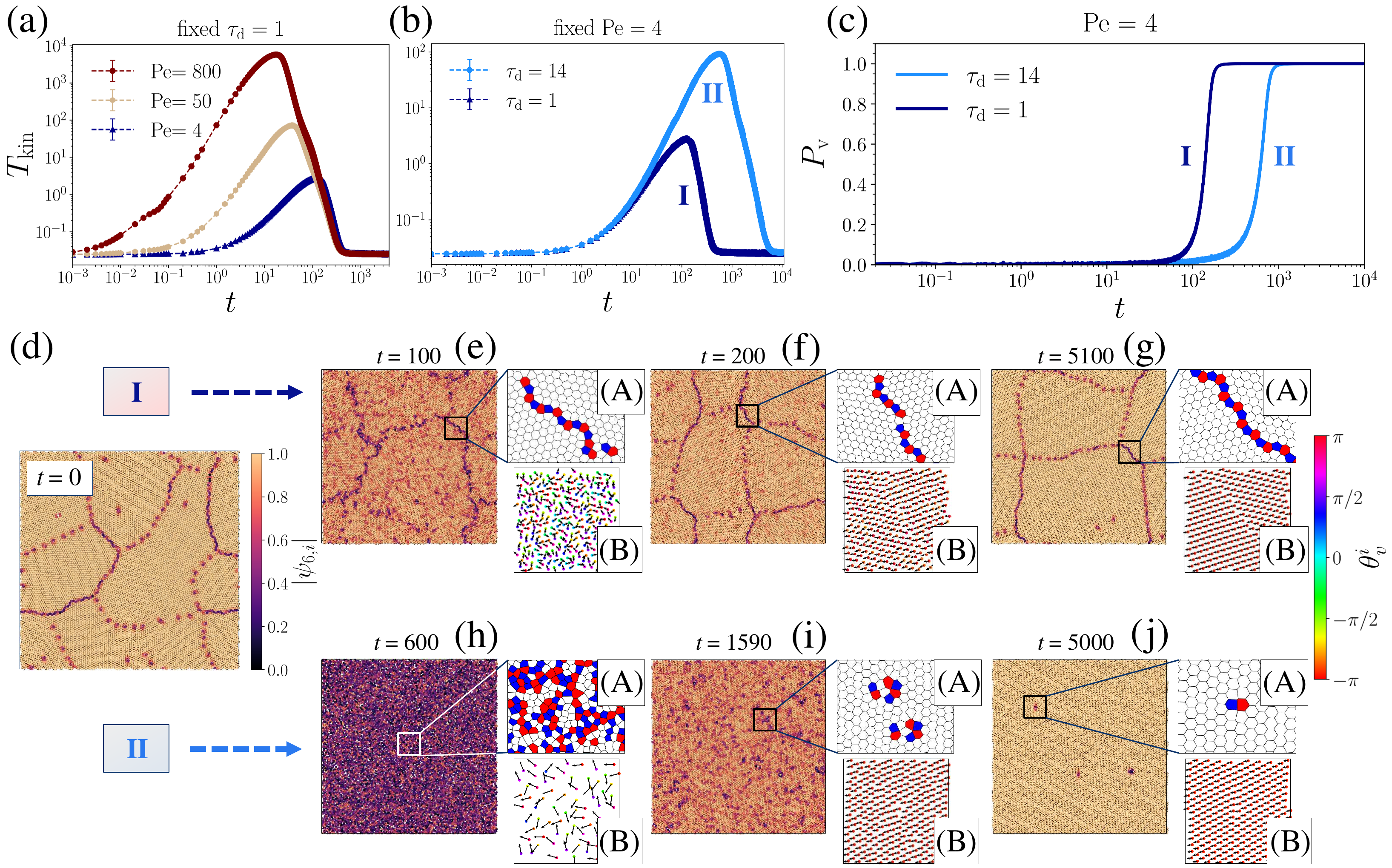}
\caption{\textbf{Defect annealing in an atomic active crystal}. \textbf{(a)} Time evolution of kinetic temperature $T_{\mathrm{kin}}$ for three different Pe = 4, 50, and 800 at fixed  dissipative timescale $\tau_\mathrm{d} (= m/\gamma_{\mathrm{d}}$) = 1. Here, $\gamma_{\mathrm{d}}$ quantifies dissipative friction between the particles (see text for details). \textbf{(b)} Time evolution of $T_{\mathrm{kin}}$, here for fixed Pe = 4 and two different $\tau_\mathrm{d}$ = 1 and 14. These results are averaged over three independent realizations each. Error bars are smaller than the marker sizes. \textbf{(c)} Time variation of the velocity alignment order parameter $P_{\mathrm{v}}$ for the two cases shown in (b), marked as \textbf{I} and \textbf{II}. They refer to fast and slow annealing, respectively (see text). \textbf{(d)} Configuration of the initial passive atomic crystal in a $|\psi_{6,i}|$-map at $t = 0$. \textbf{(e--g)} Snapshots during the fast annealing process marked as \textbf{I} in (b) and (c). Evolution of the atomic crystal is illustrated through $|\psi_{6,i}|$-maps at $t$ = 100, 200, and 5100. \textbf{(h--j)} Similar snapshots during the slow annealing process marked as \textbf{II} in (b) and (c), here at times $t$ = 600, 1590, and 5000. In each of panels (e--j), the zoomed-in parts of the crystal are detailed by (A) Voronoi neighbor maps. Five-fold and seven-fold defects are marked in blue and red, respectively. Moreover, (B) velocity directions $\mathbf{v}_i/|\mathbf{v}_i|$ in the zoomed-in regions are indicated by color maps for the orientation angles $\theta_v^i$ and directional unit vectors. }
\label{fig:atomic}
\end{figure*}

Generally, our results concern defect dynamics and structural ordering in various living active crystals. Examples are bacterial crystals in two dimensions \cite{petroff2015fast, petroff2018nucleation}, starfish embryos \cite{tan2022odd}, cells \cite{cohen2020mechanical, shen2025flocking}, as well as synthetic colloidal active crystals \cite{palacci2013living, palacci2014light, mognetti2013living, singh2017non, ramananarivo2019activity, briand2018spontaneously} and crystalline robotic swarms \cite{wang2024robo, vansaders2021sculpting}. The two distinct protocols and their efficiency in defect-annealing used in our work could be experimentally realized and tested using photo-activated synthetic colloids. $l_p$-controlled protocols could be envisaged experimentally by tuning the propulsion speed $v_0$ through the intensity of light \cite{palacci2013living, vutukuri2020light}, while keeping the rotational diffusion coefficient $D_r$ fixed \cite{ramananarivo2019activity}. Thereby, $l_p (=v_0/D_r)$ is varied at fixed $\tau_{p} (=1/D_r$). $\tau_p$-controlled protocols could likewise be realized experimentally. It was demonstrated for active Brownian colloids that propulsion velocity ($v_0$) and rotational diffusivity $D_{r}$ (=1/$\tau_{p}$) can be tuned independently through electric and magnetic fields, respectively  \cite{fernandez2020feedback}. In particular, $v_0$ can be varied proportionally with $D_{r}$ while maintaining $l_{p}$ (= $v_{0}/D_{r}$) constant.

Interestingly, anti-alignment can also lead to collective migration \cite{kursten2025emergent}. Combined experiments and simulations investigated charged active colloids that show electrostatic repulsion as well as turn-away torques \cite{das2024flocking}. At high density, a collectively migrating Wigner crystal is recovered. It undergoes a poly-to-single-crystalline transformation upon increasing velocity (activity) at fixed $D_r$, for given translational and rotational friction coefficients. Despite the distinct alignment interactions, this transformation closely resembles the one observed in our system. It can be understood within our $l_p$-controlled annealing framework and thus resembles one part of our concept. Moreover, the observed orientational fluctuations \cite{das2024flocking} may provide a similar mechanism for poly-to-single-crystal ordering, even in the presence of anti-aligning interactions. Our results may therefore extend to a broader class of active crystals exhibiting collective migration.

Finally, one may ask for possible defect annealing in collectively migrating crystals that do not possess remaining rotational degrees of freedom of active forcing on the single particle level. In other words, how crucial is the presence of remaining spatiotemporal orientational fluctuations on defect annealing once global alignment has emerged? To answer this question, we consider a minimal atomic model \cite{lobaskin2013collective} of a defect-rich active crystal. Each self-propulsion force acts along the direction of current velocity of the corresponding point particle, is therefore slaved to its velocity direction, and does not introduced an additional degree of freedom. Atoms interact through a purely repulsive WCA potential. Similarly to the above, all simulation parameters are expressed in reduced units. Specifically, temperature is measured in units of $\epsilon/k_B$, and units of time are given by $\sqrt{{m\sigma^2}/{\epsilon}}$. Dissipative and random forces act pairwise. See Methods for further details of this atomic active crystal model. Within this point-particle description, interatomic collisions are the sole contributor for structural reorganization. 

This crystal is coupled to a heat bath and maintained at a fixed temperature $T=0.025$, which is identical to that used in the previous overdamped ABP simulations. Dissipative coupling introduces a characteristic timescale $\tau_\mathrm{d}=m/\gamma_s$, where $m$ denotes the mass of each particle. $\gamma_s$ is the 
friction coefficient that controls the strength of dissipative momentum transfer within any interacting pair of particles. For this underdamped system, we can directly control and realize annealing through the kinetic temperature of the active crystal after removing its center-of-mass velocity. 

Within this framework, Figs.~\ref{fig:atomic}(a,b) present two protocols analogous to those employed above for the ABP crystal. Figure~\ref{fig:atomic}(a) shows the time evolution of the kinetic temperature $T_{\mathrm{kin}}$ \cite{hecht2024define} for varying Pe at fixed damping timescale $\tau_\mathrm{d}=1$. Conversely, Fig.~\ref{fig:atomic}(b) fixes activity at a low magnitude (Pe = 4) for fast ($\tau_\mathrm{d} = 1$) and slow ($\tau_\mathrm{d} = 14$) dissipation. At fixed $\tau_\mathrm{d}$, increasing Pe produces progressively larger transient peaks in $T_{\mathrm{kin}}$, reflecting stronger velocity fluctuations, before relaxing to the target temperature $T=0.025$. Conversely, increasing $\tau_\mathrm{d}$ delays this relaxation, defining the fast and slow quenches marked by \textbf{I} and \textbf{II} in Fig.~\ref{fig:atomic}(b), respectively. This distinction is reflected in the velocity polarization $P_\mathrm{v}$ ($= \left\langle \left|\sum_{i=1}^N \mathbf{v}_i/|\mathbf{v}_{i}|\right| \right\rangle\!/N$), which approaches $P_\mathrm{v} \approx 1$ rapidly for \textbf{I} but slowly for \textbf{II}, see Fig.~\ref{fig:atomic}(c). 

Starting from the defect-rich state in Fig.~\ref{fig:atomic}(d), the corresponding structural evolution is shown for the fast and slow quenches in Figs.~\ref{fig:atomic}(e--g) and (h--j), respectively. Pairwise dissipation suppresses relative velocity differences between the particles. Therefore, increasing $\tau_\mathrm{d}$ slows the alignment \cite{lobaskin2013collective} and delays the onset of collective motion. The fast quench thus retains a poly-crystalline structure, see Fig.~\ref{fig:atomic}(g). Conversely, the slow quench enables defect annealing and forms single-crystalline order, see Fig.~\ref{fig:atomic}(j). Voronoi maps of the corresponding defect configurations are shown in panels (A) of Fig.~\ref{fig:atomic}(e--j). Notably, the velocity directions of the particles for the same magnified crystal region remain almost rigid, tightly oriented during collective motion, see panels (B) in Figs.~\ref{fig:atomic}(g,j). In contrast to that, they are seen to fluctuate before the saturation of global velocity alignment, see panels (B) in Figs.~\ref{fig:atomic}(f,i). Thus, more generally, even without separate orientational degrees of freedom, defect annealing can be achieved by delaying collective motion through the damping timescale. This suggests that the structural defect annealing via delayed alignment is generic to active crystals, rather than tied to a specific propulsion model.

To conclude, our study reveals that continuous processes of localized spatiotemporal orientational disordering in self-aligning active crystals do not compete antagonistically with structural ordering. Instead, they can unlock the pathways through which structural defects are annealed. By tuning the timescales of persistent active motion and of alignment, these fluctuations can be regulated to promote defect motion, grain-boundary shrinkage, and ultimately the emergence of single-crystalline order. Therefore, residual orientational disorder becomes a mechanism for enhancing structural order. 

This perspective connects statistical physics of active and soft matter with the mechanics of defects and microstructure. Simultaneously, it suggests new opportunities for the design of dynamically reconfigurable materials, from photonic crystals to active metamaterials. Rather than suppressing fluctuations, one may tune them as a control parameter for defect dynamics, microstructural evolution, and mechanical functionality. Very recently, defect annealing has also been observed in delayed swarmalator crystals \cite{chanin2026aging}. There, temporal delay in orientational dynamics produces slow aging and progressive defect elimination. Despite the different microscopic mechanisms, this result features an interesting phenomenological similarity to that of our findings. Essentially, both systems highlight how temporal regulation of collective orientational dynamics can control structural relaxation. They suggest that controlled temporal memory may provide a general route to anneal defects in active crystals. 

Future work shall address defect dynamics by resolving the direct competition between velocity-independent alignment and persistence. 
If active colloidal suspensions in a carrier medium are considered, including hydrodynamic interactions will be an important further step. Potentially, they can alter defect transport and annealing. Extending these ideas across colloidal, biological, and synthetic active materials could reveal common principles governing fluctuation-assisted structural organization. More broadly, the ability to engineer orientational fluctuations and relaxation timescales may provide a general strategy for designing active materials.

\section*{Methods} \label{sec4}

\subsection*{Model and simulation details}

\noindent We study active crystals prepared in a dense poly-crystalline state as mentioned in the main text. The high packing fraction $\phi = 0.87$ is close to the two-dimensional close-packed limit, $\phi_{\mathrm{cp}} = \pi/(2\sqrt{3}) \approx 0.906$. Unless otherwise stated, simulations are performed in a two-dimensional square box of dimensions $100\sigma \times 100\sigma$, containing $N=11074$ particles. To assess finite-size effects, we additionally consider larger systems with box dimensions $200\sigma \times 200\sigma$ ($N=44183$) and $300\sigma \times 300\sigma$ ($N=99327$). We adjust the lattice constant to $a=1.02\sigma$, regardless of the system size. All simulation parameters are expressed in reduced units, where length and energy are scaled by particle diameter $\sigma$ and the WCA interaction strength between the particles $\epsilon$, respectively. Temperature is measured in units of $\epsilon/k_B$. Time is expressed in units of  $\sqrt{{m\sigma^2}/{\epsilon}}$. The units of all other quantities follow accordingly. Thus, the friction coefficients $\gamma_{t}$ and $\gamma_{r}$ are expressed in units of $\sqrt{m\epsilon}/\sigma$ and $\sigma\sqrt{m\epsilon}$, respectively. Moreover, the translational and rotational Gaussian white noises $\bm{\xi}_{i}$ and $\xi_{\theta,i}$ are measured in units of $\sqrt[4]{\epsilon/(m\sigma^2)}$ in the numerical evaluations. Thus, in these units, 
$\langle \bm{\xi}_{i} (t)\rangle = \mathbf{0}$, $\langle \xi_{\theta,i}\rangle = 0$, $\langle {\xi}_{i,m} (t) {\xi}_{j,n} (t^\prime)\rangle = \delta_{ij}\delta_{mn} \delta(t-t^\prime)$, and $\langle \xi_{\theta,i} (t) \xi_{\theta,j} (t^\prime)\rangle = \delta_{ij} \delta(t-t^\prime)$, where $\delta$ represents the Kronecker delta and the Dirac delta function. All molecular dynamics simulations are performed using \textsc{LAMMPS} \cite{thompson2022lammps, Lammps}.

\subsubsection*{Active Brownian particles (ABP) model}

\noindent Our two-dimensional system of $N$ interacting active Brownian particles (ABP) follows a self-alignment mechanism at a constant temperature $T$ as described in the main text. We start from simulating Brownian particles in hexagonal lattices under periodic boundary conditions. The particles interact by a repulsive Weeks-Chandler-Andersen (WCA) potential defined as 
\begin{equation}
\label{eq:wca}
\begin{split}
U(r)=
\begin{cases}
4\epsilon \biggl[
\left(\frac{\sigma}{r}\right)^{12}
-\left(\frac{\sigma}{r}\right)^{6}
\biggr]
+ \epsilon,
& r \le 2^{1/6}\sigma, \\ 
0,
& r > 2^{1/6}\sigma,
\end{cases}
\end{split}
\end{equation}
which is evaluated for the center-to-center distance $|\mathbf{r}_{ij}|$ between any pair of particles $i$ and $j$. Initially, using Brownian dynamics simulations, the passive system is equilibrated at a higher temperature $T$ = 10 for $10^ 7$ steps with timestep $\Delta t$ = $10^{-4}$. Defect-rich passive Brownian crystals are prepared from instantaneously quenching the system from $T = 10$ to the desired lower temperature $T = 0.025$, where it is further equilibrated for $10^ 7$ time steps. Then introducing the  self-propulsion force and a deterministic aligning torque on each particle takes us to the self-aligning active Brownian particle (ABP) crystals, the internal dynamics of which is described by Eqs.~(\ref{eq:trans}) and (\ref{eq:rot}) of the main text.

\subsubsection*{Persistence length- and persistence time-controlled protocols}

\noindent The degree of activity is quantified using the P\'eclet number Pe = $f^ {a}\sigma/k_{B} T $. Moreover, the persistent length $l_p$ is defined via $l_{p} = v_{0} \tau_{p} = \sigma\,\mathrm{Pe} /3 $. Starting from a defect-rich state, we consider two protocols. One is persistence-length $l_p$-controlled, the other one is persistence time $\tau_p$-controlled. In the former, $\tau_{p}$ is fixed, while $l_{p}$ is varied by changing $\mathrm{Pe}$ (or, equally, the active force $f^a$). In the coupled translational–rotational dynamics, this corresponds to varying $v_{0}$, while keeping the translational and rotational friction coefficients $\gamma_{t}$ and $\gamma_{r}$ fixed. In the latter protocol, $\tau_{p}$ is varied, keeping $l_{p}$, $\mathrm{Pe}$, and $f^a$ fixed. Since changing $\tau_{p}$ here requires varying both $\gamma_{r}$ and $\gamma_{t}$, this protocol simultaneously varies $\gamma_{t}$ and $v_{0}$, while maintaining $f^a=\gamma_{t} v_{0}$ and $l_p$ constant.

In other words, we explore the $l_p$–$\tau_p$ parameter space at fixed packing fraction $\phi=0.87$. Thus, in our units, we vary $1.33 \lesssim l_p \le 400$, set by $4 \le \mathrm{Pe} \le 1200$, which implies $0.1\le f^a\le 30$. For each $l_p$, or accordingly Pe, we vary $0.125 \le \tau_p \le 3.33$ ($0.3\le D_r\le8$). At large $\tau_p$, we use $\Delta t=5\times10^{-6}$ and run simulations for $3\times10^7$ – $1.6\times10^8$ timesteps, with longer runs at smaller $\mathrm{Pe}$ to assess whether defect-free states can be reached. At smaller $\tau_p$ (higher $D_r$), we reduce the timestep to $\Delta t=10^{-6}$ for numerical stability in the presence of translational and rotational noise and run for $2\times10^8$ steps. For each parameter set, we perform 4–5 independent realizations for ensemble averaging. We recall that $l_p$ and Pe are here strictly linked to each other. Therefore, we typically list both values in our rescaled units in the main text.

Rather than modeling a specific experimental realization, our objective is to systematically investigate the structural evolution of active crystals across weakly, intermediately, and strongly persistent regimes. Extending the parameter range to Pe $\sim$ 1200 allows us to access the persistence-dominated regime, where the transient velocity fluctuations and the orientation relaxation dynamics that are central to this work are most clearly manifested. The higher activity also facilitates defect annihilation within a computationally accessible simulation time, enabling the structural relaxation to be realized under practical circumstances.

\subsection*{Calculation of Voronoi neighbors and hexatic order parameter}

\noindent To characterize the local crystalline order and identify defects, we perform Voronoi tessellation with periodic boundary conditions, using the software suit freud \cite{freud2020}. The resulting Voronoi neighbor list is used to determine the coordination number of each particle and to construct an adjacency list containing the Voronoi neighbors of every particle. Particles with coordination number different from six are classified as defects, typically with five and seven neighbors. A 5-7 defect pair constitutes a dislocation. Such dislocations are identified by constructing bipartite graphs between the 5- and 7-fold defects, with an edge introduced only when the corresponding particles are Voronoi neighbors. The python package NetworkX was used for implementation \cite{SciPyProceedings_11}. Defect fractions are calculated as $n_{d}^{f} = N_{d}/N$, where $N$ is the total number of particles and $N_{d}$ represents the total number of defects. For instance, all particles with 5 or 7 neighbors, denoted by $N_{d}^{5}$ and $N_{d}^{7}$, respectively, are summed as the total number of defects $N_{d}^{\mathrm{tot}} = N_{d}^{5}+N_{d}^{7}$, when there are no other types of defects. 5-7 pairs formed by neighboring particles are identified as dislocations, where their total number is $N_{d}^{\mathrm{dislo}} = 2N_{d}^{5-7}$.   

Local structural order of each particle $i$ is captured by the hexatic order parameter $\psi_{6,i} (\mathbf{r}_{i}) =  \sum_{j=1}^{N_{i}} e^{i6\theta_{ij}}/{N_{i}}$, where particle $i$ has $N_{i}$ nearest neighbors obtained from Voronoi analysis. $\theta_{ij}$ represents the angle formed by the vector connecting particle $i$ to its neighbor $j$ and the $x$-direction. The magnitude of this complex order parameter represents local deviation of the crystal structure from hexagonal symmetry. Loss of hexagonal symmetry is indicated by $|\psi_{6,i}| < 1$, while a perfect hexagonal structure implies $|\psi_{6,i}| = 1$. The angular argument of $\psi_{6,i}$ quantifies local grain orientation $\Theta_{6,i}$ around particle $i$, where $\Theta_{6,i} = \mathrm{arg}(\psi_{6,i})/6$. Its value varies from $-\pi/6$ to $\pi/6$ due to the six-fold rotational symmetry of the hexagonal lattice.

\subsection*{Calculation of the structure factor}

\noindent We calculate the static structure factor for a given configuration from the particle positions as
$S(\mathbf q)=\left|\sum_{j=1}^{N}e^{-i\mathbf q\cdot\mathbf r_j}\right|^2/N$. To visualize the instantaneous reciprocal-space structure, we calculate $S(\mathbf q)$ on a uniform $500\times500$ grid in reciprocal space. The resulting $S(q_x,q_y)$ is plotted on a logarithmic scale to resolve both strong quasi-Bragg peaks and weaker diffuse scattering.

\subsection*{Calculation of polarization fluctuations}

\noindent The magnitude $P$ of the polar orientational order parameter quantifies the degree of collective alignment of the self-propulsion directions $\hat{\mathbf{e}}_i$ of the individual particles in the ABP crystal. It is defined as $ P = \left| \sum_{i=1}^{N} \hat{\mathbf{e}}_{i}/N\right|$. When collective migration has set in and orientational order has basically saturated as a function of time $t$, $P(t)$ saturates at values around $1$. We refer to these values as $P_{s}$. From simulations, either within the $l_p$- or $\tau_p$-controlled protocols, the probability densities of the temporal polarization fluctuations  $\delta P_{s} = P_{s} - \langle P_{s} \rangle$ are extracted from the temporal data of $P_{s}(t)$ during this orientational steady-state regime. Here $\langle P_{s} \rangle$ denotes the mean value. Each probability density is obtained from a histogram of $\delta P_{s}$ using 60 bins, with the bin counts normalized by the total number of samples and the bin width. To quantify the temporal polarization fluctuations in the polarized steady state, we calculate the polarization susceptibility as $\chi(P_{s})=N\,\mathrm{var}(P_{s})$. The resulting variances are averaged over four independent simulations.

\subsection*{Calculation of parameters for grain boundary shrinkage}

\noindent As mentioned in the main text, mechanisms regarding  grain boundary (GB) shrinkage are considered only for cases in which the structures of collectively migrating crystals have evolved into two dominant competing grains. In these cases, the smaller grain forms a GB loop. Subsequently, it shrinks completely, while defects anneal under either $l_p$- or $\tau_p$-controlled protocols. Here, we briefly discuss the algorithmic sequence of identifying and analyzing the shrinkage mechanism \cite{camerin2025disentangling, lavergne2019shrinkage}:
\begin{enumerate}
    \item First, we calculate the 
    local bond-orientational order around each particle $i$ from $\psi_{6,i}$ using $\Theta_{6,i} = \mathrm{arg}(\psi_{6,i})/6$.
    \item Next, we calculate the time evolution of the probability distribution of the bond orientations $P(\Theta_6)$.  Two dominant competing grains are identified from the two highest peaks in the $P(\Theta_6)$ distribution, further corroborated by the corresponding configurational snapshots.  
    \item Each $P(\Theta_{6})$ distribution is smoothed using one-dimensional Gaussian filters with a standard deviation of $\sigma=1.2^{\circ}$. Local maxima of the smoothed distributions are then identified, yielding their peak positions. They represent the average grain orientation. The smaller peak of the two distributions is associated with the smaller grain enclosed by the GB loop. We denote the average bond orientations of the particles inside the grain as $\Theta_{6}^{\mathrm{G}}$. Analogously, the dominant peak reflects the outer crystal matrix, with a mean orientation denoted as $\Theta_{6}^{\mathrm{C}}$. Peaks separated by less than a minimum angular separation of $\Delta \Theta_{\min}=2^{\circ}$ are treated as a single peak.
    \item The misorientation angle between the grain and the outer crystal matrix is therefore calculated as \cite{lavergne2018dislocation, lavergne2019shrinkage} $\Theta_{\mathrm{mis}} = \mathrm{min} (|\Theta_{6}^{\mathrm{G}}-\Theta_{6}^{\mathrm{C}}|, 60^{\circ} - |\Theta_{6}^{\mathrm{G}}-\Theta_{6}^{\mathrm{C}}|)$, with the bound $0^{\circ} < \Theta_{\mathrm{mis}} < 30^{\circ}$ due to the hexagonal symmetry of the crystal.
    \item Each particle $i$ is assigned to one of the two identified grain orientations, $\Theta_{6}^{\mathrm{G}}$ and $\Theta_{6}^{\mathrm{C}}$, according to the smaller angular distance between its local bond-orientational angle $\Theta_{6,i}$ and the two grain orientations. 
    \item Next, the number of particles within the two domains are determined. The radius of the GB loop is calculated using \cite{camerin2025disentangling} $R_{g} = a\sqrt{N_{g}/\pi}$. Here, $a$ is the lattice constant and $N_g$ is the number of particles inside the approximately circular grain.
    \item Finally, to identify the grain shrinkage mechanism, we extract the shear coupling factor \cite{cahn2006coupling} $\beta$. It is defined as $\beta = -d\Theta_{\mathrm{mis}}/d\mathrm{ln}(R_{g}/\sigma)$, with $\Theta_{\mathrm{mis}}$ expressed in radians and $\beta$ extracted from a linear fit.  

\end{enumerate}

\subsection*{Atomic active crystal model}

\noindent To show a counterexample of collectively migrating crystals without explicit orientational degrees of freedom for the individual self-propulsion forces, we adopt a minimal model for atomic active crystals comprised of point particles \cite{lobaskin2013collective}. In this framework, likewise, we start from a two-dimensional atomistic system of $N$ (=11074) particles of diameter $\sigma$. Initially, they are arranged on hexagonal lattices in a simulation box of dimensions 100$\sigma$ $\times$ 100$\sigma$. Again, they interact through the purely repulsive WCA potential, see Eq.~(\ref{eq:wca}), under periodic boundary conditions. The system is initially equilibrated at temperature $T = 10$ in units of $\epsilon/k_B$ with a time step of 0.001 in units of $\sqrt{m\sigma^2/\epsilon}$ for $10^7$ steps. This process is followed by an instantaneous quench to the desired bath temperature $T=0.025$. We further equilibrate the system for $10^7$ steps to obtain the defect-rich atomic passive crystal of lattice constant 1.02$\sigma$, in analogy to its ABP counterpart. Both passive and active atomic systems are thermostatted using  dissipative particle dynamics (DPD). 
We consider a simple, minimal model of velocity-dependent dissipative forces \cite{erdmann2000brownian} as the source of activity to the individual particles within the solid. 

For each particle $i$, the equation of motion can be written in the form
\begin{equation}\label{eq:atomic}
    m\dot{\mathbf{v}}_{i} ={} -\sum_{j\neq i}\bm{\nabla}_{i}U(r_{ij}) + \mathbf{F}_{i}^{a} + \sum_{j\neq i}\mathbf{F}_{ij}^{D} + \sum_{j\neq i}\mathbf{F}_{ij}^{R}.
\end{equation}
In this equation, $\mathbf{F}_{i}^{a} = f^{a} \hat{\mathbf{v}}_{i}$ is the active force on each particle $i$ along its current direction of velocity $\hat{\mathbf{v}}_{i} =\mathbf{v}_i/|\mathbf{v}_i|$, with magnitude $f^{a}$. The pairwise dissipative and random forces follow the DPD equations of motion, where the dissipative force $\mathbf{F}_{ij}^{D}$ is given by 
\begin{equation}\label{eq:diss}
   \mathbf{F}_{ij}^{D} ={} - \gamma_s \,\omega^{2}(r_{ij})(\hat{\mathbf{r}}_{ij}\cdot \mathbf{v}_{ij})\hat{\mathbf{r}}_{ij}.
\end{equation}
Here, $\gamma_s$ denotes the pairwise friction coefficient, controlling the strength of dissipative momentum transfer to the heat bath. It introduces a characteristic timescale $\tau_\mathrm{d}=m/\gamma_s$. Here, $\gamma_s$ and $\tau_\mathrm{d}$ are expressed in units of $\sqrt{m\epsilon}/\sigma$ and $\sqrt{m\sigma^2/\epsilon}$, respectively. $\mathbf{v}_{ij}$ is the relative velocity between particles $i$ and $j$, while $r_{ij}$ is their mutual distance. Moreover, $\mathbf{r}_{ij}=\mathbf{r}_i-\mathbf{r}_j$ and $\hat{\mathbf{r}}_{ij}=\mathbf{r}_{ij}/|\mathbf{r}_{ij}|$. $\omega(r_{ij}) = 1-{r_{ij}}/{ 2^{1/6}\sigma}$ is a weight factor, varying between 0 and 1. Besides, we add the random forces $\mathbf{F}_{ij}^{R}$, which are defined as 
\begin{equation}\label{eq:random}
   \mathbf{F}_{ij}^{R} = \sqrt{2\gamma_{s}k_{B}T} \omega(r_{ij})\mathbf{\xi}_{ij}( t)\hat{\mathbf{r}}_{ij}.
\end{equation}
Again, $\mathbf{\xi}_{ij} (t)$ is a random variable with Gaussian distribution of zero mean and  variance $\langle\xi_{ij}(t)\xi_{mn}(t')\rangle=\delta_{im}\delta_{jn}\delta(t-t')$, where we measure $\xi_{ij}$ in units of $\sqrt[4]{\epsilon/m\sigma^2}$. 

For our simulations, first we fix the dissipative coupling $\gamma_{s}$ and the timescale $\tau_\mathrm{d}$ to 1. We vary the strength of the active force $f^a$ from 0.1 to 20 (hence Pe from 4 to 800). Under strong activity at constant packing and temperature, the crystal undergoes transient melting with the transient loss of hexatic order, followed by the recovery of crystalline order. Higher $f^a$ induces larger velocities and drives the system quickly towards a collectively migrating state. It reduces the time available for dissipative interactions per distance to facilitate coordinated reorganization. Thus, dissipating the relative velocity differences between the particles, such systems reach steady defect-rich collectively migrating crystal states. At such steady states, the kinetic temperature $T_{\mathrm{kin}}$ reaches the desired value of $ 0.025$. It is defined as \cite{hecht2024define}
\begin{equation}\label{eq:tkin}
   T_{\mathrm{kin}} = \frac{m}{Nk_{B}\,d} \sum_{i=1}^{N}  \left\langle (\mathbf{v}_{i} - \langle \mathbf{v} \right\rangle)^{2} \rangle, 
\end{equation}
where $d$ is the dimension and $\langle \mathbf{v} \rangle$ is the center-of-mass velocity.

To study defect annealing, we therefore focus on low activity values of $f^{a} = 0.1$ (hence Pe = 4). Although lower values of $f^{a}$ can also induce transient melting of the crystal, they allow more time for dissipative relaxation to facilitate coordinated particle rearrangements. This is necessary for defect annealing during recrystallization. We test strong and weak coupling scenarios by implementing $\gamma_{s} =1$, $\tau_\mathrm{d} = 1$ and $\gamma_{s}=0.0714$, $\tau_\mathrm{d} = 14$,  respectively. Defect annealing of the active crystal during weak coupling is observed. Overall velocity polarization is computed as $P_{\mathrm{v}} =  \left\langle \left|\sum_{i=1}^N \mathbf{v}_i/|\mathbf{v}_{i}|\right| \right\rangle\!/N$. A typical simulation of an atomic active crystal runs  with a time step of 0.0001 for $2\times 10^{7}-10^8$ steps. For each parameter set, we perform 3 independent simulations for ensemble averaging.

%

\begin{acknowledgments} \label{ackn}
\noindent  We thank Henning Reinken, Muhammad Muhsin Abdul Azeez, and Mrinal Sarkar for useful discussions. 
\end{acknowledgments}

\section*{Author contributions} \label{sec7}

\noindent A.M.M.\ secured funding. K.G.\ performed the study, including developing and performing the necessary steps, numerical simulations, and necessary analyses. K.G.\ designed the manuscript. K.G.\ and A.M.M.\ revised the manuscript. A.M.M supervised the project. 

\section*{Funding} \label{sec8}

\noindent We thank the  Deutsche Forschungsgemeinschaft (German Research Foundation, DFG) for support of this work through project no.\ 541972050.

\section*{Competing interests} \label{sec9}
\noindent The authors declare no competing interests.

\end{document}


\title{Tunable annealing of defects in self-aligning active crystals:  \\
Supplementary Information}

\author{Kanka Ghosh\,\orcidlink{0000-0003-1360-7699}}
 \affiliation{Otto-von-Guericke-Universit\"at Magdeburg, Institut f\"ur Physik, Universit\"atsplatz 2, 39106 Magdeburg, Germany}
\author{Andreas M. Menzel\,\orcidlink{0000-0003-0713-4979}}%
\affiliation{Otto-von-Guericke-Universit\"at Magdeburg, Institut f\"ur Physik, Universit\"atsplatz 2, 39106 Magdeburg, Germany}%

\begin{abstract}
This Supplementary Information addresses further aspects concerning the poly-to-single-crystal transformation of self-aligned active Brownian crystals. We examine stability, finite-size effects and orientation correlations associated with defect-annealed systems. Sample-to-sample variations in post-MIPS crystals and their role at intermediate persistence times in the persistence time-controlled protocol are discussed. Additional polarization analysis includes the activity dependence of the mean polar order, identification of the polarization-susceptibility minimum in the collectively migrating state, polarization relaxation times, and spatiotemporal angular fluctuations. At the grain-boundary scale, all different grain misorientations of the grain boundary loops associated with the shrinkage process are presented. Multi-grain scenario as well as the final stage of dislocation interaction preceding complete defect annihilation are shown via representative examples.
\end{abstract}

\maketitle

\subsection*{Supplementary Note 1: Example of a completely defect-free single-crystal resulting from a defect-rich poly-crystal}

\noindent The poly-to-single-crystal transformation of active crystals, described in the main text, mostly features one or two remaining dislocations or vacancies in the final state. However, given a longer simulation time, crystals can reach a completely defect-free state. In Supplementary Figure~\ref{fig:defect_free}, we show such an example of a completely defect-free active single-crystal, obtained from a defect-rich poly-crystal. This example corresponds to Pe = 640 ($l_{p} \approx$ 213.33), and $\tau_{p} = 3.33$. Similarly to that of Fig.~3 in the main text, we show the velocity variance $\delta v^ 2=\sum_{i=1}^N (\mathbf{v}_{i}-\langle \mathbf{v} \rangle)^{2}/N $ of the particles. Again, we distinguish between the contributions arising from the mean squared velocity $\langle v^2\rangle$ and the squared mean velocity $ |\langle \mathbf{v} \rangle|^{2}$. All of these are normalized by the mean squared velocity of the passive systems $\langle v^{2} \rangle_{\mathrm{passive}}$ at the same value of $D_t=0.1$. 
Starting from a defect-rich state, the ratio $\delta v^2/\langle v^{2} \rangle_{\mathrm{passive}}$ initially rises to a peak fluctuation state, associated with MIPS. Afterwards, it relaxes towards the passive baseline $\delta v^2/\langle v^{2} \rangle_{\mathrm{passive}} = 1$. Eventually, it drops further below. See Supplementary Figure~\ref{fig:defect_free}(a). Particle configurations corresponding to five progressive time instances are marked as \textbf{A}, \textbf{B}, \textbf{C}, \textbf{D}, and \textbf{E} in Supplementary Figure~\ref{fig:defect_free}(a). These states are presented via  local hexatic order maps $|\psi_{6, i}|$ and local grain orientation maps $\Theta_{6,i}$ in Supplementary Figure~\ref{fig:defect_free}(b) and (c), respectively. At $t = 246.5$, the active system is seen to reach a perfect single-crystalline order, without any remaining defects, see panels \textbf{E} in Supplementary Figure~\ref{fig:defect_free}(b,c).

\begin{figure*}[!htb]
\centering
\includegraphics[width=\linewidth]{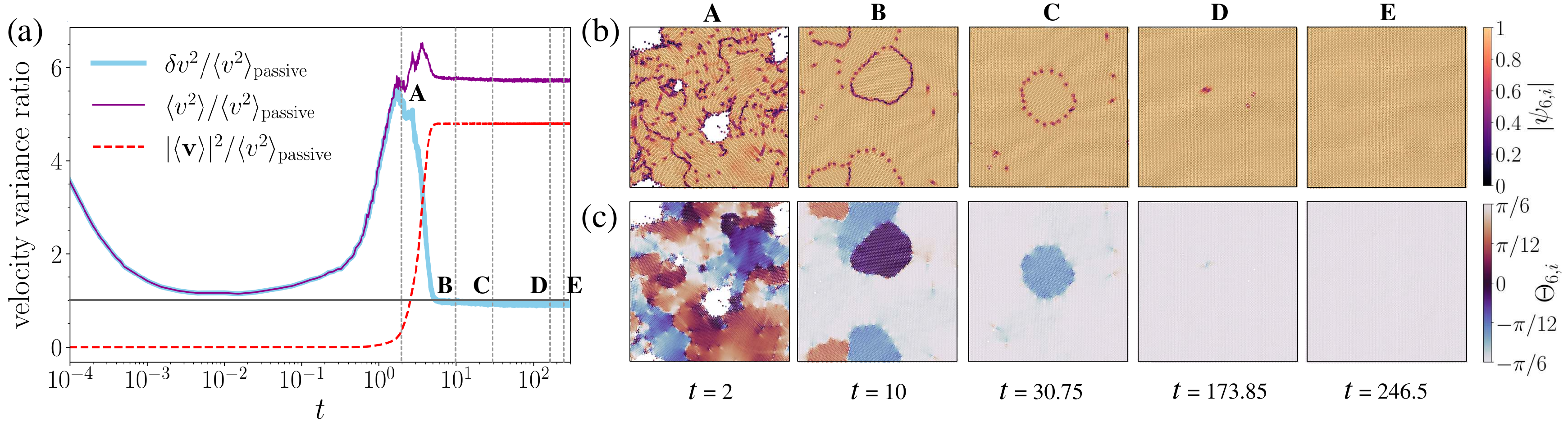}
\caption{\textbf{Demonstration of the evolution of a completely defect-free active single-crystal obtained from a defect-rich poly-crystal.} \textbf{(a)} Time evolution of the velocity variance $\delta v^ 2$ is presented together with the contributions from the mean squared velocity $\langle v^2\rangle$ and the squared mean velocity $ |\langle \mathbf{v} \rangle|^{2}$. All quantities are normalized by the mean squared velocity of the passive systems $\langle v^{2} \rangle_{\mathrm{passive}}$ at identical value of $D_t$. This example situation corresponds to Pe = 640 ($l_{p} \approx 213.33$) and $\tau_{p} =3.33$. \textbf{(b)} Particle configurations corresponding to different time instances $t = 2$ (\textbf{A}), $t =10$ (\textbf{B}), $t =30.75$ (\textbf{C}), $t =173.85$ (\textbf{D}), and $t = 246.5$ (\textbf{E}). They are marked along the velocity variance curve in {(a)}. The panels in (b) display maps of the local hexatic order $|\psi_{6, i}|$. \textbf{(c)} The same configurations are presented using maps of the local grain orientation $\Theta_{6,i}$. In both {(b)} and {(c)}, panels \textbf{E} present an example of the resulting single-crystalline state without any defect.}
\label{fig:defect_free}
\end{figure*}

\subsection*{Supplementary Note 2: Stability of the defect-free active crystal under local disturbance}

\noindent In Supplementary Figure~\ref{fig:stability}, we demonstrate the stability of the obtained defect-free active crystal under local disturbance. We consider an example situation of poly-to-single-crystal transformation at Pe = 1000 ($l_{p} \approx 333.33$) and $\tau_{p} =3.33$. At $t = 150$, the active crystal reaches the defect-free state. At this point, we reintroduce a local perturbation by randomizing particle orientations on a localized region of radius $10\sigma$ as shown in Supplementary Figure~\ref{fig:stability}(a) via $|\psi_{6, i}|$- and in Supplementary Figure~\ref{fig:stability}(e) via $\theta_i$-maps. From $t= 150.025$ to $t = 300$, the configuration snapshots in Supplementary Figure~\ref{fig:stability}(a--d) and (e--h), show the progressive annealing of defects and restoration of structural order in the active crystal. This numerical test suggests the robustness of the obtained defect-free active crystals, indicating their ability to sustain local deformation.

\begin{figure*}[!htb]
\centering
\includegraphics[width=17cm]{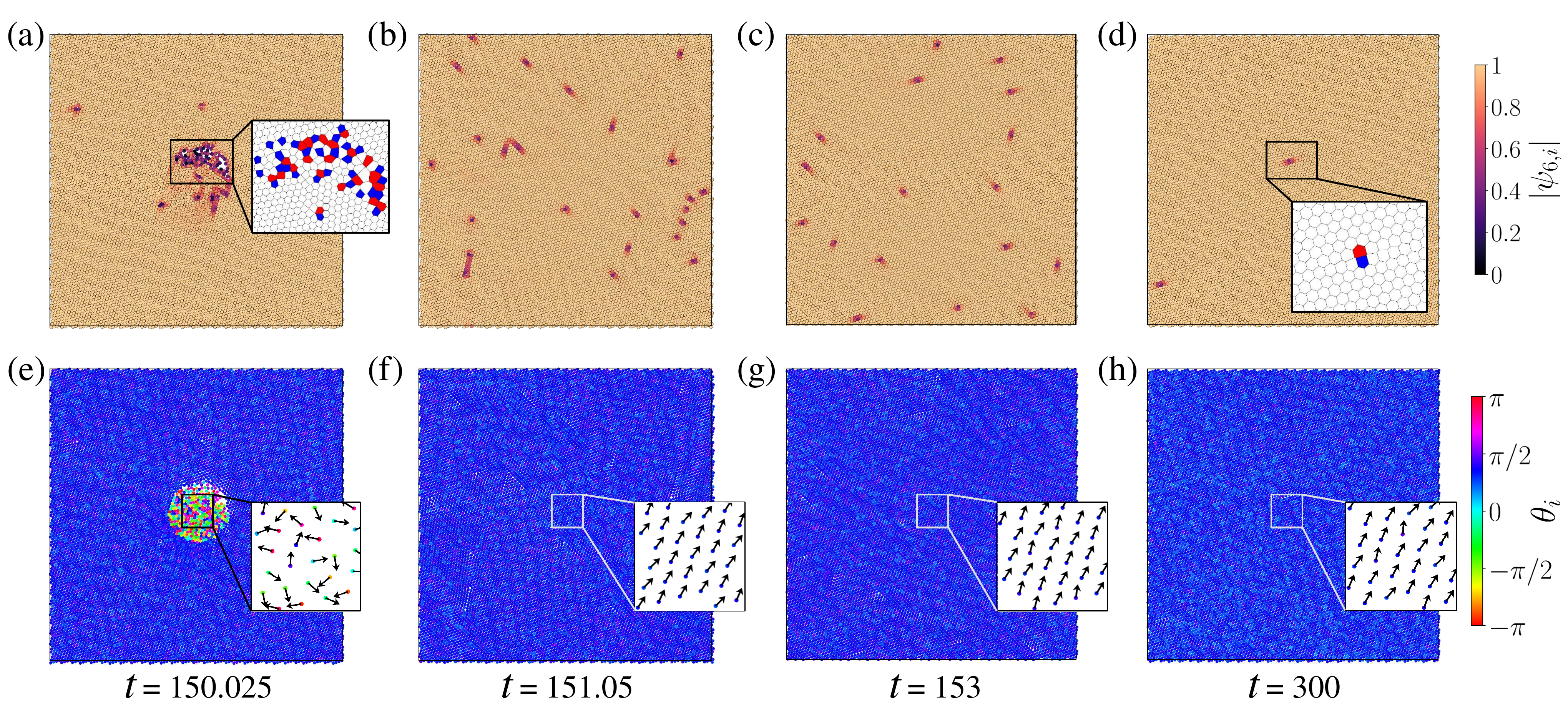}
\caption{\textbf{Stability of the defect-free active crystal under local disturbance.} \textbf{(a)} A local disturbance is introduced into a defect-free collectively migrating crystal at $t=150$, by randomizing particle orientations. The inset visualizes corresponding defects shown via Voronoi-neighbor maps, indicating a resurgence of 5- and 7-fold defects. \textbf{(b--d)} Gradual disappearance of the introduced defects, illustrated using maps of the local hexatic order parameter $|\psi_{6, i}|$. Time progresses from (b) $t = 151.05$ via (c) $t=153$ to (d) $t = 300$. \textbf{(e--h)} Gradual transformation back to a defect-free state, here illustrated using maps of the angles of particle orientation $\theta_i$, together with zoomed-in parts. This example corresponds to Pe = 1000 ($l_{p} \approx 333.33$) and $\tau_{p} =3.33$, with a simulation domain of dimension $100\sigma \times 100\sigma$ at a fixed packing fraction $\phi$ = 0.87.}
\label{fig:stability}
\end{figure*}

\subsection*{Supplementary Note 3: System-size effects on defect annealing}

\begin{figure*}[p]
\centering
\includegraphics[width=16cm]{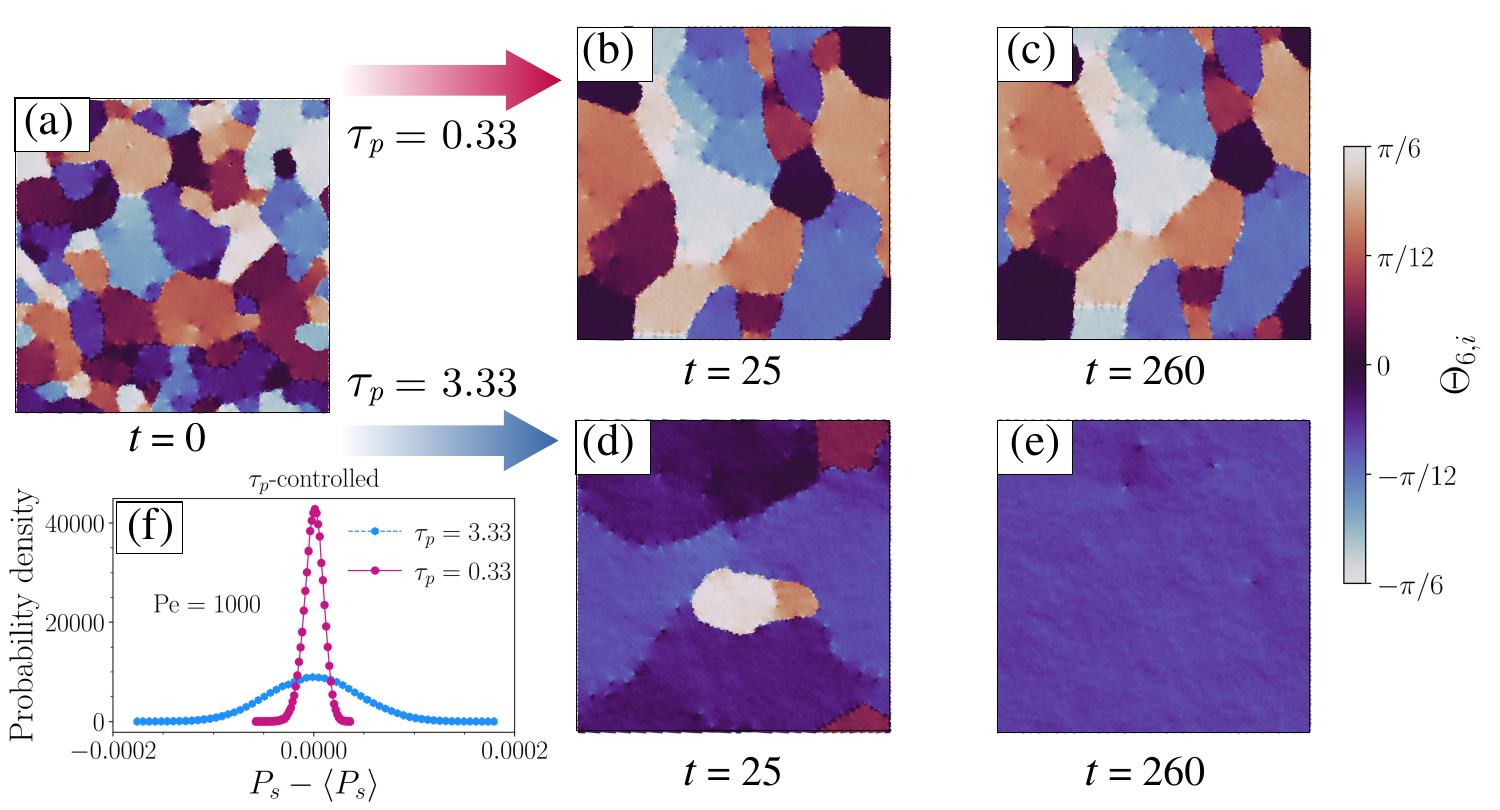}
\caption{\textbf{System-size effect on defect annealing in $\tau_p$-controlled protocols with $N = 44183$ particles.} The dimension of the simulation box is $200\sigma \times 200\sigma$. \textbf{(a)} Initial, defect-rich poly-crystalline active crystal at $t=0$. Grains are identified using colormaps of local particle orientations $\Theta_{6,i}$. \textbf{(b,c)} Time evolution for low persistence time $\tau_p=0.33$ at two progressive time instances $t = 25$ and $t=260$. \textbf{(d,e)} Same for a high value of $\tau_p=3.33$, with the defect-free collectively migrating active crystal shown in (e). \textbf{(f)} Probability density of the fluctuations of the steady-state polar orientational order parameter $P_s$, centered around its mean $\langle P_{s}\rangle$, for the same two values of $\tau_p$ at constant Pe = 1000 ($l_{p} \approx 333.33$). 
}
\label{fig:defect_free_200}
\end{figure*}

\begin{figure*}[p]
\centering
\includegraphics[width=16cm]{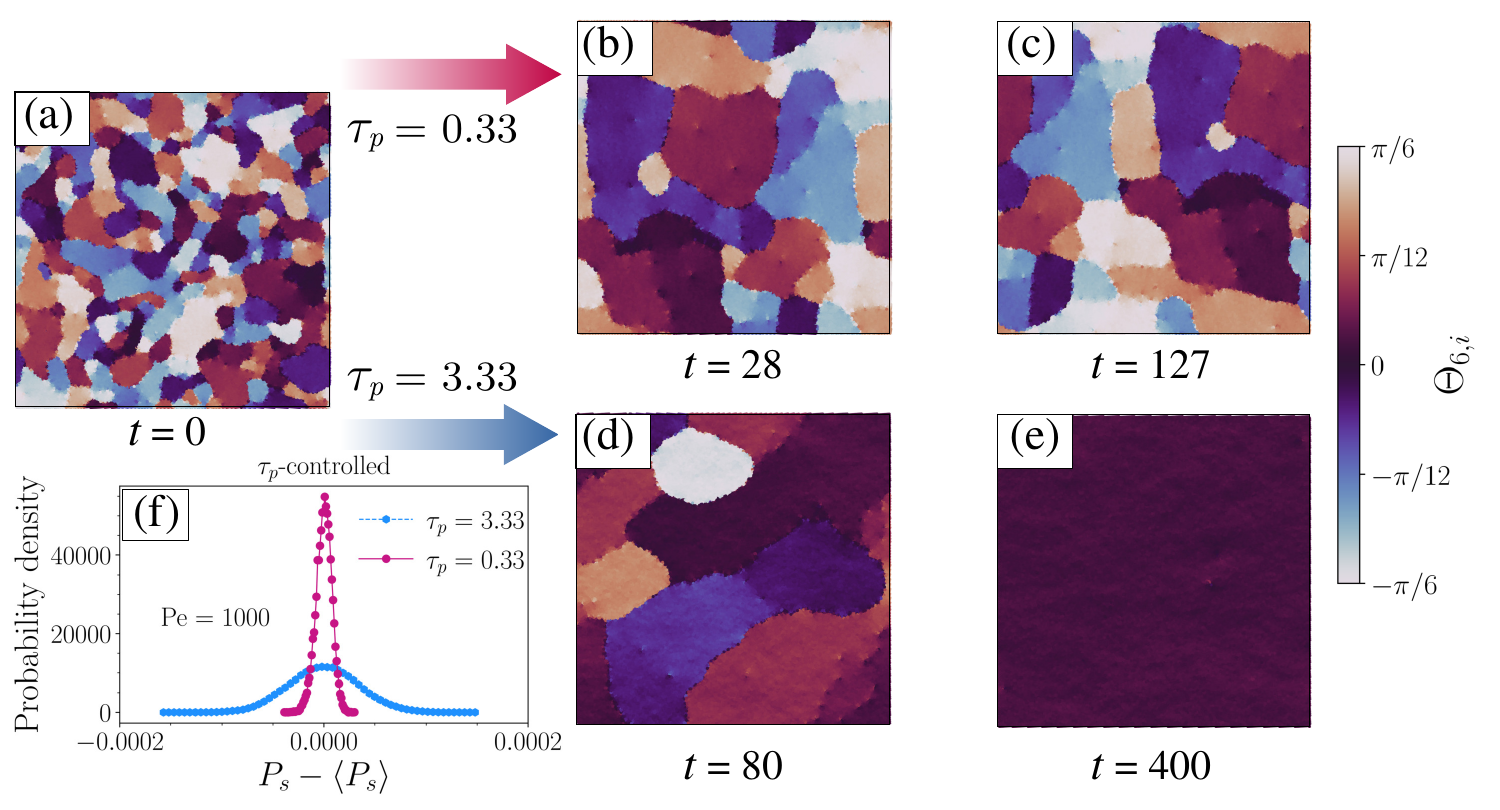}
\caption{\textbf{System-size effect on defect annealing in $\tau_p$-controlled protocols with $N = 99327$ particles.} Here, the size of the simulation box is $300\sigma \times 300\sigma$. \textbf{(a)} Initial, defect-rich poly-crystalline active crystal at $t=0$. Again, grains are identified using maps of the particle orientation angles $\Theta_{6,i}$. \textbf{(b,c)} Time evolution for low persistence time $\tau_p=0.33$, illustrated at time instances $t = 28$ and $t=127$. \textbf{(d,e)} Same for high $\tau_p=3.33$, yet at times $t = 80$ and $t=400$. A defect-free state is observed at $t=400$ in (e). \textbf{(f)} Same probability density as in Supplementary Figure~\ref{fig:defect_free_200}(f), but now for the further increased system size of $300\sigma \times 300\sigma$. 
}
\label{fig:defect_free_300}
\end{figure*}

\noindent Here, we discuss the effect of system size on our central result: tunable defect-annealing under persistence time $\tau_p$-controlled protocols, enabled by polarization fluctuations in the collectively migrating state. The results described in the main text are obtained from simulations of domain size $100\sigma \times 100\sigma$, containing $N = 11074$ particles. In Supplementary Figures~\ref{fig:defect_free_200} and \ref{fig:defect_free_300} we demonstrate defect annealing in $\tau_p$-controlled protocols at Pe = 1000 ($l_p \approx 333.33$) for larger system sizes, namely $200\sigma \times 200\sigma$ with $N = 44183$ and $300\sigma \times 300\sigma$ with $N = 99327$, respectively. The packing fraction is fixed at $\phi$ = 0.87, same as that of the system of dimension $100\sigma \times 100\sigma$. We specifically focus on two cases, $\tau_p = 0.33$ and $\tau_p = 3.33$. They correspond to low and high persistence times, respectively. Consistently, higher $\tau_p$ is seen to enable defect annealing. It yields a defect-free active crystal. Conversely, a defect-rich poly-crystalline state is obtained at low $\tau_p$. The progressive transformation of the configurations at two  time instances is illustrated, see Supplementary Figure~\ref{fig:defect_free_200}(b--e) for the system size $200\sigma \times 200\sigma$ and  Supplementary Figure~\ref{fig:defect_free_300}(b--e) for the system size $300\sigma \times 300\sigma$.  At $\tau_p = 0.33$,  configurations emerge as  poly-crystals, see the multiply oriented grains shown via $\Theta_{6,i}$-maps in  Supplementary Figure~\ref{fig:defect_free_200}(c) and in Supplementary Figure~\ref{fig:defect_free_300}(c). Single-crystalline states are observed at $\tau_p = 3.33$, exhibiting uniform values of $\Theta_{6,i}$, see Supplementary Figures~\ref{fig:defect_free_200}(e) and \ref{fig:defect_free_300}(e). Moreover, Supplementary Figures~\ref{fig:defect_free_200}(f) and \ref{fig:defect_free_300}(f) present the corresponding probability densities of the temporal fluctuations of the steady-state polar orientational order parameter $P_s$, centered around its mean $\langle P_{s}\rangle$. It is shown for the two values of $\tau_p$ mentioned above at constant Pe = 1000 ($l_{p} \approx 333.33$). In consistency with our findings in the main text, higher $\tau_p$ broadens the spatiotemporal polarization fluctuations in the collectively migrating crystal, see the blue curves in both Supplementary Figures~\ref{fig:defect_free_200}(f) and \ref{fig:defect_free_300}(f). This supports structural defect annealing. Our considerations indicate that the key result and the underlying mechanisms remain unaltered by the modified system sizes.

\subsection*{Supplementary Note 4: Orientational correlation and positional order of resulting defect-free systems}

\noindent To distinguish between solid and hexatic character of the resulting defect-free active systems, we calculate the corresponding two-point spatial correlation function $g_{6}(r)$ of the local hexatic order parameter $\psi_{6,i}$, defined as 
\begin{equation}\label{eq:g6}
    g_{6}(r) = \left\langle \frac{\sum_{i\neq j}\delta (r-\mid \mathbf{r}_{i} - \mathbf{r}_{j} \mid ) \psi_{6,i}^{*}(\mathbf{r}_{i}) \psi_{6,j}(\mathbf{r}_{j})}{\sum_{i\neq j}\delta (r-\mid \mathbf{r}_{i} - \mathbf{r}_{j} \mid )} \right\rangle .
\end{equation}
In this expression, $\langle...\rangle$ accounts for the average over several configurations at various timesteps, once the system has reached single-crystalline order. We use radial bins of width $dr = 0.1$ to evaluate Eq.~(\ref{eq:g6}). As a consequence of the Mermin-Wagner-Hohenberg theorem \cite{mermin1966absence, hohenberg1967existence, mermin1968crystalline}, two-dimensional crystalline phases in equilibrium can only exhibit quasi-long-range positional order at finite density. 

Solid phases can be identified by an algebraic decay of positional order together with long-range bond-orientational order. Conversely, the equilibrium hexatic phase is usually characterized by short-range positional order with quasi-long-range bond-orientational order. 
In the hexatic phase, bond-orientational order decays algebraically. The exponent of this decay is termed $\eta_6$, with a value of $0 <\eta_6< 1/4$ in the hexatic phase (compared to $\eta_6\approx0$ in the solid phase). 

Importantly, two-dimensional active solids have been distinguished from the hexatic phase using these two criteria \cite{klamser2018thermodynamic, paliwal2020role, shi2023extreme}.  However, positional order is found to decay with much larger exponents beyond the equilibrium BKTHNY theory \cite{shi2023extreme}.

\begin{figure}[h]
\centering
\includegraphics[width=17.5cm]{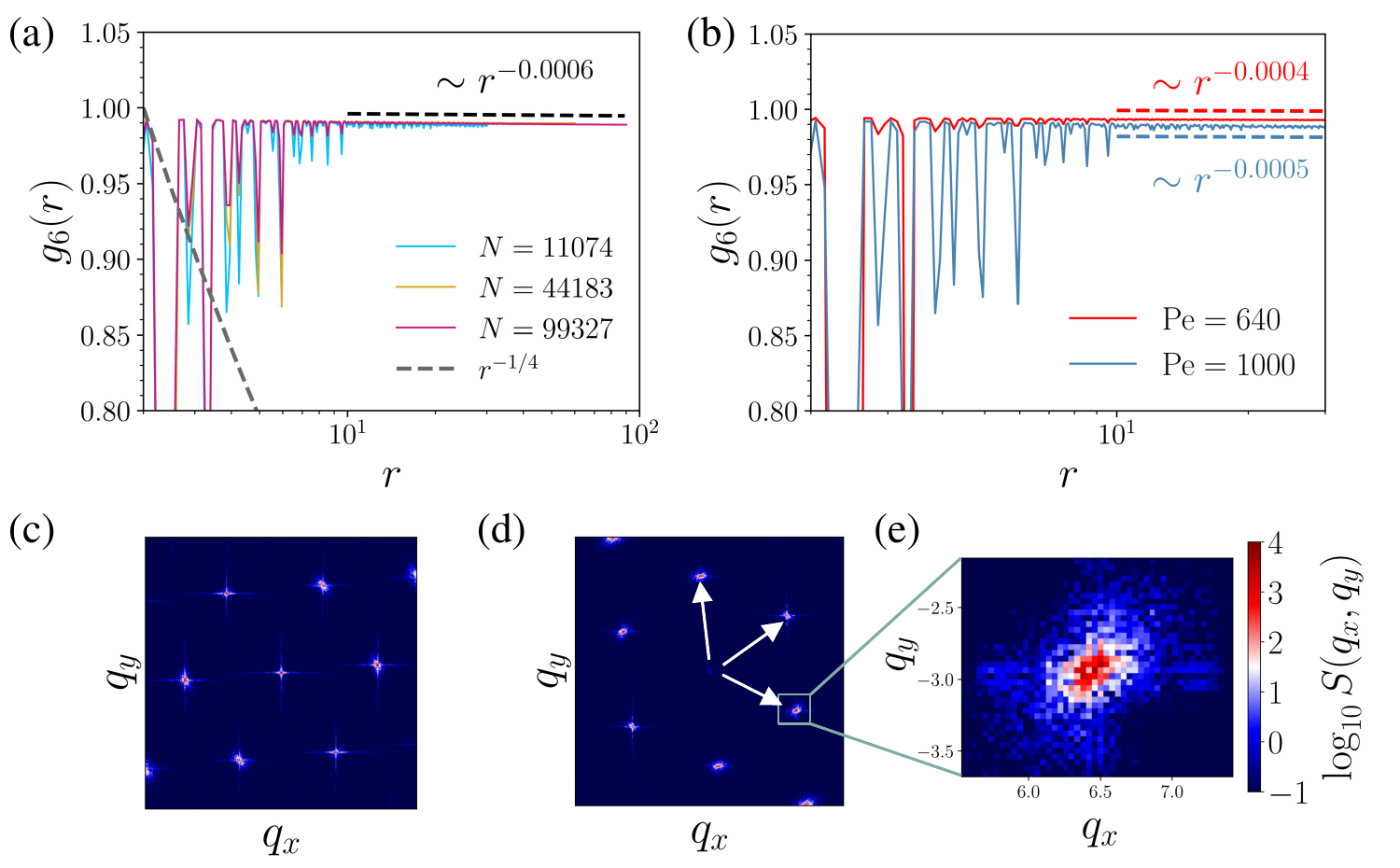}
\caption{\textbf{Bond-orientational correlation and positional order of defect-(quasi-)free states.} \textbf{(a)} Bond-orientational-order correlation functions $g_{6}(r)$ are presented for three different system sizes $N = 11074$, $N = 44183$, and $N = 99327$. Scaling is evaluated as $g_6(r)\sim r^{-\eta_6}$ with the scaling exponent $\eta_6$. The top dashed line indicates scaling as evaluated for the largest system size, $\eta_6 = 0.0006$ $(\eta_6\approx0)$, see also the Supplementary text for details. Similarly, the left dashed line indicates scaling for a hexatic phase according to the equilibrium BKTHNY theory at maximum possible algebraic decay, $\eta_6 = 1/4$. \textbf{(b)} $g_{6}(r)$ is shown for two active single-crystals, one completely without any remaining defect at Pe = 640, the other one with very few remaining defects at Pe = 1000. In both cases, the exponents $\eta_6 = 0.0004$ and 0.0005 ($\eta_6\approx0$) suggest long-range bond-orientational order, independently of the presence or absence of few leftover defects. \textbf{(c,d)} Maps of the instantaneous static structure factor $S(\mathbf{q})$, corresponding to the resulting single-crystals for the two system sizes (c) $N = 11074$ and (d) $N = 99327$, to illustrate positional order. White arrows in (d) indicate three out of six reciprocal lattice vectors. \textbf{(e)} Zoomed-in view of one of the first quasi-Bragg peaks shown in (d), exhibiting noticeable broadening.}
\label{fig:bond_order}
\end{figure}

We remark that extracting reliable positional scaling exponents poses challenges due to the strong oscillatory nature of the correlation function, when calculating it without reference to any perfect lattice, see Ref.~\cite{shi2023extreme} for details.
Yet, correlations of bond-orientational order are found to be comparatively less fluctuating. Therefore, we only focus on these orientational correlations to distinguish our resulting defect-free active crystals from the hexatic phase. We emphasize that the purpose of our study is not to compute accurate scaling exponents of the positional and bond-orientational correlation functions to characterize melting. 

Supplementary Figure~\ref{fig:bond_order}(a) presents the bond-orientational-order correlation function $g_6(r)$ for three different systems of dimensions $100\sigma\times100\sigma$ ($N=11074$), $200\sigma\times200\sigma$ ($N=44183$), and $300\sigma\times300\sigma$ ($N=99327$). The obtained exponents are $\eta_6 = 0.00051 \pm 0.00019$, $0.00049 \pm 0.00001$, and $0.00064 \pm 0.00001$ for $N = 11074$, $44183$, and $99327$, respectively, where we included standard errors of the fitting process. All of them indicate long-range bond-orientational order ($\eta_6 \approx 0$) consistent with the solid phase. To analyze the effect of remaining defects in the obtained single-crystal, we compare two active systems, see Supplementary Figure~\ref{fig:bond_order}(b). The first one does not contain any remaining defect, realized at Pe = 640. The other contains very few remaining defects at Pe = 1000. Again, in both cases, the obtained exponents $\eta_6 = 0.0004$ and $\eta_6=0.0005$ suggest a long-range bond-orientational order ($\eta_6\approx0$), independently of the presence of few leftover defects in the single-crystal.

Supplementary Figure~\ref{fig:bond_order}(c) and (d) present the effect of system size on the positional order. We show instantaneous maps of the structure factor $S(\mathbf{q})$ for $N = 11074$ and $N = 99327$, respectively. Here, only the reciprocal spaces containing first neighbors are shown. These first quasi-Bragg peaks are seen to retain six-fold symmetry, regardless of the system size. Supplementary Figure~\ref{fig:bond_order}(e) shows a magnified view of one of the first quasi-Bragg peaks. Individual peaks are found to exhibit broadening, consistent with observations reported in previous studies \cite{shi2023extreme} and as discussed in the main text.

\subsection*{Supplementary Note 5: Sample-to-sample variations in post-MIPS active crystals}

\begin{figure*}[!htb]
\centering
\includegraphics[width=\linewidth]{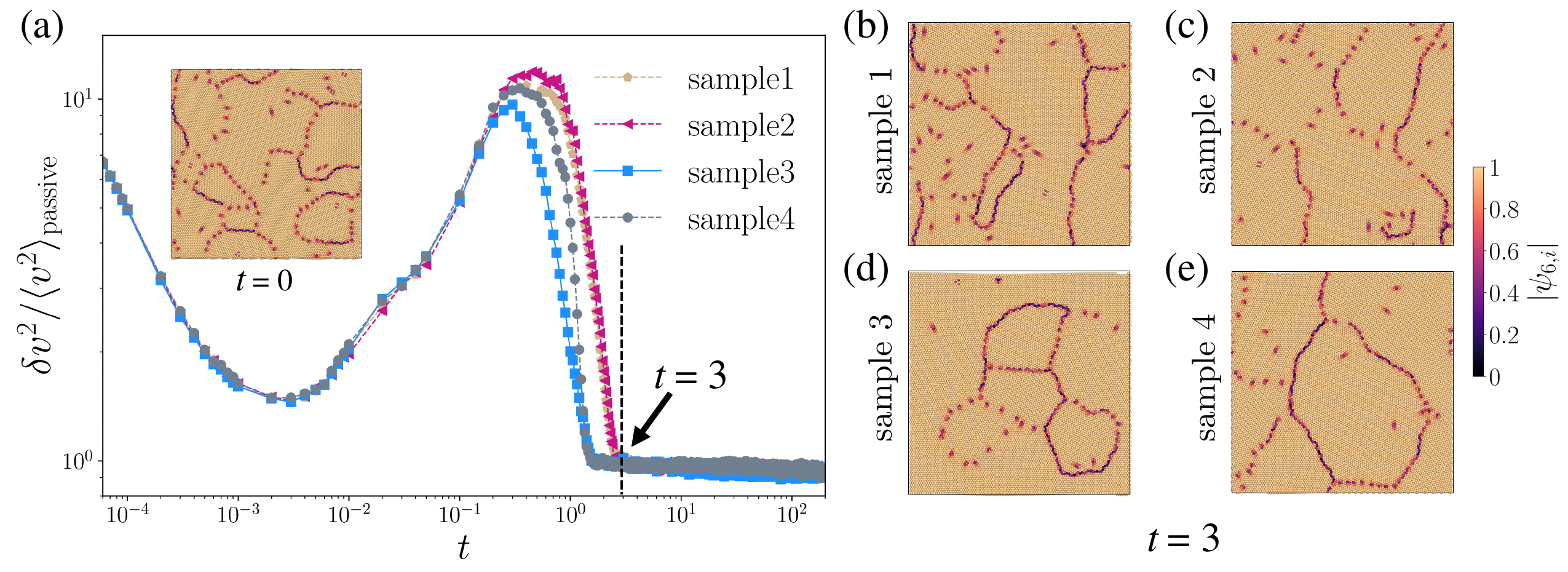}
\caption{\textbf{Examples of sample-to-sample variations in post-MIPS states.} \textbf{(a)} Time evolution of the velocity variance $\delta v^ 2$, normalized by the mean squared velocity of the corresponding passive systems $\langle v^{2} \rangle_{\mathrm{passive}}$, for four different samples. The inset shows the initial defect-rich system using a map of the degree of local hexatic order $|\psi_{6, i}|$. We here set Pe = 1000 ($l_{p} \approx 333.33$) and $\tau_{p} =1.667$. \textbf{(b--e)} Illustrations of particle configurations corresponding to the four samples introduced in (a) at time $t = 3$, using the degree of local hexatic order $|\psi_{6, i}|$.}
\label{fig:sample_var_mips}
\end{figure*}

\noindent Supplementary Figure~\ref{fig:sample_var_mips} illustrates examples of sample-to-sample variations in the post-MIPS regime of the dynamics. The samples represent independent realizations using the same parameter set at Pe = 1000 ($l_{p} \approx 333.33$) and $\tau_{p} =1.667$. As we can see from Supplementary Figure~\ref{fig:sample_var_mips}(a), the variations in  $\delta v^2 /\langle v^{2}\rangle_{\mathrm{passive}}$ for different samples start to develop notably around their peak fluctuation states, which correspond to MIPS. Consequently, all four samples follow different relaxation paths towards the passive baseline ratio $\delta v^2 /\langle v^{2}\rangle_{\mathrm{passive}} \sim 1$. This substantial sample-to-sample variation in the relaxation dynamics results in the large error bars observed in this relaxation regime in Fig.~4 of the main text. As a result, the post-MIPS emergent poly-crystals show different defect networks, as depicted in Supplementary Figure~\ref{fig:sample_var_mips}(b--e). They correspond to the four different samples.

\subsection*{Supplementary Note 6: Variation of average polar order in $l_p$- and $\tau_p$-controlled protocols}

\noindent Here, we show the dependence of the average degree of global polar orientational order $\langle P_{s}\rangle$ on Pe (hence $l_p)$ and  on $\tau_p$ in $l_p$- and $\tau_p$-controlled protocols, respectively. At fixed $\tau_p = 3.33$, $\langle P_{s}\rangle$ progressively increases from 0 towards 1 with increasing Pe, see Supplementary Figure~\ref{fig:polarsat}(a). The active crystals exhibit a collectively migrating state with $\langle P_{s}\rangle > 0.5 $ for $\mathrm{Pe} \geq 60 $ (hence $l_p \gtrapprox 20$), whereas for $\mathrm{Pe} < 60 $ (hence $l_p < 20$), they remain essentially unpolarized with $\langle P_{s}\rangle \approx 0$. On the contrary, in $\tau_p$-controlled protocols at various fixed Pe (hence fixed $l_p$) $\langle P_{s}\rangle$ depends significantly more weakly on $\tau_p$, see Supplementary Figure~\ref{fig:polarsat}(b). Even at the lowest considered Pe = 160 ($l_p \approx 53.33$), where this variation is most prominent, $\langle P_{s}\rangle$  changes only from 0.99 to 0.91 within the considered range of $\tau_p$. Thus, the degree of collective alignment is found to be primarily dependent on Pe or, equivalently, on the persistence length $l_p$.
\begin{figure}[h]
\centering
\includegraphics[width=16cm]{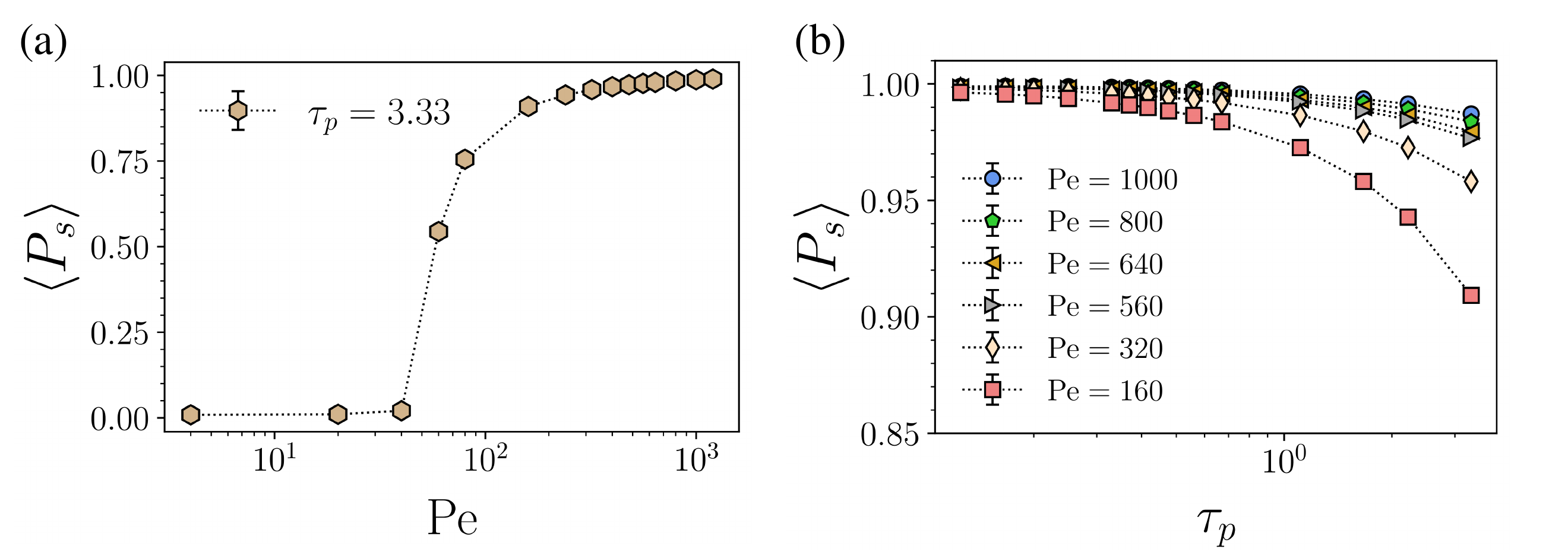}
\vspace{-.1cm}
\caption{\textbf{Variation of average polar orientational order with Pe and persistence time $\tau_p$.} \textbf{(a)} Average saturation of the degree of global polar orientational order $\langle P_s \rangle$ as a function of Pe (hence $l_p$) in $l_p$-controlled protocols at fixed $\tau_p = 3.33$. \textbf{(b)} Variation of the same degree of polar orientational order $\langle P_s \rangle$ with $\tau_p$ in $\tau_p$-controlled protocols at various fixed values of Pe ranging from 160 ($l_p \approx 53.33$) to 1000 ($l_p \approx 333.33$). All depicted data points result from averaging over four independent realizations. Error bars are smaller than the marker sizes.}
\label{fig:polarsat}
\end{figure}

\subsection*{Supplementary Note 7: Additional details on polarization relaxation times}

\noindent In Supplementary Figure \ref{fig:polar_relax_taup}, we provide additional details regarding the variation of polarization relaxation times $t_s$ with persistence time $\tau_p$, as described in Fig.~5 of the main text. We present $t_s$ as a function of $\tau_p$ for various values of Pe, ranging from 160 ($l_{p} \approx 53.33$) to 1000 ($l_{p} \approx 333.33$) under $\tau_p$-controlled protocols. All curves follow the same trend. As shown in Fig.~5 of the main text, varying $\tau_p$ enables nearly two decades of variation in $t_s$. At smaller $\tau_p$, the values of $t_s$ for different Pe nearly collapse onto a common trend. Thus, $\tau_p$-controlled protocols provide substantial control over relaxation times from fast to slow, independently of the persistence length.
\begin{figure*}[!htb]
\centering
\vspace{-.0cm}
\includegraphics[width=8cm]{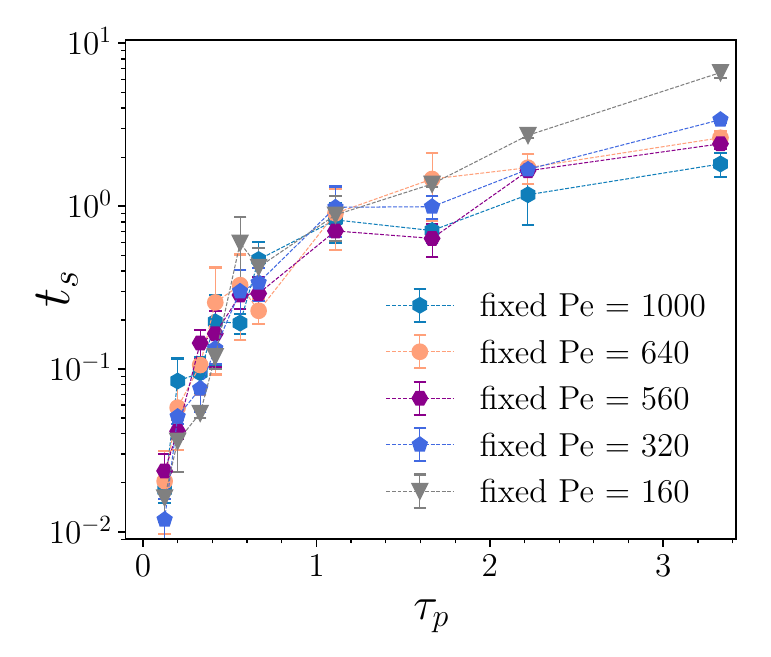}\vspace{-.3cm}
\caption{\textbf{Polarization relaxation time $t_s$ in $\tau_{p}$-controlled protocols for various fixed Pe (fixed $l_p$).} The polarization relaxation time $t_s$ is presented as a function of persistence time $\tau_p$, at various fixed values of Pe ranging from 160 to 1000 (hence fixed persistence lengths $l_{p}$ from $\approx 53.33$ to $\approx 333.33$). Error bars represent standard errors of the mean across 4$-$5 independent realizations. For each of these cases, every value of $t_s$ was extracted by fitting the global polarization curve $P(t)$ by a compressed exponential function.}
\label{fig:polar_relax_taup}
\end{figure*}

\subsection*{Supplementary Note 8: Sample-to-sample variations at intermediate $\tau_p$-values in $\tau_p$-controlled protocols}

\noindent The sample-to-sample variations in post-MIPS states, as discussed above in Supplementary Note 5, result in variations in the annealing time of the defects. Thus the number of remaining defects present in these systems can vary for identical finite simulation times. Specifically, this concerns $\tau_p$-controlled protocols at intermediate $\tau_p$ values. 

\begin{figure}[h]
\centering
\includegraphics[width=\linewidth]{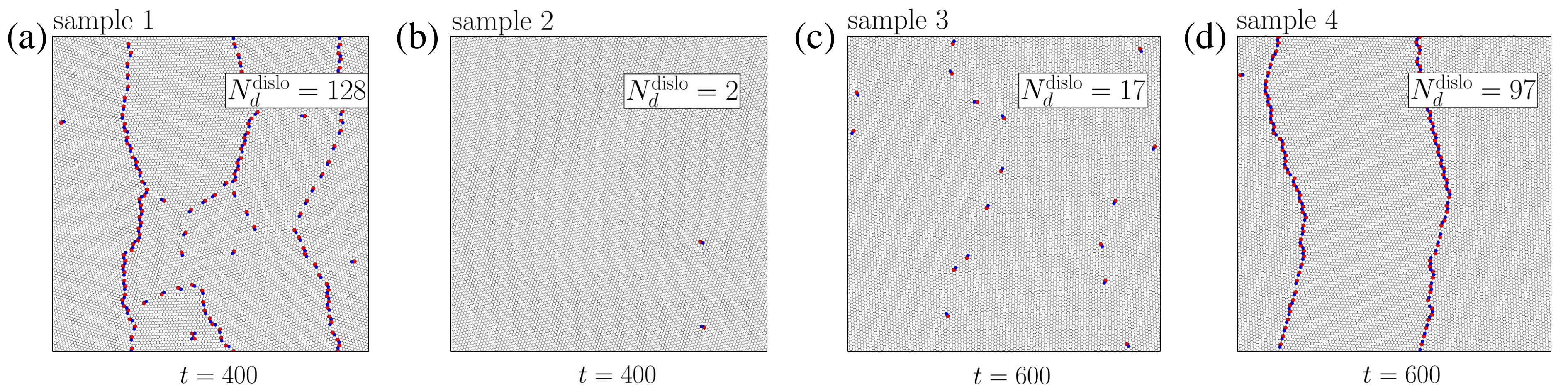}
\caption{\textbf{Sample-to-sample variations in the number of remaining defects at intermediate $\tau_p$-values.} We follow $\tau_p$-controlled protocols at Pe = 1000 ($l_{p} \approx 333.33$) and $\tau_{p}  = 1.11$. The four configuration snapshots correspond to four independent realizations at the end of the corresponding simulation times. Remaining numbers of dislocations $N_{d}^{\mathrm{dislo}}$ are displayed for each of these final states, determined from maps of Voronoi neighbors. \textbf{(a, b)} Resulting active crystals for samples 1 and 2 at $t = 400$. \textbf{(c, d)} Resulting active crystals for samples 3 and 4 at $t = 600$. The remaining numbers $N_{d}^{\mathrm{dislo}}$ of defects markedly differ between the different realizations at identical times.}  
\label{fig:intermediate}
\end{figure}

We again consider four different samples realized at identical parameter sets but different initializations. Supplementary Figure~\ref{fig:intermediate}(a--d) shows snapshots of the configurations from four independent realizations at the end of the corresponding simulation time. The number of remaining dislocations $N_{d}^{\mathrm{dislo}}$ are found to be 128 and 2 at $t = 400$ for samples 1 and 2, respectively, see Supplementary Figure~\ref{fig:intermediate}(a,b). Similarly,  Supplementary Figure~\ref{fig:intermediate}(c,d) show quite different numbers of dislocations $N_{d}^{\mathrm{dislo}} = 17$ and $N_{d}^{\mathrm{dislo}} = 97$ at $t = 600$ for samples 3 and 4, respectively. These variations amount to considerably large errorbars in the steady-state defect fractions in Fig.~6 of the main text at intermediate values of $\tau_p$. They indicate that even longer simulation times would be needed to completely anneal the defects in  samples 1 and 4.  

\subsection*{Supplementary Note 9: Instantaneous and temporal angular variances in $\tau_{p}$-controlled protocols}

\noindent In Supplementary Figure~\ref{fig:angular_variance}, we further elucidate the fluctuations of angular orientations of the active particles as a function of the persistence time $\tau_p$. To this end, from the particle orientations $\theta_i$, we calculate two parameters, namely the instantaneous and the temporal angular variances. The instantaneous angular variance $\sigma_{\theta}^2 (t)$ captures the heterogeneity in particle orientations at a particular time. It is defined as
\begin{equation}
    \sigma_{\theta}^2 (t) = \frac{1}{N} \sum_{i=1}^N \left[\theta_{i} (t) - \langle \theta(t)\rangle_i\right]^2,
\end{equation}
where $\langle \theta(t) \rangle_i$ denotes the instantaneous average collective direction of active driving, calculated using the circular mean $\langle \theta(t) \rangle_i = \mathrm{atan2}[\langle \mathrm{sin}\theta_i\rangle_i, \langle \mathrm{cos}\theta_i\rangle_i]$ over all particles $i$. Recalling that orientations of the different particles are linked to their distributed positions in space, this quantity detects spatial angular heterogeneity. 

However, in the collectively migrating state, the spatial heterogeneity of the orientations is not fixed, rather fluctuates in time. This leads to the second quantity of interest, namely temporal angular variance. It is defined as $\mathrm{var}_t[\sigma_{\theta}^2 (t)]$, where $\mathrm{var}_t$ denotes the temporal variance. Thus, this quantity measures the temporal fluctuations of the spatial distribution of particle orientations. It provides a second-order temporal statistics of the evolving orientational heterogeneity. 

Supplementary Figure~\ref{fig:angular_variance}(a) shows the temporal variance as a function of persistence time $\tau_p$ at two fixed P\'eclet numbers, Pe = 560  ($l_p \approx 186.67$) and Pe = 1000 ($l_p \approx 333.33$). We recover a trend similar to that of the polarization susceptibility curves as a function of $\tau_p$, see Fig.~7(d) of the main text. For the curve for Pe = 1000 in Supplementary Figure~\ref{fig:angular_variance}(a), the three points marked as \textbf{A}, \textbf{B}, and \textbf{C} correspond to $\tau_p = 0.125$, $0.33$, and $3.33$, respectively. These three values of $\tau_p$ capture the observed overall nonmonotonic trend in the temporal fluctuations of orientations. Supplementary Figure~\ref{fig:angular_variance}(b--d) presents explicit illustrations of the time evolution of instantaneous angular variances, subtracting their temporal average $\langle \sigma_{\theta}^2 (t)\rangle_t$, evaluated as $\delta \sigma_{\theta}^2 (t) = \sigma_{\theta}^2 (t) - \langle \sigma_{\theta}^2 (t)\rangle_t$. Here $\langle...\rangle_t$ denotes temporal average. 
\begin{figure}[h]
\centering
\includegraphics[width=17cm]{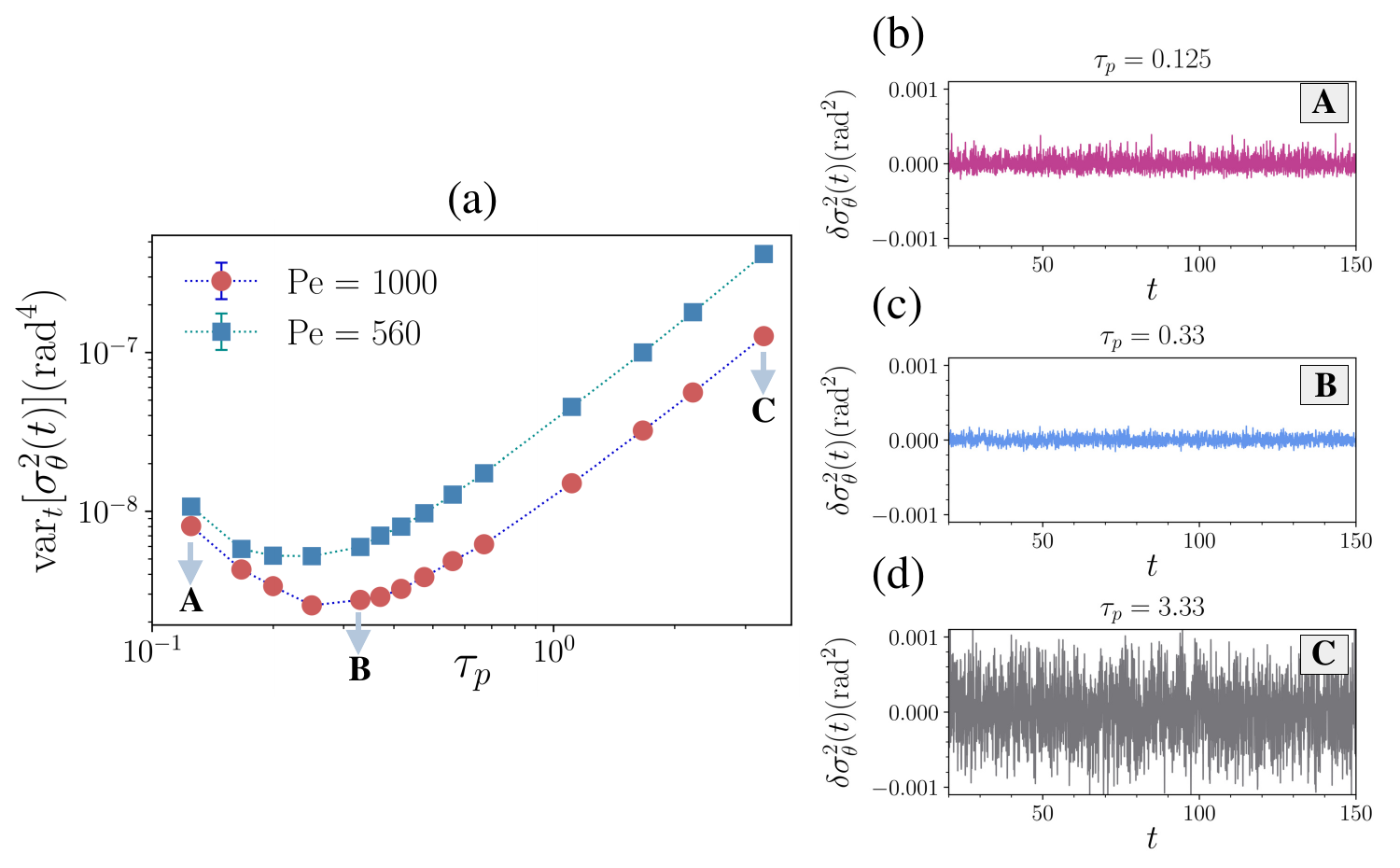}
\caption{\textbf{Instantaneous and temporal angular variances in $\tau_{p}$-controlled protocols.} \textbf{(a)} Temporal variances of the angular fluctuations $\mathrm{var}_t[\sigma_{\theta}^2 (t)]$ of the active particles are presented as a function of persistence time $\tau_p$ in $\tau_p$-controlled protocols at two fixed P\'eclet numbers, namely Pe = 560 ($l_p \approx 186.67$) and Pe = 1000 ($l_p \approx 333.33$). All data points are obtained from averaging over four independent realizations. Error bars are smaller than the marker sizes. Three data points on the curve for Pe = 1000, corresponding to $\tau_p = 0.125$, $0.33$, and $3.33$, are marked by \textbf{A}, \textbf{B}, and \textbf{C}, respectively. \textbf{(b--d)} Time series for the instantaneous angular variances $\sigma_{\theta}^2 (t)$ after subtracting its temporal average $\langle \sigma_{\theta}^2 (t)\rangle_t$, evaluated as $\delta \sigma_{\theta}^2 (t) = \sigma_{\theta}^2 (t) - \langle \sigma_{\theta}^2 (t)\rangle_t$. The three panels correspond to the three $\tau_p$-values marked as \textbf{A}, \textbf{B}, and \textbf{C} in (a). (b--d) present data for one representative realization.}
\label{fig:angular_variance}
\end{figure}

Among the depicted examples, we can visually evidence minimum temporal fluctuations at $\tau_p = 0.33$, see Supplementary Figure~\ref{fig:angular_variance}(c). Maximum temporal fluctuations of the particle orientations in these three cases occur at $\tau_p = 3.33$, see Supplementary Figure~\ref{fig:angular_variance}(d). These results further illustrate the mechanism of efficient defect annihilation in the collectively migrating state owing to enhanced spatiotemporal polarization fluctuations, as described in the main text.

\begin{figure}[h]
\centering
\includegraphics[width=10cm]{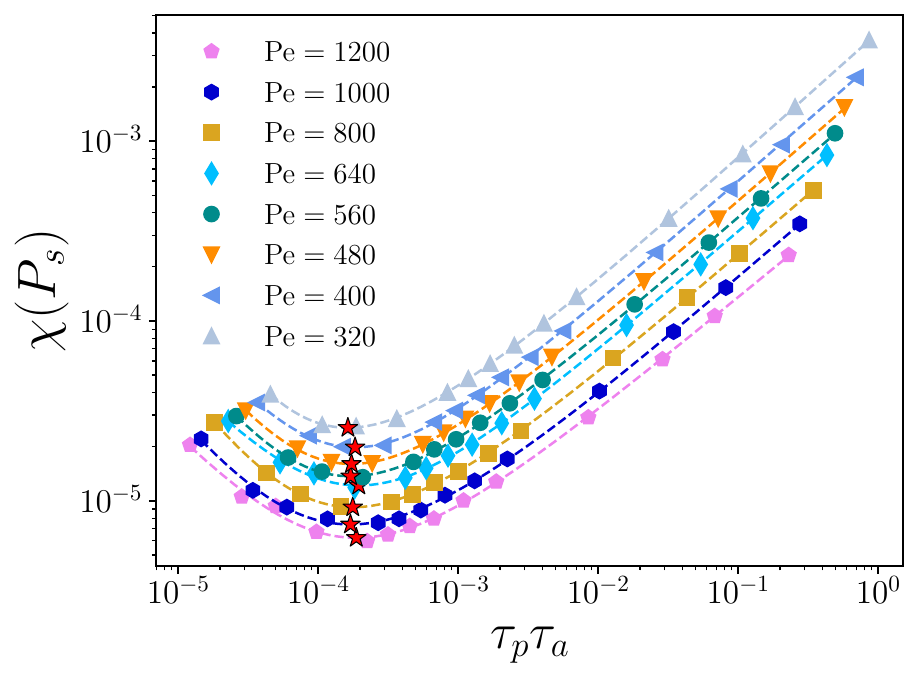}
\vspace{-.2cm}
\caption{\textbf{Identifying minima of the polarization susceptibility curves.} Polarization susceptibilities $\chi(P_s)$  corresponding to $\tau_p$-controlled protocols are presented as a function of the combined persistence-alignment timescale $\tau_p \tau_a$ for various constant values of Pe (hence $l_p$). Each data point was averaged over four independent realizations of the system. Error bars are smaller than the marker size. The corresponding minima discussed in the Supplementary text are marked as red stars on these curves.}
\label{fig:tau_min_sm}
\end{figure}
\vspace{-0.1cm}
\begin{figure*}[b]
\centering
\includegraphics[width=\linewidth]{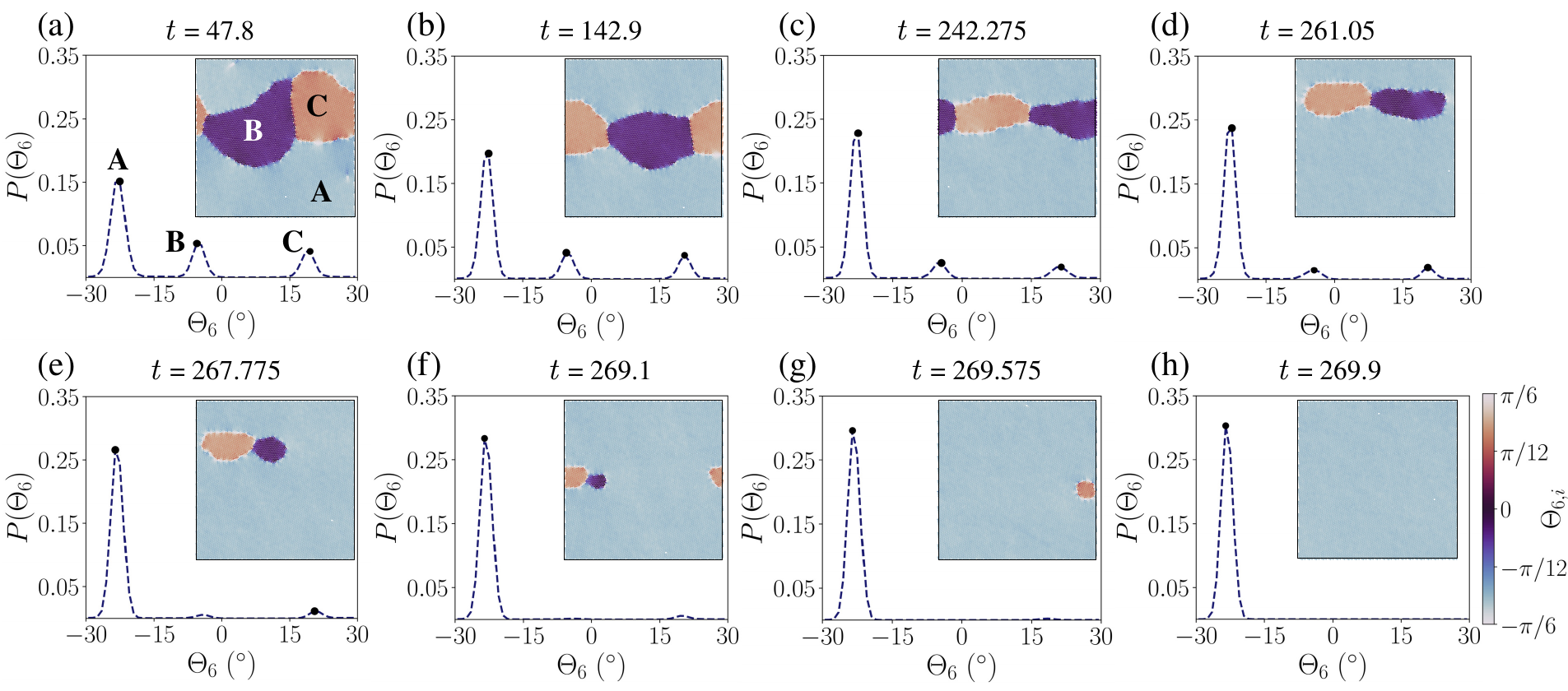}
\caption{\textbf{Multi-grain scenario before single-crystal formation.}  
Progressive time evolution of the probability distribution of local grain orientation $P(\Theta_6)$ preceding single-crystal formation. Areas of three different orientations corresponding to the peaks in $P(\Theta_6)$ are marked as \textbf{A}, \textbf{B}, and \textbf{C} in (a). While $\textbf{B}$ and \textbf{C} represent two differently oriented grains, \textbf{A} denotes the outer crystal matrix. Insets present the corresponding configurational snapshots as maps of the local orientation angle $\Theta_{6,i}$. They illustrate gradual grain shrinkage of the two grains \textbf{B} and \textbf{C}, accompanied by complete defect annihilation. This example was obtained for Pe = 240 ($l_p = 80$) and $\tau_p = 3.33$.}
\label{fig:multigrain_sm}
\end{figure*}
\begin{figure}[h]
\centering
\includegraphics[width=15.5cm]{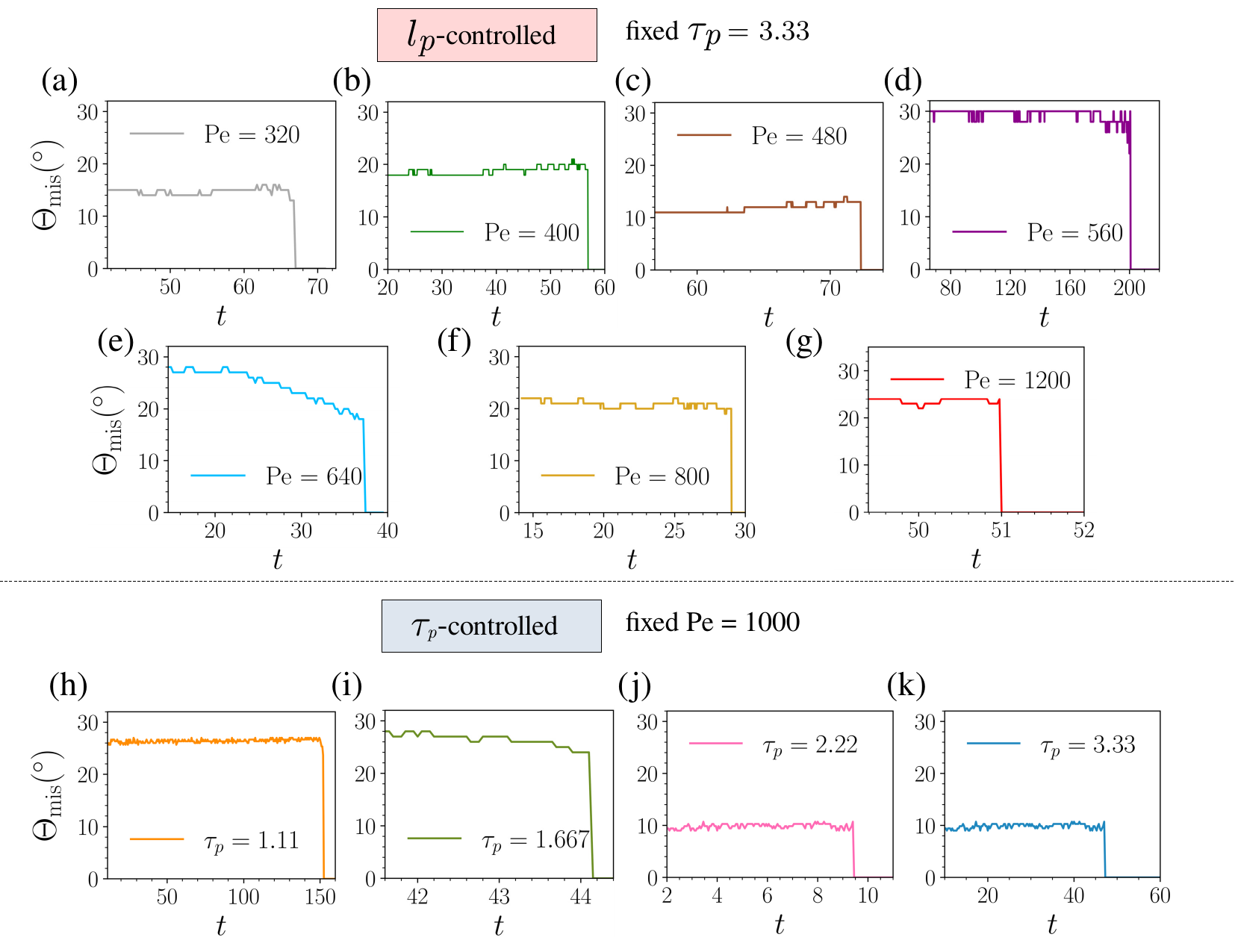}
\caption{\textbf{Time evolution of grain misorientation during grain shrinkage.} \textbf{(a--g)} Time evolution of the misorientation angle $\Theta_{\mathrm{mis}}$ between the grain and the surrounding crystal matrix are presented for $l_p$-controlled protocols at fixed $\tau_{p} = 3.33$.  \textbf{(h--k)} Same for $\tau_{p}$-controlled protocols at fixed Pe = 1000 ($l_p\approx333.33$).}
\label{fig:misorientation_sm}
\end{figure}

\subsection*{Supplementary Note 10: Extracting minima from polarization susceptibility curves in collectively migrating active crystals}

\noindent Here, we provide additional details on obtaining the minima $\tau_c^{*2}$ of the curves for the polarization susceptibility $\chi(P_s)$ in the collectively migrating state. We refer to Fig.~7(h) of the main text. Supplementary Figure~\ref{fig:tau_min_sm} presents $\chi(P_s)$ as a function of the combined persistence-alignment timescale $\tau_p \tau_a$ for various constant Pe (hence fixed $l_p$) in $\tau_p$-controlled protocols. As was discussed in the main text, $\tau_a = \gamma_r / \zeta v_0 = {3}(k_B T/\sigma^2) \tau_p ^2/{\zeta f^a}$. Thus, when we fix in rescaled units in addition to the temperature the parameter $\zeta = 0.01$ and set the constant $f^a$ for each of these curves, we obtain $\tau_a \sim \tau_p^2$. The nonmonotonic variations of $\chi(P_s)$ as a function of $\tau_p \tau_a$ were fitted by $A(\tau_{p}\tau_{a})^\alpha + B(\tau_{p}\tau_{a})^{-\beta} + C$, from which the minima can be found as $\tau_{c}^{*2} = \tau_p^{*}\tau_a^{*} = ({B\beta}/{A\alpha})^{{1}/{(\alpha + \beta)}}$. These minima are marked as red stars on the curves shown in Supplementary Figure~\ref{fig:tau_min_sm}.

\subsection*{Supplementary Note 11: Additional details on grain shrinkage and defect annihilation}

\noindent Beyond the single-grain scenario that we here mostly focused on, Supplementary Figure~\ref{fig:multigrain_sm} demonstrates an example of a multi-grain system at the last stage of grain shrinkage, before final transformation into a single-crystal. We show the time evolution of the active crystal using probability distributions of local grain orientation $P(\Theta_6)$. As insets, we include corresponding configuration snapshots as maps of the orientation angle $\Theta_{6,i}$. Even in the final stages preceding single-crystal formation, the two grains \textbf{B} and \textbf{C} continue to shrink simultaneously to the benefit of the growing crystal matrix \textbf{A}. Ultimately, \textbf{A} determines the orientation of the resulting active single-crystal.

Next, we return to only one remaining grain. We show the time evolution of the misorientation angles $\Theta_{\mathrm{mis}}$ between the approximately circular grains and the outer crystal matrices for each of the points in Fig.~8(e,f) of the main text. Corresponding time variations of the misorientation angles are shown for both $l_p$- and $\tau_p$-controlled protocols in Supplementary Figure~\ref{fig:misorientation_sm}(a--g) and in Supplementary Figure~\ref{fig:misorientation_sm}(h--k), respectively.

\begin{figure*}[h]
\centering
\includegraphics[width=\linewidth]{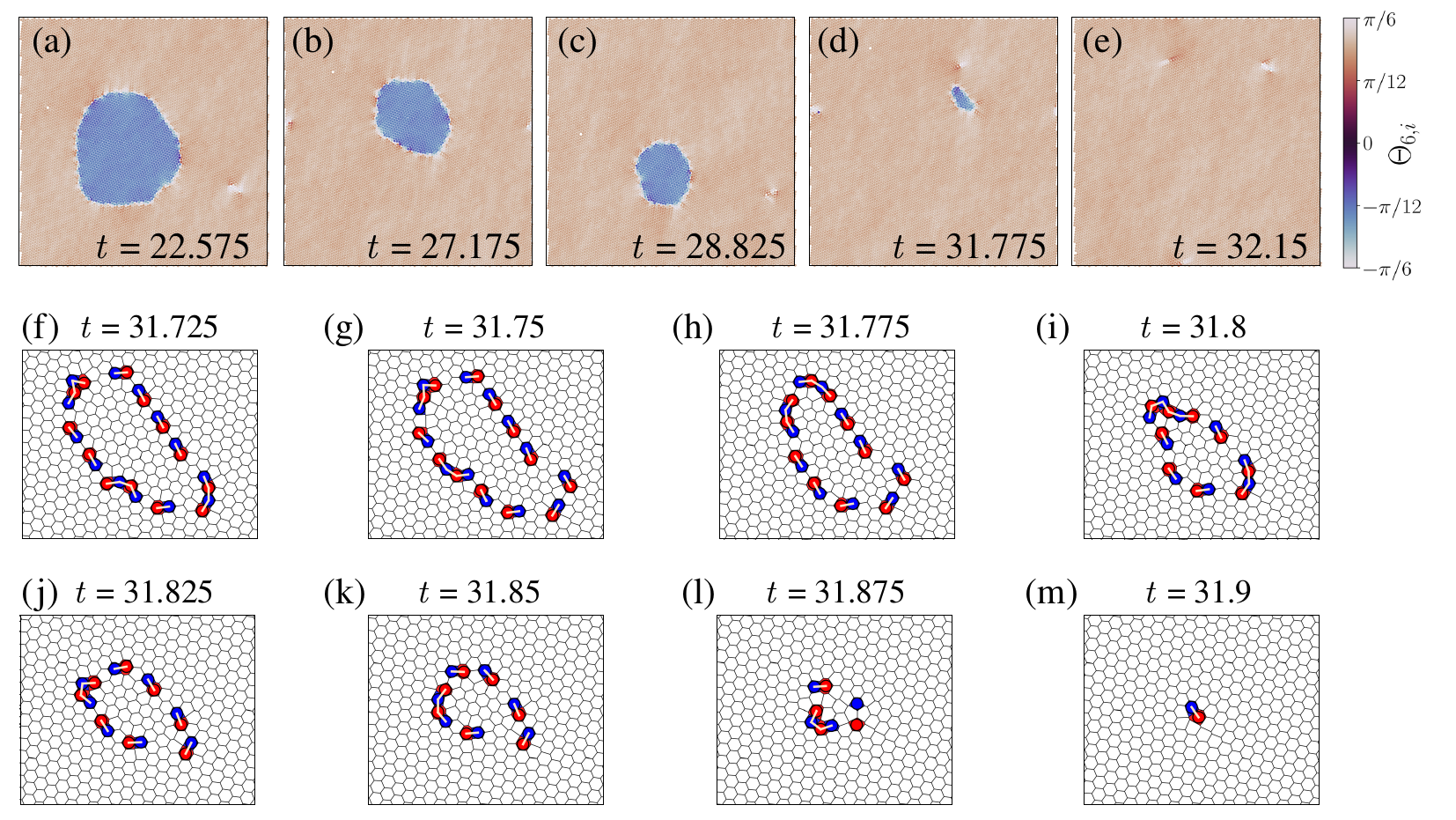}
\caption{\textbf{Dislocation annihilation associated with grain shrinkage.} \textbf{(a--e)} Example of a particle configuration containing one grain, undergoing grain shrinkage. Particle configurations are represented by maps of local grain orientations $\Theta_{6,i}$. \textbf{(f--m)} Same example scenario, now illustrating progressive annihilation of the dislocations immediately preceding complete defect annihilation and thus disappearance of the misoriented grain, using Voronoi-neighbor maps. An active, collectively migrating single-crystal results. Parameters were set to Pe = 800 ($l_p \approx 266.67$) and $\tau_p = 3.33$.}
\label{fig:dislocation_sm}
\end{figure*}

Finally, Supplementary Figure~\ref{fig:dislocation_sm}(a--e) demonstrates an example of particle configurations containing a single grain and its progressive shrinkage at the last stage of defect annihilation. During the final part of the process, Supplementary Figure~\ref{fig:dislocation_sm}(f--m) tracks the interactions between the dislocations constituting the grain-boundary loop and their eventual annihilation. From each 5-fold (blue colored particles) to the nearest 7-fold (red colored particles) defect, the shortest line is drawn. These lines identify dislocation pairs. In a typical dislocation-annihilation event, interacting dislocations induce small particle displacements in addition to collective migration. They rearrange the local neighborhood and recover the hexagonal coordination of the active crystal.
\\


%